\newif\ifShowKeys
\ShowKeystrue

\documentclass[letterpaper,10pt]{article}

\pdfoutput=1
\usepackage[no-natbib-sort]{my-jheppub}

\usepackage{amsmath, amssymb}

\usepackage{bm,environ,mathrsfs,array,arydshln,booktabs,float,slashed,hyperref}
\usepackage{tensor} 	

\usepackage{wick}

\usepackage[nodayofweek]{date time}

\usepackage{graphicx,epsfig}
\usepackage{epic}
\usepackage{tikz} 
\usetikzlibrary{decorations.pathreplacing,decorations.markings,snakes}

\tikzset{middlearrow/.style={
        decoration={markings,
            mark= at position 0.5 with {\arrow{#1}} ,
        },
        postaction={decorate}
    }
}

\usepackage{aurical}
\usepackage[T1]{fontenc}

\usepackage{framed}
\definecolor{shadecolor}{RGB}{255, 230, 204}

\allowdisplaybreaks

\newcommand{\be}{\begin{equation}}
\newcommand{\ee}{\end{equation}}
\newcommand{\mc}{\mathcal }
\newcommand{\mb}{\mathbb }

\newcommand{\la}{\label}
\newcommand{\eps}{\varepsilon}

\def \ov {\over}\def \te  {\textstyle} 
\def \ci {\cite}
\def \foot {\footnote}

\def \m {\mu}
\def \n {\nu}
\def \del{\partial}

\newcommand{\rf}[1]{(\ref{#1})}

\def \l {\lambda}
\def\z{\zeta}
\def \iffa {\iffalse}
 \def \no {\notag}\def \OO {{\cal{O}}}
 \def \ha {{\textstyle{1 \ov 2}}}
\def \sql {\sqrt{\lambda}}

\newcommand{\N}{\mathcal{N}}

\newcommand{\de}{\Delta}
\newcommand{\ysx}{\widetilde y}
\newcommand{\ty}{\widetilde{y}}
\newcommand{\KN}{\bG}

\makeatletter
\DeclareFontFamily{OMX}{MnSymbolE}{}
\DeclareSymbolFont{MnLargeSymbols}{OMX}{MnSymbolE}{m}{n}
\SetSymbolFont{MnLargeSymbols}{bold}{OMX}{MnSymbolE}{b}{n}
\DeclareFontShape{OMX}{MnSymbolE}{m}{n}{
    <-6>  MnSymbolE5
   <6-7>  MnSymbolE6
   <7-8>  MnSymbolE7
   <8-9>  MnSymbolE8
   <9-10> MnSymbolE9
  <10-12> MnSymbolE10
  <12->   MnSymbolE12
}{}
\DeclareFontShape{OMX}{MnSymbolE}{b}{n}{
    <-6>  MnSymbolE-Bold5
   <6-7>  MnSymbolE-Bold6
   <7-8>  MnSymbolE-Bold7
   <8-9>  MnSymbolE-Bold8
   <9-10> MnSymbolE-Bold9
  <10-12> MnSymbolE-Bold10
  <12->   MnSymbolE-Bold12
}{}

\let\llangle\@undefined
\let\rrangle\@undefined
\DeclareMathDelimiter{\llangle}{\mathopen}%
                     {MnLargeSymbols}{'164}{MnLargeSymbols}{'164}
\DeclareMathDelimiter{\rrangle}{\mathclose}%
                     {MnLargeSymbols}{'171}{MnLargeSymbols}{'171}
\makeatother

\title{Correlators on  non-supersymmetric  Wilson line\\ in $\N=4$ SYM  
  and AdS$_2$/CFT$_1$ }
  
\author[a]{Matteo Beccaria,} 
\author[b]{Simone Giombi} 
\author[c]{ and \ \ Arkady A. Tseytlin\footnote{Also at Lebedev Institute and ITMP, Moscow State University.}} 

\abstract{
Correlators of  local operators inserted on a straight or circular Wilson loop in a conformal gauge theory have 
the structure of a one-dimensional ``defect'' CFT. 
As was shown  in arXiv:1706.00756,  in the case of supersymmetric Wilson-Maldacena loop in $\N=4$ SYM
one  can  compute the leading strong-coupling  contributions to 4-point 
correlators of the simplest ``protected''  operators   by   starting with  the AdS$_5 \times S^5$  string  action 
expanded near the AdS$_2$ minimal surface  and evaluating  the  corresponding tree-level  AdS$_2$  Witten diagrams.
Here  we  perform  the analogous computations in the non-supersymmetric case  of  the standard  
Wilson loop with no coupling to the scalars. The  corresponding  non-supersymmetric ``defect'' CFT$_1$ 
should have an unbroken $SO(6)$ global symmetry. The elementary bosonic  operators (6 SYM scalars 
and  3 components of the SYM field strength) are  dual respectively to the $S^5$ embedding 
coordinates and AdS$_5$  coordinates transverse to the minimal surface ending on the line at the boundary. 
The $SO(6)$ symmetry is preserved  on the string side provided 
the 5-sphere coordinates satisfy Neumann boundary conditions (as opposed to Dirichlet in the supersymmetric case); 
this implies that one should integrate over the $S^5$  expansion point. The massless $S^5$ fluctuations   
then   have  logarithmic propagator, corresponding to the boundary scalar operator having  
dimension $\Delta= {5\ov \sql} + \ldots$ at strong coupling. The resulting functions  of 1d cross-ratio  
appearing in the 4-point functions turn out to have a more complicated structure than in the 
supersymmetric case, involving polylog (Li$_3$ and Li$_2$) functions. 
We also discuss  consistency with the  operator product expansion which allows   
extracting the leading strong coupling corrections to the anomalous dimensions of the 
operators appearing in the intermediate channels.}

\affiliation[a]{Dipartimento di Matematica e Fisica Ennio De Giorgi,\\
Universit\`a del Salento \& INFN, Via Arnesano, 73100 Lecce, 
Italy} 

\affiliation[b]{Department of Physics, Princeton University, Princeton, NJ 08544, USA}

\affiliation[c]{Blackett Laboratory, Imperial College, London SW7 2AZ, U.K.}
                       
\emailAdd{matteo.beccaria@le.infn.it, sgiombi@princeton.edu, tseytlin@imperial.ac.uk} 

\begin{document}



\begin{tabbing}
\hspace*{11.7cm} \=  \kill 
    \> PUPT-2583 \\
    \> Imperial-TP-AT-2019-01  
\end{tabbing}
		
\maketitle
\flushbottom


\def \ed{\bibliographystyle{JHEP}\bibliography{BT-Biblio}\end{document}}
  \def \edo {\end{document}}
\def \z  {\zeta} \def \s {\sigma} \def \P  {\Phi} 
 \def \tet {\textstyle} \def \D {\Delta} 
 \def \OO {{\mc O}}
\def \CC {{\rm C}}
  \def \GG {{\rm G}}
  \def \OO {{\mc O}} 
  \def \GN {{\rm G_N}} 
\def \KK {{\rm K}}
\def \KN {{\rm G_N}} 
\def \g {\gamma} 
\def \z {\zeta} 
\def \yz {{\rm y}}\def \rG {{\rm G}} 
 \def \rC {{\rm C}} 
 \def \brG {\GN} 
 \def\ads  {AdS$_2$ }
\def \bG {{\rm N}}
\def \FF  {{\rm F}}
\def \br  {\bm{\rho}}
\def \nD {{\rm N}}
 \def \ho {\widehat \Omega}
 \def \widebar {\overline }
\def \mG  {\mathbb G}
\def \half {{\textstyle{1\ov 2}}}
\def \V {{\rm R}} 
 
 \newpage
\section{Introduction}

Wilson loops  are  important  observables in gauge theories. 
In  addition to the standard Wilson loop (WL),    in  the $\N=4$ super Yang-Mills  theory one can define   also a special  
Wilson-Maldacena loop (WML)  which is locally-supersymmetric due to an extra coupling to  SYM scalars. 
In the  case of  a straight line or circular loop  that we shall consider below, the WML  is  also  globally supersymmetric  (BPS). 
 
Both WL and WML  
are natural objects to study \cite{Alday:2007he, Polchinski:2011im}. 
For smooth contours their expectation  values do  not have logarithmic divergences and 
thus are consistent with conformal  covariance. 
For straight  line or circular contour  they   preserve a $SL(2,R)$ subgroup of the 4d conformal group, and hence they may be viewed as examples of one-dimensional conformal defects of the 4d gauge theory. In fact, 
the WL and WML may be interpreted respectively as  the  UV and IR  fixed points of a  1d  RG  flow   
of the scalar coupling constant in the Wilson line exponent 
 \cite{Polchinski:2011im} (see also \ci{Beccaria:2017rbe}). 
In the large $N$ limit, their expectation values in the strong coupling  ($\l\gg 1 $)  expansion  
are given by the $AdS_5 \times S^5$ string path integral 
over the world surfaces ending on an infinite line (or circle) at the boundary  of $AdS_5$ and with the $S^5$   scalars 
subject to the Dirichlet   (in the WML case)  or the Neumann    (in the WL case) boundary 
conditions   \cite{Alday:2007he,Polchinski:2011im}. 

In addition to the WL  expectation value  it is  interesting also to  study  correlation functions of 
local operators  inserted along the loop 
(see, e.g., \ci{Polyakov:2000ti,Drukker:2006xg,Alday:2007he,Sakaguchi:2007ba,Drukker:2011za,Correa:2012at}).\foot{The operator insertions
 are equivalent to deformations of the Wilson line \cite{Drukker:2006xg,Cooke:2017qgm}, 
so that the knowledge of all of  the correlators should,  in principle, allow one  to compute the expectation 
value of  a general Wilson loop which is a  deformation of a  line or circle.}  
These correlators are constrained by the $SL(2,\mathbb{R})$ 1d conformal symmetry, and define an 
effective defect 1d CFT  \cite{Drukker:2006xg,Sakaguchi:2007ba, Cooke:2017qgm}.\footnote{More generally, the data of a defect CFT include additional observables, such as 
``bulk-defect'' correlators, that describe the coupling between operators on the defect and ``bulk'' operators 
inserted away from the defect. See e.g. \cite{Billo:2016cpy}.}  In the supersymmetric WML case 
this  CFT$_1$  was studied in \ci{Cooke:2017qgm,Giombi:2017cqn} (see also 
\ci{Kim:2017sju,Giombi:2018qox,Kim:2017phs,
 Drukker:2017dgn, Correa:2018fgz,Liendo:2018ukf,Beccaria:2018ocq,Bianchi:2018zpb,Giombi:2018hsx,Kiryu:2018phb,
  Mazac:2018mdx,Cavaglia:2018lxi} for  some recent discussions 
of the  1d defect CFT approach to Wilson loop computations in $\mc N=4$ SYM). 
In  \ci{Giombi:2017cqn}    it was shown how to compute some  correlation  functions on 
the supersymmetric WML   at  strong coupling   using  string theory, i.e. AdS/CFT. 
Our aim below will be to perform  analogous computations 
in the case of  the standard WL which should correspond  to a different, non-supersymmetric defect CFT. 
    
Let us  first  review  the supersymmetric  WML  case, i.e.    $W = {\rm Tr}\, {\cal P} 
e^{\int dt\left(i\dot x^{\mu}A_{\mu}+|\dot x|\theta^A \Phi_A\right)}$, 
  where  $\Phi_A$   are the SYM  scalars  ($A=1,\dots , 6$).
    For  an   infinite straight line (or circle) and $\theta^A$  being 
  a constant  vector  this operator  preserves 16 of the 32 supercharges of the $\N=4$ 
superconformal group $PSU(2,2|4)$.  Choosing  the  defining line as   the Euclidean time  $x^0=t\in (-\infty,\infty)$  and
$\theta^A$  pointing  in the 6-th direction we get
$
W = {\rm Tr}\, {\cal P }e^{\int dt \left(i A_t+\Phi_6\right)}
$. 
The   correlators   of   the gauge-theory  operators $O(x)$  inserted  along the line
(we suppress    exponential factors appearing between   the operators) 
\begin{equation}\la{1}
\llangle   \OO(t_1) \cdots \OO(t_n)  \rrangle
 \equiv 
 \langle {\rm Tr}\,{\cal P} \big[O(x(t_1))\cdots O(x(t_n))\ e^{\int dt(iA_t+ \Phi_6)}\big]\rangle
\  \end{equation}
can  be interpreted 
 as correlators   of  the corresponding  conformal operators   $\OO(t) $  in  an  effective  defect  CFT$_1$.
 We shall use the notation $\llangle \cdots \rrangle$  for correlators of operators inserted on the Wilson line. 
 We shall sometimes not distinguish between 
 $O(x(t))$ and $\OO(t)$ like in eq.\rf{2}  below.
 
 This  CFT has   $d=1$, $\N=8$  superconformal symmetry  $OSp(4^*|4) \subset PSU(2,2|4)$ which   contains: 
 (i)   $SO(5)$ subgroup of the $SO(6)$ 
  rotating  5   scalars $\Phi_a$ ($a=1, \dots, 5$) not coupled directly to the loop;  
(ii)    $SO(3) \times SO(2,1)$ subgroup of the 4d conformal group $SO(2,4)$ 
($SO(3)$  rotations around
the line  and dilatations, translation and special conformal transformation along the line); (iii) 16 
 supercharges preserved by the WML.
  The  operators $O$ on the line  belong   to representations  of 
 $OSp(4^*|4)$  (i.e.  are  labelled by the  1d scaling dimension $\Delta$ and   representation of 
 ``internal''  $SO(3)\times SO(5)$). The simplest multiplet contains 8+8 operators corresponding to a short representation of 
 $OSp(4^*|4)$ with protected  dimensions; the bosonic ones are the 
 5 scalars $\Phi_a$  (with $\Delta=1$)  and the 3  ``displacement''  operators in the directions ($i=1,2,3$)  transverse to the  line
 ${\mathbb F}_{ti}\equiv iF_{ti}+D_i \Phi_6$ (with $\Delta=2$).
 Their  2-point functions in the planar SYM theory then have the exact form
\begin{equation}
\begin{aligned}\la{2}
\llangle \Phi_a(t_1)\Phi_b(t_2) \rrangle =\delta_{ab} \frac{C_\P}{ (t_{12}) ^2}\ , 
\end{aligned}
\end{equation}
where  $C_\P(\lambda)=2B(\lambda)=   \frac{\lambda}{8\pi^{2}}-\frac{\lambda^{2}}{192\pi^{2}}+\dots    $  
is   twice  
the  Bremsstrahlung function $B(\l)= \frac{\sqrt{\lambda } \, I_2(\sqrt{\lambda })}{4\pi ^2\,  I_1(\sqrt{\lambda })} $
 \cite{Drukker:2011za,Correa:2012at}.
 Similarly, one finds $\llangle {\mathbb F}_{ti}  (t_1) {\mathbb F}_{ti}(t_2) \rrangle =\delta_{ij} 
 \frac{C_{\mathbb F}(\lambda)}{ (t_{12}) ^4}
 $, where $  C_{\mathbb F} = 12  B(\lambda)$. 
 The three-point functions of these elementary  operators $O=( \Phi_a, {\mathbb F}_{ti})$
vanish by the $SO(3)\times SO(5)$ symmetry  while 
their   four-point correlators  are  non-trivial
functions of the   1d conformal cross-ratio  $\chi$   and the 't Hooft coupling. 
For example, for 4 operators of the same dimension 
\begin{equation}
\llangle  \OO_{\Delta}(t_1)\, \OO_{\Delta}(t_2)\, \OO_{\Delta}(t_3)\, \OO_{\Delta}(t_4)\rrangle = 
\frac{1}{(t_{12}\,  t_{34})^{2\Delta}} 
\, {G}(\chi; \l)\ , \qquad \qquad \chi= \frac{t_{12}\, t_{34}}{t_{13}\, t_{24}}\ .  \la{3}
\end{equation}
Ref.  \ci{Giombi:2017cqn}    computed  these correlators   
at strong coupling using the  dual string theory in AdS$_5 \times S^5$.  
At large  string tension $T= {\sql \ov 2 \pi}$   the  minimal surface    corresponding  to the 
$\ha$ BPS Wilson line    is represented  by  
AdS$_2$ space  embedded into  AdS$_5$  and   fixed  at a point  in  the $S^5$.  
The 1d conformal group $SO(2,1)$ is then  the isometry of AdS$_2$, i.e. one  gets  a novel 
 example of  the AdS$_2$/CFT$_1$ duality.  
 This   CFT$_1$,   which is  ``induced''   from the 4d CFT  on the 1d  defect,   
 is not expected to  have a description based on a  local 1d  Lagrangian 
 (for example, representing the Wilson loop path ordered exponential in terms of a 1d  
 auxiliary fermionic path integral 
 \cite{Samuel:1978iy, Ishida:1979bc, Arefeva:1980zd, Gervais:1979fv, Gomis:2006sb, Hoyos:2018jky} 
 and integrating  out  the 4d fields would lead to a non-local 1d fermion action). 
  
The  AdS$_2$ multiplet   of  string fluctuations transverse to the string surface 
includes \cite{Drukker:2000ep}: (i)    5 massless 
scalars $y^a$ ($S^5$ fluctuations near the fixed vacuum point); (ii) 
3 massive ($m^2=2$) scalars $x^i$   (AdS$_5$  fluctuations), and (iii) 8 fermions with $m^2=1$. 
These AdS$_2$  fields   are then naturally   identified  with the 8+8   basic  CFT$_1$   operators 
 \cite{Sakaguchi:2007ba,Faraggi:2011bb,Fiol:2013iaa}. The standard relation $\Delta(\Delta -d)=m^2$  between 
 the AdS$_{d+1}$ scalar mass and the corresponding  
CFT$_d$  operator dimension implies that the massless $y^a$ fields 
should  be dual to the   scalars $\Phi^a$  with  $\Delta=1$ inserted on the line  and 
subject to the standard (Dirichlet) boundary conditions, while 
the  AdS$_5$ fluctuations $x^i$ with $m^2=2$ should be dual to
 ${\mathbb F}_{ti}$ with $\Delta=2$. 

As  was  explained in \ci{Giombi:2017cqn}, using the quartic  vertices between the $y_a$ and $x_i$ fields  appearing in the 
expansion of  the string   action   around the  AdS$_2$  minimal surface 
 one  is able to compute 
  the corresponding tree-level Witten diagrams in AdS$_2$ and extract the strong coupling   predictions for 
the four-point functions of the  protected operators  on the WML  
\begin{equation} 
\llangle \OO(t_1)  \cdots \OO(t_n)  \rrangle  ={ \langle   X(t_1) \cdots X(t_n) \rangle}_{_{\rm AdS_2}}  \ . \la{4} \
 \end{equation}
Here $\langle  \cdots \rangle{_{_{\rm AdS_2}} }$  is the  expectation value in the  2d  world-sheet  theory 
  with the  bulk-to-boundary  propagators  attached to the points  $t_1, \cdots, t_n$ at the boundary,   
$X \sim  y^a$  corresponds to   $\OO \sim  \Phi^a$  
and   $X \sim x^i$  corresponds to   $\OO\sim   {\mathbb F}_{it}$.
The   expansion parameter for the AdS$_2$ 
Witten diagrams is the inverse string tension  $T^{-1} =\frac{2 \pi}{\sqrt{\lambda}}$.\foot{As   the 2d theory 
defined by the fundamental 
superstring 
action is  to be UV finite,  the duality with the  boundary 1d CFT  should hold for any value  of $\l$, including 
world-sheet loop corrections.} 

Applying the  OPE   to 
 \rf{3}   one   can  extract  the leading  corrections to the scaling dimensions of the ``two-particle'' 
operators built out of products of two of the protected insertions  ($y_a \del^n_t y_a$, etc.).
In particular,  for the  lowest-dimension unprotected operator $y_a y_a$  at strong coupling one finds 
  \ci{Giombi:2017cqn,Beccaria:2017rbe}
\begin{equation}
\Delta = 2-\tfrac{5}{\sqrt{\lambda}}+  \OO(\tfrac{1}{ ( \sql)^2})  \,.
\label{5}
\end{equation}
The $y_a y_a$  operator  may be identified with $\P_6$  for   which at weak coupling one finds
 \ci{Alday:2007he}
\be  
\la{60}
\llangle \Phi_6 (t_1)\Phi_6(t_2) \rrangle = \frac{C_{\P_6}}{(t_{12})^{2\Delta}}\ , \qquad \qquad 
C_{\P_6} =   \tfrac{\lambda}{8\pi^{2}}+ \cdots \ , \qquad  \Delta= 1  +   \tfrac{\lambda}{4\pi^{2}}+ \cdots\ , 
\ee
so that \rf{5} is  consistent with a smooth growth  of $\Delta$  from weak to strong coupling.

Let us now  turn to  our present case of   interest --  correlators on the  standard (non-supersymmetric)  Wilson line.  
Since  here   $W= {\rm Tr}\, {\cal P} e^{i \int dt \, \dot x^{\mu}A_{\mu}}$     has  no coupling to scalars, 
the full $SO(6)$   global symmetry should be preserved, i.e. 
  the  correlators  of operators inserted on the line  should  correspond to a   non-supersymmetric  CFT$_1$  
  with the $SO(2,1)$ conformal and  $SO(3)\times SO(6)$  ``internal''  symmetry. 
Since  there is no supersymmetry,   the dimension  of  the scalars    will no longer be protected. In particular, 
 instead of \rf{2} (and \rf{60})  we should   get 
\begin{equation}
\la{6}
\llangle \Phi_A(t_1)\Phi_B(t_2) \rrangle =\delta_{AB} \frac{C'_\P}{(t_{12})^{2\Delta}}\ , \qquad \qquad 
C'_\P=   \tfrac{\lambda}{8\pi^{2}}+ \cdots \ , \qquad  \Delta= 1 -   \tfrac{\lambda}{8\pi^{2}}+ \cdots\ .
\end{equation}
The leading   weak-coupling term in  $C'_\P$ is the same as in \rf{2} or \rf{60}, as it  is determined just by the 
normalization of the free scalar propagator.
 In general, however, the 2-point function normalization  factor  like $C'_\P$ 
  is scheme dependent and hence arbitrary, since the operator gets renormalized and has non-trivial scaling 
  dimension.\foot{\la{f6}
  The reason why the  normalization  constant $C_\Phi$ in \rf{2}  
  in the   supersymmetric WML case is meaningful is  that $\Phi^a$ has   protected  dimension 
  and   is in the same  multiplet as the displacement operator  $ {\mathbb F}_{ti}= i F_{ti} + D_i \Phi_6 $; 
  this has a natural normalization due to its relation to translations in the directions orthogonal to the defect. 
  Hence the normalization constant in its 2-point function defines a meaningful observable, 
  somewhat analogous to the ``central charge'' coefficient $C_T$ in  the correlator of two stress tensors.
   In the  non-supersymmetric  WL  case  the  displacement operator   dual to $x^i$  will be  simply proportional
    to the field strength component $\FF_{ti} = i F_{ti}$ 
    \cite{Correa:2012at}
   and the  coefficient  in the   corresponding  2-point function \rf{510} will also  be a meaningful function of $\lambda$.
   However, the scalar  operator  normalization $C'_\Phi = C_Y$   will be scheme-dependent  and we shall 
   fix it  in a particular way (see \rf{402}). 
 } 
 The  leading  correction  to  $\Delta$  in \rf{6} was computed  in  \ci{Alday:2007he}.\footnote{Recently,  
 it was rederived  as a consequence of 
integrability  of  a  certain  $SO(6)$ invariant spin chain \cite{Correa:2018fgz}.
This  provides  a weak-coupling  indication that correlators on  the  standard WL   may be described by an integrable theory. 
Since  the AdS$_5\times S^5$ superstring action is an integrable  2d theory,
the approach of  \ci{Giombi:2017cqn} suggests that the same  may be expected also 
at  strong coupling (both in the  supersymmetric and non-supersymmetric  cases).  
}

Our aim will be to explore these CFT$_1$ correlators  at strong coupling 
using  similar   AdS$_2$/CFT$_1$  set-up  as  in \ci{Giombi:2017cqn}.
 The  minimal  surface   in AdS$_5$ corresponding to the 
 straight-line   WL  at the boundary has  the  same AdS$_2$  geometry
 and thus  the spectrum of  string fluctuations  will again  contain  5 massless $S^5$ 
scalars $y^a$,  3   AdS$_5$  scalars $x^i$  with $m^2=2$  and 8  fermions with $m^2=1$. 
The boundary conditions for the scalar  $x^i$  do  not change,  and this should be  
  dual     to the usual  field strength  operator ${F}_{ti}$. The latter,   being  the displacement operator in the defect CFT$_1$, 
  should have protected dimension, i.e.  $\Delta_{{F}}= 2 $ for all $\lambda$. 

In the  supersymmetric  WML  case, where the expansion is around a particular point in $S^5$, one may use  
an  explicit parametrization  of $S^5$ like  ($Y_A Y_A=1$)
\be \la{8}  Y_a= { y_a\ov 1 + {1\ov 4} y^2}, \quad  Y_6=\sqrt{1- Y_a Y_a} =  
{ 1 - {1\ov 4} y^2\ov 1 + {1\ov 4} y^2}\ , \qquad \qquad
ds^2_{S^5} = dY_A dY_A =  {dy_a dy_a \ov (1 + {1\ov 4} y^2)^2} \ . \ee
Then the expansion   in $1\ov \sql$   is equivalent to expansion  in  powers of 
 $y_a$  subject to Dirichlet  b.c. and   one   is left with $SO(5)$ as  manifest   symmetry of 
 their correlators  \ci{Giombi:2017cqn}.\foot{ Let us  note also   that in the present case of  UV  finite   AdS$_5 \times $S$^5$ 
 superstring model  
 there  will be no automatic restoration of $SO(6)$ symmetry (either in  flat 2d space  or AdS$_2$, cf. \ci{Carmi:2018qzm}).}

The  key difference with  the  supersymmetric  WML case 
   is that  now  the  $S^5$ scalars  should   be subject  to the Neumann (or ``alternative'' \ci{Klebanov:1999tb}) boundary 
   conditions  which break supersymmetry  \ci{Alday:2007he,Polchinski:2011im,Beccaria:2017rbe}. 
   This leads, in particular,   to 
    an additional   integration over  a  point in $S^5$   restoring 
   the full $SO(6)$    symmetry   in the corresponding correlators.\foot{The contribution of the $S^5$ zero modes 
   implies also  that    in contrast to the  large $\l$ asymptotics  $ \langle  W \rangle \sim  (\sql)^{-3/2} e^{\sql} $
 of the  WML \cite{Erickson:2000af},   for  the  standard WL 
 one   gets  $\langle  W\rangle \sim  \sql \,  e^{\sql}  $  \ci{Beccaria:2017rbe}.
 Let  us note also   that integration over   sphere  0-modes  is  important also in the context of  ratio  of  BPS  Wilson loops in  
 \cite{Medina-Rincon:2018wjs}.
 }
 We will assume   that the   counterparts  of  the SYM scalars $\Phi_A$   on the string side  should  be  the $S^5$ 
 embedding coordinates   $Y_A$  \, ($Y_A Y_A=1$)  on which $SO(6)$ acts linearly. 
 For a massless AdS$_2$ scalar    one has $\Delta( \Delta -1) =0$ which gives    $\Delta=0$   for the 
 Neumann (N)  boundary conditions. 
 The first non-vanishing    strong-coupling correction to $\Delta$ in this case     was argued to be \ci{Alday:2007he}
 \be \la{7} 
 \Delta =  \tfrac{5}{ \sql} +   \OO(\tfrac{1}{ ( \sql)^2}) \ . \ee 
 The same result was found also  in \ci{Beccaria:2017rbe}, following   \ci{Polchinski:2011im}.\foot{As $Y_6 = 1 - \ha y_a y_a  + \cdots $ (see \rf{8})     at strong coupling $\P_6$   may  be  identified  with $y_a y_a$    and thus  should   have  the dimension $ 2-\tfrac{5}{ \sql} +...$  as in 
     \rf{5}. 
  Since in the WL case all 6 scalars have the same dimension, \rf{5} and \rf{7}  are   then consistent \ci{Beccaria:2017rbe}  with  the fact 
  that  the dimensions of  scalars  with the standard (D)  and   alternative (N)   boundary conditions  in AdS$_2$  should  sum up  to 2.} 
 We  will  reproduce \rf{7}  directly   by computing  the 2-point function  \rf{6}   interpreted as   the scalar 
 correlator  $\langle   Y_{A}   (t_1) Y_{B} (t_2)   \rangle_{_{\rm AdS_2}} $
   below.

 In the case of Neumann boundary  conditions  on $y_a$ in \rf{8}    one is to integrate over their zero mode or  position of the expansion point 
 on $S^5$. 
 This  is equivalent  to  integrating over the embedding coordinates $Y_A$  without  breaking
  $SO(6)$. Then  we should have  the following analog of \rf{4},\rf{6}
  \begin{equation}
\la{9}
\llangle \Phi_{A_1}   (t_1)\cdots\Phi_{A_n} (t_n) \rrangle = { \langle   Y_{A_1}   (t_1)\cdots 
Y_{A_n} (t_n)   \rangle}_{_{\rm AdS_2}} 
\ .
\end{equation}
    The  computation of \rf{9} 
     can be implemented in  a  manifestly  $SO(6)$ covariant way by setting $Y_A = n_A + \z_A (\s) + ...\   (n_A \z_A =0)$   
 and integrating over the  fluctuations $\z_A$ and  the constant direction $n_A$. 
 In practice, it  is sufficient to consider the   $SO(6)$   singlets
   like $\langle Y_{A}(t_1)  Y_{A}(t_2)  Y_{B}(t_3)  Y_{B}(t_4) 
  \rangle $  which   will not depend  on the position of the expansion point 
  $n_A$ and thus averaging over $n_A$ will not be required.
  Such 
  $SO(6)$  singlets will also be IR   finite in the quantum theory  \ci{Elitzur:1978ww,David:1980rr,Miramontes:1990nf}.

 \
 
The rest of the paper is organized as follows.  
In section 2 we shall  first review the  computation of   4-point correlators  on the 
 supersymmetric  Wilson line   at strong coupling, following 
 \cite{Giombi:2017cqn}. The starting point   is the   bosonic part of the AdS$_5 \times S^5$ 
  string action expanded near  the AdS$_2$   minimal  surface 
that defines  the corresponding quartic vertices between the $x^i$ and $y^a$ fields. After 
 summarizing   some general relations   for 4-point functions in  CFT$_1$  we will present 
the  expressions for the leading-order strong-coupling terms  in the  $G(\chi) $ functions 
in the scalar  4-point  correlators   in \rf{2.35} and \rf{153}. In section \ref{chaos} we make some comments on the analytic continuation to the out of time order correlators relevant for chaos \cite{Maldacena:2015waa}, which appear to display a maximal Lyapunov exponent.  

In section 3  we will turn to the non-supersymmetric Wilson line case  and describe the  general $SO(6)$ invariant 
computational scheme, based on using the Neumann  propagator for the  fluctuations of the $Y^A$    fields 
and  averaging  over the $S^5$ expansion point $n^A$. 
In section 4 we shall use it to compute the 2-point function \rf{6}   at strong coupling or  $\langle Y_A (t_1) Y_B (t_2) \rangle$ 
for $SO(6)$ scalars in AdS$_2$ (see \rf{4.1},\rf{41}). We shall  reproduce the leading  term in the dimension $\Delta$ in \rf{7}
and also  demonstrate   (in section 4.2)    that the  subleading ${1\ov ( \sql)^2}\log^2$ corrections  ``exponentiate'', i.e. have the right coefficient 
to be consistent with the 1d conformally invariant form of the 2-point function in \rf{4.1}. The  subleading ${1\ov ( \sql)^2}\log$ correction in \rf{4.1}   corresponding  next to leading coefficient  $d_2$  in 
$\Delta = {5\ov  \sql}   +  {d_2\ov ( \sql)^2}  + ...$  should  receive contributions from the fermionic 
 1-loop graphs (cf. Fig. \ref{fig:loop}) 
 and we will   not attempt to compute it here. 

In section 5 we will compute the 
mixed   correlator  $\llangle { \FF}\indices{_{t}^{i}} (t_1)  { \FF}\indices{_{t}^{i}} (t_2)   \P_A(t_3) \P_B(t_4) \rrangle $  at strong coupling or the leading  contribution to  the $G(\chi)$  function  in $ \langle x^{i} (t_1)  x^{j} (t_2)    Y_A(t_3) Y_B(t_4) \rangle $    in \rf{51}
coming from the diagrams in Figs. \ref{fig:XXZZ} and \ref{fig:XXZZbis}. The resulting connected contribution  to $G$  is given by  \rf{5.13},\rf{5222}  and happens to be    simply proportional to the  expression in 
 the supersymmetric case in \rf{151},\rf{153}. The reason for this  relation is explained in section 6.2.
 
 Section 6   is devoted to the computation of the $Y$-scalar  4-point function \rf{6.2},\rf{6622}. 
 We shall  first determine the leading order  $1\ov (\sql )^2$  contribution to the  singlet  function $G_S(\chi)$ \rf{6.799},\rf{6.7} 
 coming from tree-level graphs in Fig. \ref{fig:disc2}  and  graphs with 1-loop propagator corrections like  in Fig. \ref{fig:dash}.
 The corresponding functions in the traceless symmetric  $G_T$  and  antisymmetric $G_A$  parts of the correlator 
 are given in \rf{6.9},\rf{610}.  We shall  then turn to the order  $1\ov (\sql )^3$  contribution  coming from the 
 tree-level graph with contact bulk vertex in Fig. \ref{fig:zzzz}. 
 
 In section 6.2 we  will  explain   how one can by-pass  the complication 
 of directly computing  the AdS$_2$ bulk integrals of the products of  four  logarithmic Neumann  propagators 
 by first differentiating the correlator over the boundary points, then relating  it to correlators in the theory with  standard Dirichlet 
 propagators and finally  integrating back. In addition to the   contact diagram  contribution  there is also  the  order 
 $1\ov (\sql )^3$  contribution  coming from ``reducible''  tree diagrams  in Fig. \ref{fig:disc3}  and  similar diagrams with 1-loop 
``dressed'' propagators which are computed in Appendix \ref{app:other} (see \rf{G9}, \rf{88}). It is only the   sum of all $1\ov (\sql )^3$ 
 corrections that is  conformally invariant with the resulting singlet function   given in \rf{6599}.  Similar expressions are found for $G_T$ and $G_A$
 functions. 
 Compared  to the supersymmetric case 
 expressions in \rf{2.35}  they are  more complicated    containing  polylog (Li$_3$ and Li$_2$) functions of  $\chi$. 
 In section 5 and section 6.4   we also   comment on  the consistency of the results for the $G$-functions  with the OPE in \rf{2.15} 
 extracting the  leading-order strong-coupling corrections  to  the dimensions  of  composite operators appearing in the intermediate channels (cf. Appendix \ref{app:BPSanom}).  
We also include several  other Appendices  reviewing some  general relations  and discussing  technical points. 

There are a number of   interesting  directions to explore in the future. One is  how 
the   classical  integrability of the AdS$_5 \times S^5$   string    theory is  reflected  in the  correlation functions  like \rf{9}.
Some  connection  to integrability is  expected  since, on  the one hand, 
the  knowledge of  tree-level  correlators   is related to the 
value of  string action on  world sheets   ending on more general wavy contours, while, on the other hand, 
the  classical string integrability   allows  one to find  more general Wilson-line type    solutions (cf., e.g.,  \cite{Dekel:2015bla}   and \ci{Ishizeki:2011bf}).
It would  be   important to identify more direct correspondence   at the level of particular  correlators  (and 
the associated   AdS$_2$ Witten diagrams) possibly   analogous to  constraints on flat-space  S-matrix in 
integrable 2d models.  Another  is an extension of the computations  in \cite{Giombi:2017cqn}  and the present paper to  AdS$_2$ world-sheet loop level including also  the 
Green-Schwarz fermions. 
Finally, it would  be interesting to establish   a  connection between  the strong-coupling results for the correlators found 
in this  paper and general  results obtained in the framework of 1d bootstrap 
(generalizing  the  analysis  of  \ci{Liendo:2018ukf} in the supersymmetric  case).

\section{Correlators  on  supersymmetric Wilson line  at strong coupling }

Before turning to the  non-supersymmetric 
WL case let us  start with a review  of the  computation of   4-point correlators  on 
the supersymmetric Wilson line   at strong coupling  following 
 \cite{Giombi:2017cqn}.

\subsection{$\text{AdS}_5\times S^5$ string action in static gauge as AdS$_2$ bulk theory action}

The bosonic part of the superstring action on $\text{AdS}_5\times S^5$ may be written as 
\be\la{21}
S_B =\ha T \int d^2\sigma\,\sqrt{h}\,h^{\mu\nu}\Big[\frac{1}{z^2}(\partial_\mu x^0
\partial_\nu x^0+       \partial_\mu x^i
\partial_\nu x^i+\partial_\mu z\,\partial_\nu z)+\frac{\partial_\mu y^a\partial_\nu y^a}{(1+\frac{1}{4}
y^2)^2}\Big],\qquad T=  \frac{\sqrt\lambda}{2\pi} \ , 
\ee
where $\sigma^\mu = (t,s)$ are Euclidean world-sheet coordinates, $r=(0,i) = (0,1,2,3)$  label 
4-boundary coordinates  and  $a=1, \dots, 5$ --  the  $S^5$  coordinates. 
The minimal surface  ending on the  straight line $x^0 = t$ at the  boundary  is 
\be\la{22}
z=s,\qquad x^0 = t, \qquad x^i=0, \ \qquad y^a = 0,
\ee
with the  induced  metric being the AdS$_2$ metric 
\be \la{210}
g_{\mu\nu}d\sigma^\mu d\sigma^\nu = \frac{1}{s^2}(dt^2+ds^2) \ ,  \qquad \qquad  g_{\mu\nu} = \frac{1}{s^2}\delta_{\mu\nu} \ . \ee
The embedding  of $\text{AdS}_2$ into $ \text{AdS}_5$ can be made 
explicit  using  the coordinates  (here $x^2=x^ix^i$, $i=1,2,3$)
\be\la{23}
ds_{{\rm AdS}_5}^2 = \frac{(1+\frac{1}{4}x^2)^2}{(1-\frac{1}{4}x^2)^2}\,ds_2^2+\frac{dx^idx^i}{(1-\frac{1}{4}x^2)^2},
\qquad \qquad ds_2^2 = \frac{1}{z^2}(dx_0^2+dz^2).
\ee
Then perturbation theory near the above minimal  surface   can be described by 
 the  string  action  in the  Nambu form  taken  in  the  static gauge $z=s$, $x^0 = t$  
\be
\la{2.9}
S_B = T\,\int d^2\sigma\sqrt{
\det\Big[
\frac{(1+\frac{1}{4}x^2)^2}{(1-\frac{1}{4}x^2)^2}\,g_{\mu\nu}(\sigma)
+\frac{\partial_\mu x^i \partial_\nu x^i}{(1-\frac{1}{4}x^2)^2}
+\frac{\partial_\mu y^a \partial_\nu y^a}{(1+\frac{1}{4}y^2)^2}
\Big]
} = T\,\int d^2\sigma\sqrt{g}\,L_B \ , 
\ee
where $g_{\mu\nu}$  is the background AdS$_2$ metric \rf{210}. 
This  action representing    a straight
fundamental string in $\text{AdS}_5\times S^5$ stretched along $z$
may be interpreted as  a 2d field theory
of 3+5 scalars ($x^i, y^a)$  propagating in AdS$_2$ geometry. It has   manifest (linearly-realised)  symmetry 
$SO(2,1)\times SO(3)\times SO(5)$.

Expanding  \rf{2.9} 
 in powers of  $x^i$ and $y^a$ we get  an interacting theory   for 3  massive  ($m^2=2$)  scalars $x^i$ 
 and   5  massless  scalars $y^a$  propagating  in AdS$_2$ described by 
 $ L_B =\  L_2  + L_{4x}   + L_{2x,2y} + L_{4y}   + \cdots $:  
 \begin{align}   
L_2=&\tet  \frac{1}{2}  g^{\m\n}\del_\m x^i \del_\n x^i  +  x^i x^i + \frac{1}{2} g^{\m\n}\del_\m y^a \del_\n y^a\ ,  \la{2.6} 
\\[2pt]
L_{4x} =&\ \tet  \frac{1}{8} (g^{\m\n}\del_\m x^i \del_\n x^i )^2  
                 - \frac{1}{4} (g^{\m\n} \del_\m x^i \del_\n x^j) \; (g^{\rho\kappa} \del_\rho x^i \del_\kappa  x^j)
\nonumber     \\
             &\ \tet + \frac{1}{4}  x^i x^i  (g^{\m\n} \del_\m x^j \del_\n x^j) + \frac{1}{2} x^i x^i\, x^j x^j \ , 
\label{2.7}            
\\[2pt]
L_{2x,2y}=&\ \tet  \frac{1}{4} (g^{\m\n}\del_\m x^i \del_\n x^i )\,(g^{\rho\kappa} \del_\rho y^a \del_\kappa  y^a) 
          - \frac{1}{2}  (g^{\m\n} \del_\m x^i \del_\n y^a) \; (g^{\rho\kappa} \del_\rho x^i \del_\kappa  y^a)\ , \la{2.8}
\\[2pt]
L_{4y} =&\  \tet
-\frac{1}{4} (y^b y^b) (g^{\m\n} \del_\m y^a \del_\n y^a) 
+\frac{1}{8}  (g^{\m\n}\del_\m y^a \del_\n y^a)^2
-\frac{1}{4}  (g^{\m\n} \del_\m y^a \del_\n y^b) \; (g^{\rho\kappa} \del_\rho y^a \del_\kappa  y^b)\ . \la{2.99}
\end{align}
Assuming that both   scalars   are subject to the  standard (Dirichlet) boundary conditions at $z=s=0$ 
and applying the standard AdS/CFT  relation  ($\Delta (\Delta -1)=m^2$) we conclude that  $x^i$  and $y^a$ 
should be dual, respectively,  to  the $\Delta=2$  and $\D=1$  operators at  the 1d  boundary $x^0=t$.
There are also 
8 fermionic fields 
transforming  in the $(\bm{2}, \bm{4})$ representation of 
$SU(2)\times Sp(4)\simeq SO(3)\times SO(5)$.


Starting with  the 2d  bulk  theory (\ref{2.9})  and computing   Witten diagrams with bulk-to-boundary 
propagators attached to the points $\{t_{n}\}$ on the boundary  will give  us  correlators  in the
  boundary CFT$_1$  and thus  the strong-coupling expansion of the  SYM correlators 
of the  corresponding   gauge-theory operators ($ x_i  \leftrightarrow {\mathbb F}_{ti}   $, $y_a \leftrightarrow  \P_a$) 
inserted along the WML   (see \rf{1},\rf{4}). 
As the Lagrangian $L_{B}$ has no cubic terms,
 the first  non-trivial  contribution to the simplest 4-point  correlation functions of $x^{i}$ and $y^{a}$ is 
 given just by the contact 4-point vertices  in \rf{2.7}--\rf{2.99}.

\subsection{Conformal invariance and crossing constraints on 4-point functions in CFT$_1$}
\la{sec:crossing}

The 4-point function of  primary operators $\OO$   with the same  dimension $\Delta$ is constrained by the $SO(2,1)$
conformal invariance to take the form
\be
\la{2.13}
\llangle \mc O_{\Delta}(t_{1})\,\mc O_{\Delta}(t_{2})\,
\mc O_{\Delta}(t_{3})\,\mc O_{\Delta}(t_{4})\rrangle = \frac{1}{(t_{12}\, t_{34})^{2\Delta}}
\, G(\chi),\qquad \qquad \chi = \frac{t_{12}\, t_{34}}{t_{13}\, t_{24}} \ . 
\ee
The function $ G(\chi)$ in (\ref{2.13}) admits the OPE   (see, e.g., \cite{Dolan:2011dv})
\be
\la{2.15}
 G(\chi) = \sum_{h}c_{\Delta, \Delta; h}\,\chi^{h}\,  F_h (\chi)  \ , \qquad \qquad F_h \equiv    {}_{2}F_{1}(h,h,2h,\chi),
\ee
associated with the s-channel exchange of fields with conformal dimension $h$.
The OPE coefficients in (\ref{2.15}) may be expressed in terms of the  coefficients in the 2-point  and 3-point functions as 
$
c_{\Delta, \Delta; h} = \frac{(C_{\Delta,\Delta,h})^2}{(C_{\Delta,\Delta})^2 (C_{h,h})^2}
$.  
For the 4-point function with two pairwise equal dimensions, one has
\begin{align}
& \llangle \mc O_{\Delta_{1}}(t_{1})\,\mc O_{\Delta_{2}}(t_{2})\,
\mc O_{\Delta_{1}}(t_{3})\,\mc O_{\Delta_{2}}(t_{4})\rrangle = 
\frac{1}{(t_{12}t_{34})^{\Delta_{1}+\Delta_{2}}}\,\left|
\frac{t_{24}}{t_{13}}\right|^{\Delta_{12}}\, G(\chi),\la{2.17} \\
& G(\chi) = \sum_{h} c_{\Delta_{1}, \Delta_{2}; h}\,\chi^{h}\,
_{2}F_{1}(h+\Delta_{12}, h-\Delta_{12}, 2h, \chi) ,  \qquad \qquad \Delta_{12}
= \Delta_{1}-\Delta_{2},  \la{217} 
\end{align}
The expressions   for the $ G(\chi)$ functions in \rf{2.13},\rf{2.17}   in the  case of the 
(generalized) free field theory are summarized in Appendix \ref{app:freefield}.



Together with the conformal invariance, we should also take into account the crossing
invariance of the 4-point function. 
Having in mind  applications to  the cases  of  $SO(5)$  or $SO(6)$ invariant scalar correlators in   defect  CFT$_1$'s  associated with the 
WML or  WL, let us  discuss crossing  for the   general  $SO(N)$
flavour symmetry. 
Let us consider a  primary  operator $\OO_A$ with  dimension $\de$ and vector  index $A=1,\dots, N$ 
 of $SO(N)$. 
Then  the  analog of the  correlator  (\ref{2.13})  will be 
\begin{align}
\la{2.18}
& \llangle\OO^{A}(t_{1})\OO^{B}(t_{2})\OO^{C}(t_{3})\OO^{D}(t_{4})\rrangle
= \frac{[C_\D(\lambda)]^{2}}{t_{12}^{2\de}t_{34}^{2\de}}\,G^{ABCD}(\chi) \ .
\end{align}
where we separated the factor   related to the  normalization factor $C_\D$   in the 2-point function. 
$G^{ABCD}$
can be decomposed into  singlet,  symmetric traceless tensor  and  antisymmetric tensor parts 
as
\begin{align}
\la{2.19}
G^{ABCD} &= G_{S}(\chi)\,\delta^{AB}\delta^{CD}
+G_{T}(\chi)\Big[\delta^{AC}\delta^{BD}+
\delta^{BC}\delta^{AD}-\tfrac{2}{N}\,\delta^{AB}\delta^{CD}
\Big]
+G_{A}(\chi)\Big[\delta^{AC}\delta^{BD}-\delta^{BC}\delta^{AD}\Big]\ , 
\end{align}
so that
\begin{align}
\la{2.20}
G^{AABB} = N^{2}\,G_{S}, \qquad \qquad 
&G^{ABAB} = N\,G_{S}+(N+2)(N-1)\,G_{T}+N(N-1)\,G_{A},\notag \\
&G^{ABBA} = N\,G_{S}+(N+2)(N-1)\,G_{T}-N(N-1)\,G_{A} \ . 
\end{align}
Thus  $G_{S},G_{T},G_{A}$ can  be   found as  combinations of  invariant  contractions
\begin{align}
\la{2.21}
 G_{S} = \tfrac{1}{N^{2}}\,G^{AABB}, \qquad \qquad 
&G_{T} = \tfrac{1}{2\,(N+2)(N-1)}\Big[G^{ABAB}+G^{ABBA}-\tfrac{2}{N}G^{AABB}\Big],\\
& G_{A} = \tfrac{1}{2\,N(N-1)}\Big[G^{ABAB}-G^{ABBA}\Big]. \la{221}
\end{align}
Crossing transformations are generated by the leg exchanges 
$3\leftrightarrow 4$ and $1\leftrightarrow 3$  in  \rf{2.18}  which, in addition to exchanging the 
corresponding flavour indices,    amount to $t_3\leftrightarrow t_4$ and $t_1\leftrightarrow t_3$  or, equivalently, 
\be\la{2119}
\chi\stackrel{3\leftrightarrow 4}{\to} \frac{\chi}{\chi-1},\qquad\qquad 
\chi\stackrel{1\leftrightarrow 3}{\to} 1-\chi.
\ee
From (\ref{2.21})  one finds that under $3\leftrightarrow 4$ 
\be
\la{2.23}
G_{S}(\chi) = G_{S}(\tfrac{\chi}{\chi-1}),\qquad\qquad
G_{T}(\chi) = G_{T}(\tfrac{\chi}{\chi-1}),\qquad\qquad
G_{A}(\chi) = -G_{A}(\tfrac{\chi}{\chi-1}).
\ee
The $1\leftrightarrow 3$ exchange leaves invariant $G^{ABAB}$ and 
swaps $G^{AABB}\leftrightarrow G^{ABBA}$. 
Taking into account the transformation of the 
prefactor $\frac{1}{t_{12}^{2\de}\,t_{34}^{2\de}}$ in \rf{2.18}, this gives
\be
\la{2.24}
G^{AABB}(\chi) = \big(\tfrac{\chi}{\chi-1}\big)^{2\de}\,G^{ABBA}(1-\chi),\qquad \qquad 
G^{ABAB}(\chi) = \big(\tfrac{\chi}{\chi-1}\big)^{2\de}\,G^{ABAB}(1-\chi).
\ee
Using (\ref{2.23}) and (\ref{2.24}) we  observe that instead  of three   functions in \rf{2.19} 
we have only one    independent, i.e. we
can express
the $G_T$ and $G_A $    in terms of   $G_S$.
 Explicitly, we have
\begin{align}
G^{ABAB}(\chi) = \chi^{2\de}\,G^{AABB}\big(\tfrac{1}{1-\chi}\big),\qquad\qquad 
G^{ABBA}(\chi) = \big(\tfrac{\chi}{\chi-1}\big)^{2\de}
\,G^{AABB}(1-\chi),\la{222}
\end{align}
and therefore 
\begin{align}
G_{T}(\chi) &= -\tfrac{N}{(N+2)(N-1)}\,G_{S}(\chi)
+\tfrac{N^{2}}{2(N+2)(N-1)}\Big[\chi^{2\de}\,G_{S}\big(\tfrac{1}{1-\chi}\big)
+\big(\tfrac{\chi}{\chi-1}\big)^{2\de}\,G_{S}(1-\chi)\Big], \la{2.26} \\
G_{A}(\chi) &= \tfrac{N}{2(N-1)}\Big[\chi^{2\de}\,G_{S}\big(\tfrac{1}{1-\chi}\big)
-\big(\tfrac{\chi}{\chi-1}\big)^{2\de}\,G_{S}(1-\chi)\Big]. \la{226}
\end{align}

\subsection{Strong-coupling expansion of the $SO(5)$ scalar 4-point  function} 

Let us now review the result of 
 \cite{Giombi:2017cqn} for the 
 tree-level 4-point  correlator of the  $S^{5}$ fluctuations  $y^{a}$
 dual to the 5 SYM scalars $\Phi^{a}$, $a=1, \dots, 5$  not coupled to the Wilson-Maldacena loop in \rf{1}.
Since the dimensions of the operators $\Phi^{a}$ are protected by supersymmetry, we should have\foot{In what follows we shall 
for simplicity 
omit  the label  AdS$_2$ in the corresponding correlators, i.e. 
 $\langle y^{a}(t_{1}) y^{b}(t_{2})\rangle_{\text{AdS}_{2}} \equiv \langle y^{a}(t_{1}) y^{b}(t_{2})\rangle $, etc. }
\begin{align}
& 
\llangle\Phi^{a}(t_{1})\Phi^{b}(t_{2})\rrangle
=\langle y^{a}(t_{1}) y^{b}(t_{2})\rangle = \delta^{ab}  \frac{C_\P}{(t_{12})^{2}},\la{225}\\
& 
\llangle\Phi^{a}(t_{1})  \Phi^{b}(t_{2})\Phi^{c}(t_{3})  \Phi^{d}(t_{4})\rrangle
= \langle y^{a}(t_{1}) y^{b}(t_{2})y^{c}(t_{3})y^{d}(t_{4})\rangle =  \frac{C_\P^{2}}{(t_{12}\, t_{34})^{2}}\,G^{abcd}(\chi).\la{2.25}
\end{align}
With the   normalization coefficient $[C_{\Phi}(\lambda)]^{2}$ extracted we will have 
$G^{a_{1}a_{2}a_{3}a_{4}}(\chi) = \delta^{a_{1}a_{2}}\delta^{a_{3}a_{4}}+\mc O(\chi)$.
The tensor $G^{a_{1}a_{2}a_{3}a_{4}}$  can  be  split into the $S, T, A$   parts  according to
(\ref{2.19}) with $N=5$.
Expanding at strong coupling (i.e.   small $\frac{1}{\sqrt\lambda}$)  we will have 
\be\la{227}
G_c (\lambda) = G_c ^{(0)}+\tfrac{1}{\sqrt\lambda}\,G_c^{(1)}+\dots \ , \qquad  c= S, T , A \ . 
\ee
The leading order contributions $G^{(0)}$  comes from  with disconnected diagrams 
like  in Fig.~\ref{fig:bps-disc}. Here and below  for simplicity we draw the 1d boundary  as a circle  rather than a line.
\begin{figure}[ht]
\centering
\begin{tikzpicture}[line width=1 pt, scale=0.5]
\draw[densely dashed] (0,0) circle (2);;
\draw (30:2)--(150:2);
\draw (-30:2)--(-150:2);
\node[right] at (30:2) {$y$}; \node[left] at (150:2) {$y$}; 
\node[right] at (-30:2) {$y$}; \node[left] at (-150:2) {$y$};
\draw[fill=white] (30:2) circle (0.12);
\draw[fill=white] (-30:2) circle (0.12);
\draw[fill=white] (150:2) circle (0.12);
\draw[fill=white] (-150:2) circle (0.12);
\end{tikzpicture}
\caption{Leading order disconnected contribution $G^{(0)}$ with other similar  diagrams obtained by crossing.
}\label{fig:bps-disc}
\end{figure}
It is  thus given by the free-field contribution (cf. \rf{A.1})
\begin{align}
\la{2.29}
\llangle \Phi^{a}(t_{1})   \Phi^{b}(t_{2}) \Phi^{c}(t_{3}) \dots\Phi^{d}(t_{4})\rrangle_{\text{disc.}} &= C_{\Phi}^{2}
\Big[\frac{\delta^{ab}\delta^{cd}}
{t_{12}^{2}t_{34}^{2}}+\frac{\delta^{ac}\delta^{ad}}
{t_{13}^{2}t_{24}^{2}}+\frac{\delta^{ad}\delta^{bc}}
{t_{14}^{2}t_{23}^{2}}\Big]\notag \\
&= \frac{ C_{\Phi}^{2}}
{(t_{12}t_{34})^{2}}\Big[\delta^{ab}\delta^{cd}+\chi^{2}
\delta^{ac}\delta^{bd}+\frac{\chi^{2}}{(1-\chi)^{2}}\delta^{ad}
\delta^{bc}\Big].
\end{align}
Comparing with (\ref{2.19}) gives
\be
\la{2.30}
G_{S}^{(0)}(\chi) = 1+\tfrac{2}{5}\,G_{T}^{(0)}(\chi),\qquad
G_{T}^{(0)}(\chi) = \tfrac{1}{2}\,\Big[\chi^{2}+\frac{\chi^{2}}{(1-\chi)^{2}}\Big],\qquad
G_{A}^{(0)}(\chi) = \tfrac{1}{2}\,\Big[\chi^{2}-\frac{\chi^{2}}{(1-\chi)^{2}}\Big].
\ee
The first subleading  correction comes from the contact diagram in Fig.~\ref{fig:bps-conn}
\begin{figure}[ht]
\centering
\begin{tikzpicture}[line width=1 pt, scale=0.5]
\draw[densely dashed] (0,0) circle (2);;
\draw (40:2)--(-140:2);
\draw (-40:2)--(140:2);
\node[right] at (40:2) {$y$}; \node[left] at (140:2) {$y$};
\node[right] at (-40:2) {$y$}; \node[left] at (-140:2) {$y$};
\draw[fill=white] (40:2) circle (0.12);
\draw[fill=white] (-40:2) circle (0.12);
\draw[fill=white] (140:2) circle (0.12);
\draw[fill=white] (-140:2) circle (0.12);
\draw[fill=black] (0,0) circle (0.12);
\end{tikzpicture}
\caption{Contact diagram contributing  to first subleading strong-coupling   correction 
$G^{(1)}$. 
}\label{fig:bps-conn}
\end{figure}
where the 4-point vertex comes  from \rf{2.99}. 
The bulk-to-boundary   propagator    corresponding to a 
massive scalar  in   AdS$_{d+1}$ is $(\Delta ( \Delta - d) = m^2$)  
\begin{align}
\la{2.32}
K_{\Delta}(z, x; x') &= \mc C_{\Delta}\,  \KK_{\Delta}(z, x; x') \ , \qquad   \KK_{\Delta}(z, x; x') \equiv 
\Big[\frac{z}{z^{2}+(x-x')^{2}}\Big]^{\Delta}\,,  \\
\llangle \mc O_{\Delta}(x)\mc O_{\Delta}(x')\rrangle &= \frac{\mc C_{\Delta}}{|x-x'|^{2\Delta}},
\qquad \qquad \mc C_{\Delta} = \frac{\Gamma(\Delta)}{2\pi^{d/2}\,\Gamma(\Delta+1-\frac{d}{2})} \ , \la{232}
\end{align}
where we have assumed a particular normalization of the 2-point function of the  associated boundary field.\footnote{Explicitly, in this normalization $C_{\Phi}(\lambda)=\mc C_{1}(1-\frac{3}{2\sqrt{\lambda}}+\ldots)=\frac{4\pi}{\sqrt{\lambda}}B(\lambda)$, with $\mc C_{1}$ given in (\ref{2.30x}). The higher order corrections in $\lambda$ are determined by the Bremsstrahlung function $B(\lambda)$ and should be reproduced by computing loop corrections to the boundary-to-boundary propagators in Figure \ref{fig:bps-disc}.}
For $d=1$ and $\Delta=1$ this gives  
\be
d=1,\  \Delta=1: \qquad \qquad   K_{1}(z, t; t') = \frac{1}{\pi}  \KK_1 \ , \qquad \KK_1 = \frac{z}{z^{2}+(t-t')^{2}},\qquad \qquad 
\mc C_{1} = \frac{1}{\pi}.\la{2.30x} 
\ee
The contribution of the  connected diagram  corresponding to the vertex in \rf{2.99}  is then 
\begin{align}
\la{2.34}
\llangle\Phi^{a} (t_{1})\Phi^{b}(t_{2})\Phi^{c}(t_{3})\Phi^{d}(t_{4})\rrangle_{\text{conn}}
&=
\langle y^{a}(t_{1})y^{b}(t_{2})y^{c}(t_{3})y^{d}(t_{4})
\rangle_{{\rm conn}}\notag \\
&= \frac{(\mc C_{1})^{2}}
{(t_{12}\, t_{34})^{2}}\,\tfrac{1}{\sqrt\lambda}\,(G^{(1)})^{abcd}\ , 
\end{align}
where the corresponding  functions in \rf{2.19}  are then
\begin{align}
\la{2.35}
G^{(1)}_{S}(\chi) = &
-2\tfrac{\chi ^4-4 \chi ^3+9 \chi ^2-10 \chi +5}{5 (\chi -1)^2}
+\tfrac{(2 \chi ^4-11 \chi ^3+21 \chi ^2-20 \chi +10) \chi ^2 \
}{5 (\chi -1)^3}\log \chi  -\tfrac{(2 \chi ^4-5 \chi ^3-5 \chi +10)}{5 \chi} \log (1-\chi )  ,\notag \\
G^{(1)}_{T}(\chi) =&-\tfrac{(2 \chi ^2-3 \chi +3) \chi ^2}{2 (\chi \
-1)^2}+\tfrac{(\chi ^2-3 \chi +3) \chi ^4}{(\chi -1)^3} \log \chi  -\chi ^3 \log (1-\chi )  , \\
G^{(1)}_{A}(\chi) =& -\tfrac{(\chi -2) (2 \chi ^2-\chi +1) \chi }{2 (\chi \
-1)^2}+\tfrac{(\chi -2) (\chi ^2-2 \chi +2) \chi ^3 }{(\chi -1)^3}\log \chi 
-(\chi ^3-\chi ^2-1) \log (1-\chi )\ .\notag
\end{align}
These expressions are  found by  computing  AdS$_2$  integrals as  discussed in 
Appendix \ref{app:Dfunctions}.
Here and in what follows 
we assume as in \cite{Giombi:2017cqn}  that $\log \chi \equiv \log |\chi|, \  \log (1-\chi) \equiv \log |1-\chi|$ so that 
the resulting  expressions are defined as real on the whole line $\chi \in (-\infty, \infty)$.


The  leading order  terms \rf{2.30} in  $G_{S,T,A}(\chi)$ are given by  the free-field  expressions 
 associated with the exchange of 2-particle states $\Phi^{a}\partial_{t}^{k}\Phi^{b}$ that can be  decomposed as 
\be\la{2.346} 
[\Phi\Phi]_{2n}^{S}\sim \Phi^{a}\partial_{t}^{2n}\Phi^{a},\qquad
[\Phi\Phi]_{2n}^{T}\sim \Phi^{(a}\partial_{t}^{2n}\Phi^{b)},\qquad
[\Phi\Phi]_{2n+1}^{A}\sim \Phi^{[a}\partial_{t}^{2n+1}\Phi^{b]}.
\ee
The connected  contributions \rf{2.35} 
 provide  the $1\ov \sqrt\lambda$ corrections to the OPE coefficients and scaling dimensions $ h_n
 = 2 + 2 n + { 1 \ov \sql}  \gamma^{(1)}+\cdots$   of these operators. 
In general, there is a mixing between $[\Phi\Phi]^{S}_{2n}$
(with  $n>0$) and $\mb F \mb F$ and 2-fermion operators, while $[\Phi\Phi]_{2n+1}^{A}$
mixes with 2-fermion states in the $(\bm{1},\bm{10})$ of $SU(2)\times Sp(4)\simeq SO(3)\times SO(5)$.
The mixing is absent   for  $[\Phi\Phi]^{S}_{0}$ or $[\Phi\Phi]^{T}_{2n}$  and for 
 these operators one finds  (see Appendix \ref{app:BPSanom})
 \be
\la{2.37}  h_n  = 2 + 2 n + \tfrac{1}{\sql}  \gamma^{(1)}+\cdots \ , \qquad 
\gamma^{(1)} _{[\Phi\Phi]^{T}_{2n}} = -3\,n-2\,n^{2},\qquad 
\gamma^{(1)}_{[\Phi\Phi]^{S}_{0}}  = -5\ . 
\ee
Assuming   that one can identify the   scalar $\Phi_{6}$ coupled to the WML   with the singlet composite  field 
$y^{a}y^{a}\sim [\Phi\Phi]_{n=0}^{S}$    one finds that   strong coupling  expansion of its dimension should be given by \rf{5}.

Finally, let us mention   that one can similarly compute  the strong-coupling expansion  of the 
 correlation functions 
involving  AdS$_5$  coordinates $x^i$    dual to  the  dimension $\Delta =2$ operator 
${\mathbb F}_{it}$  inserted on the Wilson line.  In particular, one finds  for the connected part of the mixed correlator of two AdS and two sphere fluctuations \cite{Giombi:2017cqn}
\begin{align} 
&\llangle   {\mathbb F}^{i}_{t}  (t_1)\,   {\mathbb F}^{j}_{t}  (t_2) \, \Phi ^{a}(t_3)\, \Phi^{b}(t_4)\, \rrangle_{\rm conn}=\langle x^{i}(t_1)x^{j}(t_2)y^{a}(t_3)y^{b}(t_4)\rangle_{\rm conn} 
= \delta^{ij}\delta^{ab} \frac{G_{\rm conn} (\chi)}{t_{12}^4\,  t_{34}^2}  \ ,\la{151}\\
&
G_{\rm conn}(\chi) = \tfrac{1}{\sql} {\cal C}_{1}  {\cal C}_{2} \, G^{(1)} (\chi)   =\tfrac{1}{\sql} \tfrac{2}{3\pi^2} G^{(1)} (\chi)  \ , \qquad \qquad G^{(1)}
=- {4}\,\Big[   1   -   \big(\half - \tfrac{1 }{ \chi} \big)   \ln   (1 - \chi) \Big] \ . \la{153}
\end{align}
The explicit expression for the 4-point correlator of $x^i\sim {\mathbb F}^{i}_{t}$ can also be found in \cite{Giombi:2017cqn}. 

\subsection{Analytic continuation to the ``chaos configuration''}
\label{chaos}

It is interesting to consider the analytic continuation of the above results to the out of time order correlator relevant to chaos \cite{Maldacena:2015waa, Maldacena:2016hyu}. Let us focus on the $SO(5)$ singlet part of the 4-point function of sphere coordinates, which is given by the contracted correlation function 
$\langle y^a(t_1)y^a(t_2) y^b(t_3) y^c(t_4)\rangle$. Following \cite{Maldacena:2016hyu}, in order to obtain the relevant thermal out of time order configuration $y^a(t)y^b(0)y^a(t)y^b(0)$, one can map the line to the thermal circle by $t_i=\tan(\pi\tau_i/\beta)$, $i=1,\ldots,4$, and then continue to real time.\footnote{Equivalently, one should also be able to obtain the result by computing the 4-point functions directly in AdS Rindler coordinates $ds^2 = -(r^2/r^2_h-1)dt^2+\frac{dr^2}{r^2/r^2_h-1}$.} A convenient configuration considered in \cite{Maldacena:2015waa} is given by taking the four operators to be equally spaced along the thermal circle. This configuration can be obtained by setting $\tau_1=i t, \tau_2= i t+\beta/2, \tau_3=\beta/4, \tau_4=-\beta/4$, which corresponds to a value of the cross ratio 
\begin{equation}
\chi = \frac{2}{1-i\sinh(\frac{2\pi t}{\beta})}\,.
\end{equation}
In order to reach this configuration, one has to start from the expression for $G_S(\chi)$ valid in the region $\chi>1$, which can be simply obtained from (\ref{2.35}) by letting $\log(1-\chi)\rightarrow \log(\chi-1)$. Then, one may take a large $t$ limit (corresponding to the formal small $\chi$ limit of the $\chi>1$ expression) to probe the chaotic behavior. Applying this procedure to the result for $G_S(\chi)$ given in (\ref{2.35}), we find for the out of time order correlator
\begin{equation}
\frac{\langle y^a(t)y^b(0)y^a(t)y^b(0)\rangle}{\langle y^a y^a \rangle \langle y^b y^b \rangle} \simeq 1-\tfrac{\pi}{2\sqrt{\lambda}}\, e^{\frac{2\pi t}{\beta}}\, , 
\label{Lyap}
\end{equation}
where we have normalized by the product of 2-point functions 
(omitting  the explicit positions along the thermal circle). The behavior (\ref{Lyap}) corresponds to a maximal Lyapunov exponent $
\frac{2\pi}{\beta}$. The same behavior can be seen to arise from the $\langle xx yy\rangle$ correlator in (\ref{153}) and the $\langle xxxx\rangle$ correlator that can be found in \cite{Giombi:2017cqn}. This maximally chaotic behavior for correlators on the string worldsheet was also found previously in \cite{Maldacena:2017axo, deBoer:2017xdk}. 

In our static gauge approach, this result can be seen to be essentially due to the 4-derivative vertices in the Nambu-Goto action: these lead to terms in the 4-point functions of the form $\simeq \chi^{-1}\log(1-\chi)$, which are responsible for (\ref{Lyap}). We will see below that the same behavior persists for the correlators on the non-supersymmetric Wilson line, indicating that it is not sensitive to the boundary conditions. This should be due to the fact that the limit relevant to chaos is captured by the near horizon region, which is essentially flat space.\footnote{We thank Juan Maldacena for discussions on these points.} The chaotic behavior (\ref{Lyap}) should then also be related to the ``gravitational-type'' phase shift found in \cite{Dubovsky:2012wk} for the S-matrix on a long string in flat space. It would be interesting to further clarify the relation of our calculations to the exact flat space S-matrix of \cite{Dubovsky:2012wk}. 

\section{Non-supersymmetric Wilson line  case:  $SO(6)$ invariant  correlators } 
\la{sec:main}

Let us now  turn to the case of strong-coupling description of  correlators   on non-supersymmetric WL. 
As discussed  in the introduction, the corresponding non-supersymmetric 
CFT$_1$  should be dual to the   AdS$_2$  theory   defined by the  same string  action \rf{2.9}--\rf{2.99} 
but now with Neumann boundary conditions   for the $S^5$ fluctuations \ci{Alday:2007he,Polchinski:2011im,Beccaria:2017rbe}: 
$\del_s  y^a\big|_{s=0}=0$ (cf. \rf{22},\rf{210}). Then the    $SO(6) $  symmetry  of scalar  correlators 
will be restored  due to the remaining integration over the  unfixed  ``zero mode''   part of $y^a$. 

This may be implemented systematically   using the embedding coordinates $Y_A$ (without choosing 
explicitly  a particular parametrization or  solution of $Y_A Y_A =1$  as in \rf{8}).  Ignoring  the  dependence   on the transverse AdS$_5$ fluctuations 
 $x_i$  in the string action \rf{2.9} the  bosonic Lagrangian in the static gauge will take the form 
\begin{align}
\la{3.2}
 L_B   &= \sqrt{ \det ( g_{\m\n} + \del_\m Y_A \del_\n Y_A ) } = \sqrt g\, (1  + L_2 + L_4 + \cdots ) \ , \\
L_{2} &= \tfrac{1}{2} \partial^\m Y_{A} \partial_\m  Y_{A}, \qquad \qquad 
L_{4} = \tfrac{1}{8}\,(\partial ^\m Y_{A} \partial_\m  Y_{A})^{2}
-\tfrac{1}{4}(\partial^\m  Y_{A} \partial_\m  Y_{B})^{2}\ ,\la{32} 
\end{align}
so that the  path integral over   $S^5$    will   be ($Y^2\equiv  Y_A Y_A$,\  $T= \tfrac{\sql} { 2 \pi }$) 
\be
Z = \int \mc DY\,\delta({Y}^{2}-1)\,\exp\Big(-T\,\int d^{2}\sigma\sqrt g   \big[L_{2}(Y)+L_{4}(Y)+\dots\big]\Big) \ . \la{33} 
\ee
Let us separate the constant part  $n^A$  of $Y^A$   that   selects  a particular point on $S^5$  as
\be \la{333}
Y^{A}=n^{A}+\ysx^{A}(\s) \ , \qquad\qquad   n^2=1\ .\ee
 Then \rf{33} takes the form 
\be\la{3.3} 
Z = \int [dn] \int \mc D\ysx\ \delta\big ({n}_A {\ysx}_A +\tfrac{1}{2} {\ysx}_A \ysx_A\big)\,
\exp\Big(-T\,\int d^{2}\sigma\sqrt g \big[L_{2}(\ysx)+L_{4}(\ysx)
+\dots\big]\Big)  \ , 
\ee
where $\int [dn] ...\equiv  \int d^6 n\ \delta( n^2-1)...$  is the integral over $S^5$. 
The $\delta$-function constraint on $\ysx^{A}$   can be  solved perturbatively in powers of an independent fluctuation 
$\yz^A$ orthogonal  to $n^A$   as\foot{In the special case of the $y^a$ parametrization in \rf{8},\rf{2.9} 
we   had $n^A=(0,0,0,0,0,1)$  and  $\z^6=0,\  \yz^a= y^a $.} 
\begin{align}  \la{3.4}
\ty^{A}  &=  f( \yz^2)\,  n^A   + h(\yz^2)\,  \yz^A \ , \ \ \ \ \qquad     n_A \yz_A =0 \ , \\
f&= -\tfrac{1}{2}\, \yz^{2}- (\text{a}+\tfrac{1}{8})\,({\yz}^{2})^{2}+\dots,    \qquad 
h= 1+\text{a}\,{\yz}^{2}+\dots \ , \qquad  \yz^2 = \yz_A \yz_A \ ,   
\end{align}
where a is an arbitrary coefficient. We can always  choose a=0 or  redefine\foot{Such local field  redefinitions   should preserve the  ``on-shell''  correlators in AdS$_2$, see Appendix \ref{app:alternative}.} 
 $h(\yz^2)\,  \yz^A \to \z^A$. This is equivalent to 
defining  $\z^A$   as the  part of $Y^A$     orthogonal  to $n^A$.  This is what  we shall do  below, i.e. set 
\be
\la{3.5}
Y^{A} =\sqrt{ 1 - \z^2}  \, n^A + \z^A = 
    \big[1-\tfrac{1}{2}{\z}^{2}-\tfrac{1}{8}({\z}^{2})^{2}+\dots\big] n^A + \z^A \ , \qquad \qquad n^A \z^A =0 \ . 
\ee
Then  the path integral \rf{33}  or \rf{3.3} takes the form 
\begin{align}\la{330} 
Z &= \int [dn] \int \mc D\z \ \delta\big({n}_A \z_A \big)\,
\exp\Big(-T\,\int d^{2}\sigma\sqrt g \big[L_{2}(\z)+L_{4}(\z)
+\dots\big]\Big)  \ , \\
L_{2} &= \tfrac{1}{2} \partial^\m \zeta^{A}\, \partial_\m  \zeta^{A}\ , \qquad 
L_{4} = \tfrac{1}{2}\,\zeta^{A}\zeta^{B}\, \partial^\m \zeta^{A} \partial_\m\zeta^{B} 
+\tfrac{1}{8}\,(\partial^\m \zeta^{A}\, \partial_\m \zeta^{A})^{2}
-\tfrac{1}{4}(\partial^\m \zeta^{A}\,
\partial_\m \zeta^{B})^{2}\ ,  \la{3.7}
\end{align}
where we have substituted  \rf{3.5} into    \rf{32} keeping only terms up to quartic order in $\z^A$. 

The  propagator for  the massless scalar  field  $\zeta^{A}$ (with 5 independent components for fixed $n^A$) 
 is then  given   by 
\be\la{3.11} 
\langle \zeta^{A}(\sigma)\zeta^{B}(\sigma')\rangle = P^{AB}(n)\,\rG_{\rm N}(\sigma, \sigma'),\qquad 
\qquad P^{AB} = \delta^{AB}-n^{A}\,n^{B},
\ee
where $P^{AB}$ is the projector  orthogonal to $n^A$ and $\rG_{\rm N}$ is 
 the bulk  Green's  function  in AdS$_2$   \rf{210}  corresponding   to the Neumann  boundary conditions 
 (see Appendix~\ref{app:propagator})
 \be
\la{3.9}
\rG_{\rm N}(\sigma, \sigma') =  -\tfrac{1}{4\pi }
\Big( \log[  (t-t')^2+(z-z')^2 ] 
+\log[ (t-t')^2+(z+z')^2] \Big) \ .
\ee
The corresponding bulk-to-boundary propagator   will be  also denoted as $\rG_{\rm N}$:
\be\la{399}
\rG_{\rm N}(t,z; t') \equiv \rG_{\rm N}(t,z; t',0) = -\tfrac{1}{2\pi }  \log[  (t-t')^2+ z^2 ] 
 \ .
\ee
We will  also use  boundary-to-boundary propagator 
\be\la{400}
 \rG_{\rm N} (t_1,t_2) \equiv  \rG_{\rm N}(t_1,0; t_2,0) =  -\tfrac{1}{2\pi }   \nD_{12} \ , \qquad \qquad 
 \nD_{12} \equiv \log (t_{12} ^2) \ . 
\ee
As in the static   gauge (used  in \rf{2.9},\rf{33})   which is adapted to the expansion  near the  WL minimal  surface 
   the  target-space AdS coordinate $z$ is identified with  the world-sheet  coordinate $s$  (see \rf{22}) we shall often use 
 $\sigma^\m=(t,z)$ as the  coordinates in the AdS$_2$  bulk theory. 
 The propagator \rf{3.9}   is the standard  Neumann  one on a  half-plane  ($z\geq 0$) 
 with a  conformally-flat  metric  (the dependence on   conformal factor drops out due to the   conformal invariance of the 
 massless  scalar kinetic term in \rf{330}.
 The conformal factor re-enters   via  a covariant  UV cutoff, e.g., 
 after    the replacement $ (t-t')^2+(z-z')^2 \to [ {(t-t')^2+(z-z')^2\ov zz'} +\eps^{2}] \,z\,z'$  (see  Appendix~\ref{app:propagator}).

 In what follows we shall   use this  $SO(6)$ covariant set-up \rf{330}--\rf{3.9} 
 to compute correlation functions of the $S^5$  embedding coordinates 
 that should   give  as in \rf{9}   the  corresponding  scalar   correlators in the boundary CFT$_1$. 
 The expectation value  $ { \langle   Y_{A_1}   (t_1)\cdots Y_{A_n} (t_n)   \rangle}_{_{\rm AdS_2}} $ 
 will   be  computed according to \rf{3.5},\rf{330},\rf{3.7}, i.e. will include   integrating over $\z^A$ as well as 
 averaging over  the  $S^5$ direction $n^A$. From now on we shall  
   denote the AdS$_2$   expectation value  simply by $\langle \cdots \rangle$. 

The averaging over $S^5$  can be done using 
\begin{align} 
 \la{310}
\langle  n^A n^B   \rangle  &= \tfrac{1}{ 6} \delta^{AB} \ , \qquad 
\langle  n^A n^B n^C n^D    \rangle = \tfrac{1}{48}  \big( \delta^{AB} \delta^{CD}  +  \delta^{AC} \delta^{BD}  +
 \delta^{AD} \delta^{BC} \big) \ , \\ \la{3100}
\langle  P^{AB}   \rangle &=  \tfrac{5}{ 6} \delta^{AB} \ , \qquad 
\langle  P^{AB}  P^{CD}  \rangle  =  \tfrac{33}{48} \delta^{AB} \delta^{CD}  +   \tfrac{1}{48}  \big( \delta^{AC} \delta^{BD}  +
 \delta^{AD} \delta^{BC} \big) \ , \ \ \   etc.  
\end{align}
This averaging   restores $SO(6)$ symmetry and implies  that all 
correlators with odd  number of $Y_A$ should vanish, i.e. non-vanishing  ones should  be $\langle  YY  \rangle, 
\langle  xx YY  \rangle, \langle  YYYY   \rangle  $, etc.

\section{Two-point function $\langle Y^{A}Y^{B}\rangle$}
\la{sec:2-point}

The  2-point   boundary-to-boundary   correlator of $Y_A$ is supposed to reproduce  the strong-coupling expansion 
of the 2-point function of the $SO(6)$ scalars \rf{6}. Its structure is fixed by 1d conformal invariance to be  
($Y^A(t) \equiv Y^A(t, z=0)$) 
\begin{align}
\langle  Y^A(t_{1}) \,  Y^B(t_{2})\rangle = \delta^{AB} \frac{C_Y}{|t_{12}|^{2\D} }
= \delta^{AB}  C_Y \Big[ 1 &- \big( \tfrac{d_1 }{\sqrt\lambda}  +   \tfrac{d_2 }{(\sqrt\lambda)^2} +...\big) \,\log(t_{12}^{2}) \no \\
&+  \la{4.1} 
(\tfrac{d_1^2}{2\, (\sql)^2}  + ...) \log^2 (t_{12}^{2})+\cdots\Big], \\
\Delta = {\te {d_1\ov \sql}} + {\te{d_2\ov (\sql)^2 }} + {\te{d_3\ov (\sql)^3 }} +  \cdots \ ,& \ \ \ \qquad\ \     d_1 =5 \ ,   \la{41} 
\end{align}
where  the $d_1=5$ is the  expected   value of the leading  anomalous dimension coefficient \rf{7}.
 The  subleading  $d_2\ov  (\sql)^2 $ contribution to $\log$ term and thus to $\Delta $  should    come 
 from the 1-loop diagrams involving  also  the fermions  (see below). 
 
Note that the normalization of the 2-point function of the conformal operator dual to $Y^A$ is scheme dependent and hence arbitrary. On the string side, since the two-point function starts with $\langle  n^A  n^B\rangle = {1\ov 6} \delta^{AB}$, it appears to be natural to choose a scheme where     to all orders 
\be \la{402} C_Y ={\te  {1\ov 6}} \ \ee
so that the condition $Y^A Y^A=1$ at coincident points is preserved.\foot{One may 
ensure  the expected normalization of \rf{4.1}  at the coinciding points  $ \langle  Y^A(t) \,  Y^A(t )\rangle  =1$
by explicitly  keeping track of  the  boundary UV   cutoff  dependence as in 
$\langle  Y^A(t_{1}) \,  Y^B(t_{2})\rangle = {1\ov 6} \delta^{AB} \big[\frac{\epsilon^2}{|t_{12}|^2 + \epsilon^2}\big]^{\D}  $.
We will not do this  below.} This should correspond to fixing a particular choice of 2-point function normalization of the dual operator $\Phi^A$ inserted on the WL. 

\subsection{Leading  logarithmic correction}
Using \rf{3.5}, \rf{3.11}  and \rf{3100}    we find ($T^{-1}  = \tfrac{ 2 \pi }{\sql} $)
\begin{align}
\la{4.2}
\langle Y^{A}(\s_1) \, Y^{B}(\s_2) \rangle = \langle \big[ n^{A} +\z^{A} +\cdots\big]  \big[n^{B} +\z^{B}+\cdots\big]\rangle
=  \tfrac{1}{6}\delta^{AB}\,\big[ 1  + 5\,T^{-1}  \rG_{\rm N}(\sigma_1, \sigma_2)  + \cdots\big] \ . 
\end{align}
Setting $z_1,z_2\to 0$ in the propagator  in \rf{3.9},\rf{399} 
 we thus  readily reproduce  the value  $d_1=5$
in \rf{4.1}.
   We have ignored  the contribution of  the $-\ha \z^2 n^A$ term in $Y^A$  in \rf{3.5}
 as it leads (to the leading order) only to a  cutoff-dependent  constant.

As discussed in \ci{Beccaria:2017rbe},  this value is the $J=1$    case of the $J(J+4) $ eigenvalue 
corresponding to the   $S^5$  scalar spherical
harmonic  with angular momentum $J$. One may, indeed, generalize 
the computation in \rf{4.2}  to the correlator 
$\langle V^{A_1 \dots A_J} (\s_1) \,  V^{B_1 \dots B_J}  (\s_2) \rangle $  where 
$V^{A_1 \dots A_J} = Y^{\{A_1} \cdots Y^{A_J\}}$  is a totally symmetric traceless   tensor.  
It is sufficient to consider  the correlator of two primary fields 
$\langle Z^J \bar Z^J \rangle$  where $Z= u_A  Y_A$ with constant complex  null vector $u_A$   ($  u^2=0$). 
For example,  we may use  $Z=  Y_1 + i Y_2$    and then 
\be 
\langle Z^J (\s_1)  \bar Z^J (\s_2)  \rangle
= \langle \big[  M_{J}-J^{2}\, \, (M_{J}-2M_{J-1}) T^{-1}  \rG_{\rm N} (\s_1,\s_2)\big]  \rangle +\dots ,\qquad 
M_{J} \equiv  |n_{1}+i\,n_{2}|^{2J}
 \ , \la{4.4}
\ee
where   the remaining  $S^5$ average  can   be done, e.g.,  by using  the explicit spherical  angle parametrization of $n^A$.\foot{Explicitly, 
$
\langle M_{J}\rangle = \frac{1}{\pi^{3}}\int_{0}^{2\pi}d\phi
\int_{0}^{\pi}d\theta_{1}\dots d\theta_{4}\,\sin^{4}\theta_{1}\sin^{3}\theta_{2}\sin^{2}\theta_{3}
\sin\theta_{4}\ |\cos\theta_{1}+i\,\sin\theta_{1}\cos\theta_{2}|^{2J}
 = \tfrac{2}{(J+1)(J+2)}. \no $
}
  As a result,  $\langle M_{J}\rangle = \frac{2}{(J+1)(J+2)}$ and  thus 
\be\la{4.5} 
\langle (Y_{1}+i\,Y_{2})^{J} (\s_1) \ (Y_{1}-i\,Y_{2})^{J}(\s_2) \rangle = \tfrac{2}{(J+1)(J+2)}\big[
1+J(J+4)\, T^{-1}  \rG_{\rm N} (\s_1,\s_2) \big] +\cdots\ ,
\ee
with $J(J+4)$ thus replacing 5 in \rf{4.2}.

\subsection{Subleading  corrections } 
\la{sec:loop-corrections}

The order $1\ov (\sql)^2$ corrections to the 2-point function will   be given by the sum of the $\log$ and $\log^2$  terms in \rf{4.1}.
 The $d_2 \log$ term  should   originate  from the bosonic  
 ($\zeta^A$   and $x^i$, cf. \rf{2.8})  and fermionic  1-loop   diagrams --
the second  and third diagrams  in Fig. \ref{fig:loop}. We will not
  systematically include fermions  and thus will not determine $d_2$ here.

The 1d  conformal invariance of \rf{4.1}   implies   that the leading logs  at each order in $\frac{1}{(\sql)^{n}}$ should exponentiate. Thus 
 at order $\frac{1}{(\sql)^{2}}$   we should find the  $\log^2(t^2_{12}) $   term with  the coefficient being precisely 
 ${d^2_1\ov 2} = {25\ov 2}$.
To demonstrate  this requires  to go beyond the tree approximation and 
 include  the  loop contributions  of the interacting vertices in \rf{330}.\foot{It is useful 
 to compare  the present case  with that  of a free  scalar  2d  theory  which also has a logarithmic propagator, 
 $\langle XX\rangle \sim \log |z_{12}| $. Here  a  primary operator without derivatives    which will have 
 $\langle OO \rangle \sim |z_{12}|^{-2\Delta} $
  is $O= e^{a X}$.  The  choice of the exponential function is  essential for  the right combinatorics. 
  One may of course  redefine $ X\to  X', \  X=  a^{-1} \log(1 + a X') $ so that $O=1 + a X'$ 
   but then  the required contributions will come from the expansion of the redefined action $L= (\del X)^2 = { (\del X' )^2 \ov (1+ a  X')^2}$. 
  Similarly, in  the present  case  of    $Y^A(\z) = \sqrt{1-z^2} n^A + \z^A =  n^A + \z^A - \ha z^2  n^A + ...$ 
  with the  propagator of $\z$    given by \rf{400} we will not get  the  correct exponentiation 
  of $\log t_{12}^2 $   without  including extra  contributions  from loop diagrams with the interacting 
  vertices from the  action. }
  
 At order $\frac{1}{(\sql)^{2}}$   
we need to  consider  the  1-loop   contributions from the vertices   in $L_4$ in \rf{3.7}   and these 
require UV regularization. In general,  the coefficients   in the finite contributions  will  depend on a  scheme
and, as usual,  the scheme  should be chosen  so that to  preserve the required (world-sheet  and target space) 
symmetries (cf. Appendix~\ref{app:propagator}). 
\begin{figure}[ht]
\centering
\begin{tikzpicture}[line width=1 pt, scale=0.5]
\draw[densely dashed] (0,0) circle (2);
\draw (0:2) to [out=120,in=60] (180:2);
\draw (0:2) to [out=-120,in=-60] (180:2);
\node at (0,1.5) {$\z$};
\node at (0,-1.5) {$\z$};
\draw[fill=white] (0:2) circle (0.12);    \draw[fill=white] (180:2) circle (0.12);
\node at (3,0){$\ $};   
\end{tikzpicture}
\begin{tikzpicture}[line width=1 pt, scale=0.5]
\draw[densely dashed] (0,0) circle (2);
\draw (0:2)--(180:2);
\draw (0,0.6) circle (0.6);
\node at (3,0){$\ $};   
\node at (1.2,0.5) {$\z$};
\node at (-1.2,0.5) {$\z$};
\node at (0,1.6) {${ \z, x } $}; 
\draw[fill=white] (0:2) circle (0.12);    \draw[fill=white] (180:2) circle (0.12);
\draw[fill=black] (0,0) circle (0.12); 
\end{tikzpicture}
\begin{tikzpicture}[line width=1 pt, scale=0.5]
\draw[densely dashed] (0,0) circle (2);
\draw (0:2)--(0:1);
\draw (180:2)--(180:1);
\draw[middlearrow={>}] (1,0) arc (0:180:1);
\draw[middlearrow={>}] (-1,0) arc (-180:0:1);
\node at (1.4,0.5) {$\z$};
\node at (-1.4,0.5) {$\z$};
\node at (0,1.5) {$\psi$};
\node at (0,-1.5) {$\psi$};
\draw[fill=white] (0:2) circle (0.12);    \draw[fill=white] (180:2) circle (0.12);
\draw[fill=black] (1,0) circle (0.12); \draw[fill=black] (-1,0) circle (0.12); 
\end{tikzpicture}
\caption{Diagrams contributing the 2-point function $\langle YY\rangle$ at order $\frac{1}{(\sql)^{2}}$.}\label{fig:loop}
\end{figure}

There are three types of diagrams   contributing to  the 2-point function \rf{4.1} 
at order $\frac{1}{(\sql)^{2}}$  are shown in  Fig.~\ref{fig:loop}:   (i) the tree-level  one with the 
 contraction of  the $\z^2 n^A$ terms in $Y^A$ in \rf{3.5}
that  does not involve interaction vertices;
(ii)  bosonic 1-loop diagrams  with quartic  vertices   from  $L_4$ in \rf{3.7};   
(iii)  fermionic 1-loop   diagrams   with vertices  from the  fermionic terms in the full AdS$_5 \times S^5$ 
superstring action (which were ignored in \rf{2.9}). 

While the  fermionic loop contribution is  important  for  computing the subleading $d_2$ coefficient in the 
scaling dimension \rf{41},   given that $d_1$ in  \rf{41}   receives   contribution  only from bosons it  might  be  natural to expect 
that  finite  $\frac{1}{(\sql)^{2}}\log^2(t^2_{12}) $ terms in \rf{4.1} should also come  only from the    bosonic  1-loop contributions. 
Still, given that the fermionic contribution is  crucial for ensuring  the UV finiteness  of the 2d theory
(and  given  that, in general, there are power-like divergences   in the purely bosonic theory) 
 this  issue  may be regularization scheme dependent. Below we shall  
 assume that    there is no $\log^2(t^2_{12}) $ term coming  from the  fermionic loop in  Fig.~\ref{fig:loop}
 and  concentrate only on the bosonic  contributions, 
i.e. the first two diagrams  in  Fig.~\ref{fig:loop}. 

The contribution of the   first  diagram  in  Fig.~\ref{fig:loop}  is 
$\langle \frac{1}{2}\,{\zeta}^{2}(t_1 )\,  n^A \    \frac{1}{2} {\zeta}^{2}(t_2)\,n^{B}\rangle $  so it  should 
 correct \rf{4.2} (restricted to the boundary points
$\s_a=(t_a, 0)$)  by  $\g_2 \,  T^{-2} [ \rG_{\rm N} (t_1,t_2) ]^2 $ term.  In general,   we should find   ($\rG_{\rm N} (t_1,t_2) = - {1\ov 2 \pi} \nD_{12} $,  see \rf{400})
\begin{align}
\langle Y^{A}(t_1)\, Y^{B}(t_2) \rangle &=  
 \tfrac{1}{6}\,\delta^{AB}\,\big[1+ \tfrac{\g_1}{\sql} \nD_{12}   + \tfrac{\g_2}{(\sql)^2}   (\nD_{12})^2    +   \tfrac{\g_3}{(\sql)^3}   (\nD_{12})^3     +\cdots\big] \la{4.11} \ , \quad  \nD_{12} =\log(t_{12}^2)  \\
 &\g_1 = - d_1 = -5, \qquad \qquad   \g_2=\g_2^{(0)} + \g_2^{(1)}, \ \ \    \ \ \ \  \g_2^{(0)}=   \tfrac{5}{2}\ .  \la{4.123}
\end{align} 
The tree-level  contribution  $ \g_2^{(0)}= {5\ov 2}$ here   should be part of the  
total coefficient $\g_2= {d^2_1\ov 2} = {25\ov 2}$ in \rf{4.1};  the additional  term 
 $\g_2^{(1)}= {20\ov 2} =10$ should  come from the  1-loop diagrams. 

As we shall   see below,  it is only 
 the  first (``sigma-model'')  quartic vertex in $L_4$ in \rf{3.7}  that  will  contribute to the  leading $\log^2$  term in \rf{4.11}. 
It  will lead  
to several   1-loop contributions to the  correlator 
\be \la{4113} \langle \z^A(t_1) \, \z^B(t_2) \rangle=  \tfrac{1}{6} \delta^{AB} \Pi(t_{12}) \ . \ee
 One  comes from the  contraction
$ \int d^{2}\sigma\,\sqrt{g} \, 
\wick{121}{\langle 
<1\zeta^{A}(t_{1})<2\zeta^{B}(t_{2})\ >1\zeta^{C}>2\zeta^{D}\,(\partial <3\zeta^{C}\cdot\partial>3\zeta^{D})
\rangle}$ (plus  permutations).  Its  contribution is found to  be 
\begin{align} \la{4.8} 
& \Pi_{1} = -\tfrac{\sql}{2\pi} \times {5} \times   (\tfrac{2\pi}{\sql})^3\times  X_2 \times       I_2 \ , \qquad I_2= 
\int\frac{dzdt}{z^{2}}\ \GN(t, z; t_1)\, \GN( t, z; t_2)=  \tfrac{1}{4\pi} \,\log^{2}(t_{12}^{2}) \ , \\
&\qquad \qquad \qquad \qquad  X_2 = \lim_{\sigma'\to\sigma}  g^{\m\m'} \partial_{\mu}\partial'_{\mu} \GN(\sigma, \sigma') = 
 \tfrac{k}{4\pi}    \ ,   \la{4.15} 
\end{align}
where $\rG_{\rm N}(t,z;t' )$  is the bulk-to-boundary propagator  \rf{399}.
$X_2$ originates from $\langle \partial\zeta(\sigma)\cdot\partial\zeta(\sigma)\rangle$ 
and its value, in general, depends on a scheme: such correlators are,  in general, 
power  divergent and in \rf{4.15}   we dropped quadratic divergence   (cf. \rf{3366},\rf{D.12}). 
The value   of $k$  \rf{4.15}  found using the naive  point-splitting is $k=1$   but  in  AdS$_2$ case (in the presence of the boundary) 
   a  more   natural  value is $k=2$ (see discussion at the end of Appendix \ref{app:propagator}   and \rf{ddd}). 
   
The bulk integral  $I_2$  in \rf{4.8}   is computed 
using  that $\log X = -  \lim_{\eps\to 0}   X^{-\eps}$   and thus 
    $I_2 = {1\ov (2\pi)^2}  \lim_{\eps_{1,2}\to 0}  \bar I_2$, where 
\be
\la{4.16}
\bar I_2= 
\frac{\partial^2}{\partial\eps_{1}\partial \eps_2}
\frac{\Gamma(\eps_{1}+\eps_{2})}{\Gamma(\eps_{1})\Gamma(\eps_{2})}\int_{-\infty}^{\infty}dt \int_{0}^{\infty}\frac{dz}{z^{2}}\int_{0}^{1}dx
\frac{x^{\eps_{1}-1}(1-x)^{\eps_{2}-1}}{[x\,((t-t_{1})^{2}+z^{2})+(1-x)\,((t-t_{2})^{2}+z^{2})]^{\eps_{1}+\eps_{2}}}
\ee
The resulting contribution to $\g_2$ in \rf{4.11} is $(\g^{(1)}_2)_1=- {5\ov 4}k $. 

Another   contribution originates   from the  contractions 
$\int d^{2}\sigma\,\sqrt{g} 
\wick{211}{\langle 
<1\zeta^{A}(t_{1})<2\zeta^{B}(t_{2})\ >2\zeta^{C}<3\zeta^{D}\,
(\partial >1\zeta^{C}\cdot\partial>3\zeta^{D})
\rangle}$ and  $\int d^{2}\sigma\,\sqrt{g} 
\wick{211}{\langle 
<1\zeta^{A}(t_{1})<2\zeta^{B}(t_{2})\ >2\zeta^{C}<3\zeta^{D}\,(\partial >1\zeta^{D}\cdot\partial>3\zeta^{C})
\rangle}
$. 
Using that  from (\ref{3.9})
\be\la{4.12} 
\Big[\partial_{\mu} \GN(\s, \s') \Big]_{\s=\s'}   = \begin{cases}  0, & \mu=0\\ -\frac{1}{2 \pi\, z}, & \mu=1 \ , 
\end{cases}
\ee
   we get the following analog of \rf{4.8} 
(with  averaging over $n_A$ computed using \rf{3100} and $P^{CC}=5$)
\be\la{413}
\Pi_{2} = -\tfrac{\sql}{2\pi}\times \tfrac{1}{2}\times 2^{2} \times 30  \times   \tfrac{1}{\sql}\,  (\tfrac{2\pi}{\sql})^2\times  I_2'  \ , \ee 
where the bulk integral $I_2'$  is  related to $I_2$ in \rf{4.8}  via integration by parts
\be\la{414}
 I_2'= -
\int  \frac{dzdt}{z^{2}} \ z \,   \partial_{z} \GN(t, z; t_1)\, \GN( t, z; t_2)=  
  -\tfrac{1}{2 }\int \frac{dzdt}{z} \, \partial_{z}\Big[    \GN(t, z; t_1)\, \GN( t, z; t_2)       \Big] = -\tfrac{1}{2 }  I_2 \ . \ee 
  As a result, we get  an extra contribution  to  $\g_2$ in \rf{4.11}:  $(\g^{(1)}_2)_2 = {15}$. 

The remaining  term   from the first vertex in  (\ref{3.7}) 
$ \int d^{2}\sigma\,\sqrt{g} \,
\wick{231}{\langle 
<1\zeta^{A}(t_{1})<2\zeta^{B}(t_{2})\ <3\zeta^{C}>3\zeta^{D}\,(\partial >1\zeta^{C}\cdot\partial>2\zeta^{D})
\rangle}
$
contains  the logarithmically divergent contribution ($\eps$ is the  covariant  bulk UV cutoff, see \rf{3.9},\rf{D.7})
\be\la{415} 
\GN(\s, \s)  = -\tfrac{1}{2\pi }\log ( 2 \eps^2)   -\tfrac{1}{\pi}\,\log  z \ .
\ee
The   UV divergent   term should be  absorbed   into  
the renormalization of  the radius  of $S^5$    in the purely bosonic model 
but    should   be  cancelled   by   the fermionic loop  contribution   in the   superstring case. 
If we assume that the fermionic contribution cancels $\log \eps^2$  term   but 
   does not change   the  coefficient of the finite $\log z$ 
term in \rf{415} 
we will get  the following additional contribution to \rf{4113} 
\begin{align}
\Pi_{3} = -\tfrac{\sql}{2\pi}\times \tfrac{1}{2}\times 2\times 5 \times 
(-\tfrac{2}{\sql}) (\tfrac{2\pi}{\sql})^2  \,\int\frac{dzdt}{z^{2}}\log z\,z^{2}\,\sum_{\m=1}^{2}
\partial_{\m }\GN(t,z; t_{1})\ \partial_{\m } \GN(t,z; t_{2})\la{417}\ . 
\end{align}
Integrating   by parts and using   that $\del_\m \del_\m \GN =0$  we get as in \rf{4.8},\rf{414} 
\begin{align}\la{418}
\Pi_{3} = - \tfrac{5}{6\,\pi}\,(\tfrac{2\pi}{\sql})^2 \int \frac{dz dt}{z}\,  \GN(t,z; t_{1})\ \partial_{z } \GN(t,z; t_{2})
&= - \tfrac{5}{12\pi}\,(\tfrac{2\pi}{\sql})^2  \int \frac{dz dt}{z}\,\partial_{z}\Big[ \GN(t,z; t_{1})\ \GN(t,z; t_{2})\Big]\notag \\
&=
-\tfrac{5}{12(\sql)^2 }\,\log^{2}(t_{12}^{2}) \ . 
\end{align}
The additional contribution to   $\g_2$ in \rf{4.11}  is thus   $(\g^{(1)}_2)_3 = - {5\ov 2}$. 

Thus   in total we get (adding also the ``tree-level'' contribution   $\g^{(0)}_2= {5\ov 2}$) 
\be \la{419}
\g_2 = \g^{(0)}_2 +  \g^{(1)}_2 = \te {5\ov 2}  +   \big(   - {5\ov 4}k  + 15 - {5\ov 2}\big)\Big|_{k=2}  =  \te {5\ov 2} + 10 =       {25\ov 2}  \ ,  \ee 
which agrees  with \rf{4.1}    in the scheme where  $k=2$ in \rf{4.15}.\foot{That this value is indeed  the natural one 
can be seen  by generalizing the bosonic $SO(6)$    computation to the  $SO(N)$ case. 
Then $d_1$   in \rf{4.1} becomes $N-1$   and thus ${d^2_1\ov 2} = {(N-1)^2\ov 2}$. 
The corresponding  analog of \rf{419}   is then 
$\g_2= \frac{N-1}{2} -  \frac{N-1}{4} k  + \frac{N (N-1)}{2} -\frac{N-1}{2}  $  which is 
   equal to   $ \frac{(N-1)^{2}}{2} $ precisely if $k=2$.}

Finally, let us   check   that 1-loop diagrams   with 
the  other two (4-derivative) vertices  in $L_4$ in \rf{3.7} 
  do  not contribute to the $\log^2$ terms in \rf{4.11},\rf{4113}. 
The second  vertex  in (\ref{3.7})  leads to two types of contractions. 
  The first is 
$\int d^{2}\sigma\,\sqrt{g} \, 
\wick{211}{
\langle 
<1\zeta^{A}(t_{1})\, <2\zeta^{B}(t_{2})\ \partial>2\zeta^{C}\cdot\partial>1\zeta^{C}
\ \partial<3\zeta^{D}\cdot\partial>3\zeta^{D}
\rangle} $; 
using (\ref{4.15}) and doing the bulk integral we find its contribution to \rf{4113}  to be 
\be\la{420}
 \Pi_4=  -\tfrac{\sql}{2\pi}\times \tfrac{1}{8}\times k \times 2^{2}\times  25 \times    
\left[-\tfrac{\pi}{(\sql)^3}\log(t_{12}^{2})\right] \ .
\ee
It thus  contributes to  the first power of log, i.e. to the coefficient $d_2$ in the scaling dimension \rf{41}. 
In the second contraction $\int d^{2}\sigma\,\sqrt{g} \, 
\wick{211}{
\langle 
<1\zeta^{A}(t_{1})\, <2\zeta^{B}(t_{2})\ \partial>2\zeta^{C}\cdot\partial<3\zeta^{C}
\ \partial>3\zeta^{D}\cdot\partial>1\zeta^{D}
\rangle}
$
we need to use    that (see \rf{D.11},\rf{D.12}) 
\begin{align} \la{3355}
&{\rm G}_{{\rm N}}  (\sigma,\sigma') = -  \tfrac{1}{4\pi}  \log{u (u+1)}    \ , \ \qquad \qquad 
u= \ha   \tfrac{ (t-t')^2 + (z-z')^2 }{2 z z'}  +  \eps^2  \ , \\
&\la{3366}
 \partial_{\mu}\partial_\nu' \GN (\sigma,\sigma') \Big|_{\s\to \s'}  =   \tfrac{1}{8 \pi z^{2}}  \big( \tfrac{1}{\eps^2}   + 1\big)
\delta_{\mu\nu}\ .
\end{align}
Then the  bulk integral gives again  only a  $\log$ term. The third   vertex  in (\ref{3.7}) 
 that has a different 
$SO(6)$ contraction structure leads to the same  bulk integral and thus also does not 
produce  $\log^{2}$ contributions to \rf{4113}. 

Similar conclusions  are reached for the 1-loop diagrams   with the $x^i$ loop 
coming from the $\del x \del x \del y \del  y$   vertex in  \rf{2.8} (where one can replace $y^a \to Y^A$). 
Here we will need to use that  the  bulk-to-bulk  AdS$_2$ Green's function   for the massive scalar $x^i$ satisfies (cf. \rf{3355},\rf{D.7})
\begin{align} \la{4201}
&{\rm G}_{{\rm D}}^{(m^2=2)}  (\sigma,\sigma') = -  \tfrac{1}{4\pi} \Big[  ( 2u + 1) \log\tfrac{u }{ u+1}   + 2 \Big] 
 \ , \\
& \partial_{\mu}\partial_\nu'  {\rm G}_{{\rm D}}^{(m^2=2)}  (\sigma,\sigma') \Big|_{\s\to \s'}  = 
 \tfrac{1}{8 \pi z^{2}}  \big( \tfrac{1}{\eps^2}   + 1 + 2 \log \eps^2 \big) \delta_{\mu\nu} \ . 
 \la{4501}
\end{align} 
As \rf{4501}  scales   with $z$  in the same   way  as \rf{3355}  the corresponding 1-loop diagram  also does not contribute
to $\log^{2}$ term (while the  UV log divergence  should   cancel  against the contribution of the fermionic loop).


At the  next ${1 \ov (\sql)^3}$  order the $(\nD_{12})^3= \log^3 ( t_{12}^2 )$  term in  \rf{4.1},\rf{4.11}     should have   the coefficient 
$\g_3= -  {d^3_1\ov 3!} =-  { 125 \ov 6}$.  As  the expansion of $Y^A$ in \rf{3.5}   does not contain   a  $\z^3$ term
(while    the 
$\z^4$ term in $Y^A$  
will start contributing  only at order $1\ov (\sql)^4$)
all   contributions to $\g_3$   should come  from  loop diagrams. 
 The first  type of them 
 is  the first   diagram  in Fig. \ref{fig:loop} where one of  the two  tree propagators  is replaced 
by the 1-loop corrected one (i.e. the one with the corrections  from the 1-loop graphs  in Fig. \ref{fig:loop}  included), see
Fig. \ref{fig:xxx}(a).
In view  of the above discussion   this 1-loop  ``self-energy'' dressing  amounts  to the following replacement 
of each log   factor in \rf{4.11}  (cf. the first  and the second terms in \rf{4.11}  
with $\g^{(1)}_2=5\times 2 $ according to \rf{419})\foot{This shift accounts just for the leading log contributions; 
in  addition,  there will be also subleading ones  that can be accounted for by a shift like in  \rf{GG2}.}
\be \la{401}
\nD_{12} \to \nD_{12} -   \tfrac{2}{\sql} (\nD_{12})^2  \ . \ee 
Applied to the tree-level $\g^{(0)}_2$  term  in \rf{4.11}  this   will give the  following contribution to $\g_3$: 
$\g_3^{(1)}= { 5 \ov 2} \times 2 \times (-2) = -10.$
\DeclareRobustCommand\myfeynA{%
 \begin{tikzpicture}[line width=0.6 pt, scale=0.06]
\draw (-4,0)--(4,0);
\draw (0,0) circle (2); 
\end{tikzpicture}}
\DeclareRobustCommand\myfeynB{%
 \begin{tikzpicture}[line width=0.6 pt, scale=0.06]
\draw (-5,-2)--(13,-2);
\draw (0,0) circle (2); \draw (8,0) circle (2);
\end{tikzpicture}}
\DeclareRobustCommand\myfeynC{%
 \begin{tikzpicture}[line width=1 pt, scale=0.06]
\draw (-5,0)--(5,0);
\draw[fill=lightgray] (0,0) circle (2);
\end{tikzpicture}}
\DeclareRobustCommand\myfeynD{%
 \begin{tikzpicture}[line width=1 pt, scale=0.06]
\draw (-4,-2)--(4,-2);
\draw (0,0) circle (2);
\end{tikzpicture}}
\begin{figure}[H]
\centering
\begin{tikzpicture}[line width=1 pt, scale=0.5]
\node at (0,-2.6) {(a)};
\coordinate (A1) at (180:2);
\coordinate (A2) at (0:2);
\node[left]    at (A1) {$-\frac{1}{2}n^{A}\bm{\z}^{2}\ $}; 
\node[right]  at (A2) {$\ -\frac{1}{2}n^{B}\bm{\z}^{2}\ $}; 
\draw[densely dashed] (0,0) circle (2);
\draw (A2) to [out=140,in=40] (A1);
\draw (A2) to [out=-140,in=-40] (A1);
\draw[fill=white] (0:2) circle (0.15);    \draw[fill=white] (180:2) circle (0.15);
\draw[fill=lightgray] (0,0.7) circle (0.75);
\draw[fill=white,white] (0,0.7) circle (0.3);
\node at (0,0.7) {1};
\end{tikzpicture}
\hskip 1cm
\begin{tikzpicture}[line width=1 pt, scale=0.5]
\node at (0,-2.6) {(b)};
\coordinate (A1) at (180:2);
\coordinate (A2) at (0:2);
\node[left]    at (A1) {$\z^{A}$};
\node[right]  at (A2) {$\z^{B}$};
\draw[densely dashed] (0,0) circle (2);
\draw (A1) to (A2);
\draw[fill=white] (0:2) circle (0.15);    \draw[fill=white] (180:2) circle (0.15);
\draw[fill=lightgray] (0,0) circle (0.9);
\draw[fill=white,white] (0,0) circle (0.3);
\node at (0,0) {\footnotesize 2};
\end{tikzpicture}
\caption{Loop diagrams  contributing to  $\frac{1}{(\sql)^{3}}\log^3$  term in the 2-point correlator. 
In  (a) the  blob 
 stands for the  bosonic and fermionic 
one-loop diagrams in Fig. \ref{fig:loop}.
 In (b)  it stands
for the two-loop irreducible contributions like 
 \myfeynA \ or reducible iterations of one-loop diagrams as in  \myfeynB\ . 
}
\label{fig:xxx}
\end{figure}
 The second type  of  contributions   should   come from  the  2-loop  corrections to the $\z^A$-propagator  which are: 
(i) irreducible 2-loop generalizations of  the second and third  graphs in Fig. \ref{fig:loop}; (ii) reducible iterations of 
these 1-loop  graphs, see Fig.\ref{fig:xxx}(b).
These 2-loop  corrections  (which we will not compute here) 
 should   produce the  remaining contribution $ \g_3^{(2)}$
\be 
\la{4022}
\g_3= \g_3^{(1)} + \g_3^{(2)} = -  \tfrac{ 125}{ 6}\ , \qquad \qquad 
\g_3^{(1)}=  -10\ , \qquad  \g_3^{(2)}=  -  \tfrac{65}{ 6}  \ .\ee

\section{Mixed four-point function $\langle x^{i}x^{j}Y^{A}Y^{B}\rangle$ \la{mixx}}

As was   mentioned in the Introduction,   the correlators     of the three   AdS$_5$ transverse 
fluctuations $x_i$ (scalars with $m^2=2$)  dual to the  correlator  of  the field strengths   $ {F}_{ti}$ 
 at leading  order in strong-coupling expansion 
  should be the same   in both WML and WL cases  as they are described by the same classical string action 
\rf{2.9}  with the same (Dirichlet) boundary conditions for $x_i$. 
The corresponding tree-level 
2- and 4-point functions 
$\langle xx\rangle$ or  $\langle xxxx\rangle$    were computed in \ci{Giombi:2017cqn}. 
As  the boundary  operator $  {\FF}\indices{_{t}^{i}}\equiv i F\indices{_{t}^{i}}  $ dual to $x^{i}$  has the interpretation of  the displacement operator, its 
dimension $\Delta=2$ will  be   protected   also in the non-supersymmetric WL   case, i.e.  it  
 should not receive corrections    in the  strong-coupling expansion
\be \la{510} 
\llangle {\FF}\indices{_{t}^{i}} (t_1)\,   {\FF}\indices{_{t}^{j}} (t_2)   \rrangle 
= \langle x^{i} (t_1)  \, x^{j} (t_2)  \rangle    =\delta^{ij}     { C'_{x}   \ov (t_{12})^{4 } }
\ . \ee
While  in the WML  case  the  normalization factor   $C_x=C_{{\mathbb F}}(\l) $ in the analog of \rf{510}
is known exactly (being equal to 12 times the Bremsstrahlung  function), the expression for  $C'_x=C_{F}(\l)$
at strong coupling (which should have a scheme-independent meaning, see footnote \ref{f6})
 is  not known at present.\foot{It should be easy to compute  the leading 
strong-coupling correction to it as $C'_x - C_x = C_{F}-C_{\mathbb F}$     should be given by the loop 
of $S^5$ scalars  with the   internal   line being the difference   of the  Neumann   and Dirichlet   propagators.} 
The  4-point correlators $\langle xxxx\rangle$    in the  supersymmetric and non-supersymmetric cases 
may start to differ   at  the  first subleading order  in $1\ov \sql$. 

In the   case of the     4-point correlator of  two  AdS  fluctuations and two 
$S^5$   fluctuations    the difference should appear already at  the leading order at 
strong coupling.   In the   supersymmetric WML   case   when $S^5$ coordinates  were subject 
to the Dirichlet b.c.   it was computed in \ci{Giombi:2017cqn}. 
 In the WL case  with Neumann b.c. in $S^5$  directions  this correlator  should  have 
$SO(3) \times SO(6)$    symmetry and should  represent the strong-coupling limit  of  the 4-point function  of 
two displacement   operators  and two 6-scalars  (cf. \rf{3})
\begin{align}
 \la{51} 
&\llangle { \FF}\indices{_{t}^{i}} (t_1) \,  { \FF}\indices{_{t}^{i}} (t_2)   \P_A(t_3) \P_B(t_4) \rrangle 
= \langle x^{i} (t_1)  x^{j} (t_2)    Y_A(t_3) Y_B(t_4) \rangle  
 \no\\
 &\qquad \qquad =\tfrac{1}{6}\delta^{ij} \delta_{AB}   { C'_x 
  \ov (t_{12})^{4} \, (t_{34})^{2\D} }\, G (\chi)  \ ,  \qquad \qquad 
 G (\chi) = 1 + \tfrac{1}{\sql}  G^{(1)} + \tfrac{1}{(\sql)^2}  G^{(2)}   \cdots \  ,
 \end{align}
where $\Delta= {5 \ov \sql} + \cdots$ is given by \rf{41}  and as in \rf{402}  we  choose a  scheme where  $C_Y={1\ov 6}$.  

Recalling that $Y_A= n_A + \z_A - \ha \z^2 n_A + \cdots$ (see \rf{3.5})  the leading order  contributions to   \rf{51}  will come from 
the  disconnected diagrams   $   \langle xx \rangle \langle  YY  \rangle$  (see   Fig.~\ref{fig:XXYY0})
that will  contribute to  the prefactor $ {1 \ov (t_{12})^{4} \, (t_{34})^{2\D} }$  in \rf{51}. 
\begin{figure}[ht]
\centering
\begin{tikzpicture}[line width=1 pt, scale=0.5]
\draw[densely dashed] (0,0) circle (2);
\draw (-120:2) to  (120:2);
\node[right] at (60:2.1) {$n^{A}$}; \node[left] at (120:2) {$x^{i}$};
\node[right] at (-60:2.1) {$n^{B}$}; \node[left] at (-120:2) {$x^{j}$};
\draw[fill=white] (60:2) circle (0.12);      \draw[fill=white] (-60:2) circle (0.12);
\draw[fill=white] (120:2) circle (0.12);    \draw[fill=white] (-120:2) circle (0.12);
\node at (4,0) {$+$};
\end{tikzpicture}
\begin{tikzpicture}[line width=1 pt, scale=0.5]
\draw[densely dashed] (0,0) circle (2);;
\draw (60:2) to  (-60:2);
\draw (-120:2) to (120:2);
\node[right] at (60:2.1) {$\z^{A}$}; \node[left] at (120:2) {$x^{i}$};
\node[right] at (-60:2.1) {$\z^{B}$}; \node[left] at (-120:2) {$x^{j}$};
\draw[fill=white] (60:2) circle (0.12);      \draw[fill=white] (-60:2) circle (0.12);
\draw[fill=white] (120:2) circle (0.12);    \draw[fill=white] (-120:2) circle (0.12);
\end{tikzpicture}
\caption{Leading order disconnected   contributions to  $\langle x^{i}x^{j}Y^{A}Y^{B}\rangle$.}\label{fig:XXYY0}
\end{figure}
Here the      bulk-to-boundary propagator  for $x$ (given   by \rf{2.32}  with $\Delta=2$) 
and the   bulk-to-boundary propagator for  the massless   field $\z$  given by \rf{399}, i.e.
\begin{align} \la{56}
K_{2}(t,z; t') &= \mc C_{2}\,  \KK_{2}(t,z; t') \ , \qquad\,\,   \KK_{2}(t,z;t') \equiv 
\Big[\frac{z}{(t-t')^2 + z^{2}}\Big]^{2}\,,\qquad \qquad\ \,\mc C_2 = {\te{ 2 \ov 3 \pi}} \ ,   \\
\rG_{\rm N}(t,z; t')&= \mc C_{\rm N}\,   \bG (t,z; t')\ , \qquad \bG  (t,z; t') \equiv  
  \log[  (t-t')^2+ z^2 ] 
  \,,\qquad \qquad \mc C_{\rm N}  \equiv -  {\te { 1 \ov  2 \pi}} \ , 
\la{57}
\end{align}
so that (ignoring  an infinite rescaling of $x^i$ by a $z\to 0$ factor) 
\be\la{588}
 \langle x^{i} (t_1)  x^{j} (t_2) \rangle =   { C'_x\ov (t_{12})^{4}}\ , \qquad   \qquad 
 C'_x = \tfrac {2\pi}{\sql} \mc C_2  + \OO(\tfrac {1}{(\sql)^2} )  \ . \ee
 One  may  normalize  the 4-point function   on the 2-point function of $x^i$,  i.e.  absorb   the 
 factor of $C'_x$   into a redefinition of  the operator $x$; we will not do this here.  
 
To compute   the non-trivial    correction to   $\langle xxYY \rangle$  we need to use the 4-vertices  in \rf{2.8} 
where    we may replace    $\del_\m y_a \del_\n  y_a \to \del_\m   Y_A  \del_\n  Y_A$  (the two   expressions 
are   the same to  quartic order in the fields). 
The leading  connected  contribution to $G (\chi)$ will  come from the connected diagram 
in  Fig.~(\ref{fig:XXZZ}). 
\begin{figure}[ht]
\centering
\begin{tikzpicture}[line width=1 pt, scale=0.5]
\draw[densely dashed] (0,0) circle (2);
\draw (40:2)--(-140:2);
\draw (-40:2)--(140:2);
\node[right] at (40:2.1) {$\z^{A}$}; \node[left] at (140:2) {$x^{i}$};
\node[right] at (-40:2.1) {$\z^{B}$}; \node[left] at (-140:2) {$x^{j}$};
\draw[fill=black] (0,0) circle (0.12);
\draw[fill=white] (40:2) circle (0.12);      \draw[fill=white] (-40:2) circle (0.12);
\draw[fill=white] (140:2) circle (0.12);    \draw[fill=white] (-140:2) circle (0.12);
\end{tikzpicture}
\caption{Connected  contribution to $\langle x^{i}x^{j}Y^{A}Y^{B}\rangle$.
The 4-vertex   comes from the quartic Lagrangian \rf{2.8}.    }\label{fig:XXZZ}
\end{figure}

There is   also 
another  connected contribution  to $\langle x^{i}x^{j}Y^{A}Y^{B}\rangle$  when  $Y^A$ is replaced 
by $n^A$ and $Y^B$  by $ -\ha  \z^2 n^B$ (or vice versa), see Fig.~\ref{fig:XXZZbis}.
\begin{figure}[ht]
\centering
\begin{tikzpicture}[line width=1 pt, scale=0.5]
\coordinate (A1) at (140:2);
\coordinate (A2) at (40:2);
\coordinate (A3) at (-140:2);
\coordinate (A4) at (-40:2);
\draw[densely dashed] (0,0) circle (2);
\draw (A1)--(0,0);
\draw (A3)--(0,0);
\draw (A2) to [in=60,out=180] (0,0);
\draw (A2) to [in=0,out=-110] (0,0);
\node[right] at (A2) {$-\frac{1}{2}\,{\z}^{2}\,n^{A}$};
\node[right = 0.1cm] at (A4) {$n^{B}$};
\node[left] at (A1) {$x^{i}$};
\node[left] at (A3) {$x^{j}$};
\draw[fill=black] (0,0) circle (0.12);
\draw[fill=white] (A1) circle (0.12);      \draw[fill=white] (A2) circle (0.12);
\draw[fill=white] (A3) circle (0.12);      \draw[fill=white] (A4) circle (0.12);
\end{tikzpicture}
\caption{Connected contribution with $Y^A$ replaced by $ -\ha  \z^2 n^A$. There is a similar diagram with 
$A\leftrightarrow B$. 
}\label{fig:XXZZbis}
\end{figure}

We  get  for the tree-level connected contribution of  the  diagram  in Fig.~(\ref{fig:XXZZ}) to the correlator in \rf{51}\foot{Here 
the vertex \rf{2.8}  in  the  string action  \rf{2.9} 
contributes $\sql\ov 2 \pi $ and four propagators $ ({2\pi\ov \sql})^4$. 
One power of  normalization  factor $ \tfrac{2\pi}{\sqrt\lambda} \mc C_{2}$  of  the $x$-propagator   is extracted to represent $C'_x$  in \rf{51}.} 
\begin{align}
\la{5.3}
&\frac{G_{\text{conn}}(\chi)}{t_{12}^{4}\ t_{34}^{2\Delta}} = -  5 
\times  (\tfrac{2\pi}{\sqrt\lambda})^2 \,
  \mc C_{2}\, (\mc C_{\rm N})^{2}\, \mathsf{Q}_{xy}  \ , \\
&\mathsf{Q}_{xy} \equiv  \int\frac{dtdz}{z^{2}}\Big[
\partial\KK_{2}(t_1)\cdot\partial\KK_{2}(t_2)
\,\partial\bG(t_3)\cdot\partial\bG(t_4)
-\partial\KK_{2}(t_1)\cdot\partial\bG(t_3)
\,\partial\KK_{2}(t_2)\cdot\partial\bG(t_4)\notag \\
&\qquad \qquad \qquad\ \ \  -\partial\KK_{2}(t_1)\cdot\partial\bG(t_4)
\,\partial\KK_{2}(t_2)\cdot\partial\bG(t_3)\Big],  \la{566}
\end{align}
where the factor $5$   came from \rf{3100},  $ \del  A \cdot \del B \equiv   g^{\m\n} \del_\m  A \cdot \del_\n B$, and 
$\KK_{2}(t_1) \equiv \KK_{2}(t,z;t_1)$, etc.  The expression \rf{566} can be simplified using 
the relations (cf. \rf{B.2})
\begin{align}
\partial &\KK_{2}(t_1)\cdot \partial \KK_{t_2}(t_2) = 
4\,\big[\KK_{2}(t_1)\KK_{2}(t_2)-2\,(t_{12})^{2}\,
\KK_{3}(t_1)\KK_{3}(t_2)
\big],
\notag \\
\partial &\bG(t_1)\cdot \partial \bG(t_2) = 
2\,z\,\big[ \KK_{1}(t_1)+\KK_{1}(t_2)\big]-2\,(t_{12})^{2}\,
\KK_{1}(t_1)\,\KK_{1}(t_2),
\la{5.6}\\
\partial &\KK_{2}(t_1)\cdot \partial \bG(t_2) = 
-4\,z\,\KK_{3}(t_1)+4\, (t_{12})^{2}\,\KK_{3}(t_1)\,\KK_{1}
(t_2), \qquad \quad \KK_{n} (t_1)  \equiv \KK_{n} (t,z; t_1)= 
\Big[\frac{z}{(t-t')^2 + z^{2}}\Big]^{n} . \no
\end{align}
The contribution of the diagram in Fig.~\ref{fig:XXZZbis}  is similar: 
including it    gives the total connected contribution 
  by  replacing $\mathsf{Q}_{xy}(1,2,3,4)$   with 
\begin{align}
\la{5.7}
{\mathsf{Q}}^{\rm (tot)}_{xy}(t_1,t_2,t_3,t_4) &= \mathsf{Q}_{xy}(t_1,t_2,t_3,t_4) -\tfrac{1}{2}\mathsf{Q}_{xy}(t_1,t_2,t_3,t_3)
-\tfrac{1}{2}\mathsf{Q}_{xy}(t_1,t_2,t_4,t_4).
\end{align}
This  results in the following replacement in \rf{566}\footnote{Since 
this depends only on the difference 
$\bG(t_3)-\bG(t_4) = \log \frac{(t-t_3)^2 + z^2 }{ (t-t_4)^2 + z^2 }$
the same result is found if we start with the  manifestly AdS$_2$ (or conformally)  invariant bulk-to-boundary propagator 
corresponding to \rf{D.11}, i.e. $\bG (t,z; t') =  \log\frac{(t-t')^2 + z^{2}}{z}$. This ensures that the resulting integral is conformally invariant. }
\be
\la{5.8}
\partial_{(\mu}\bG(t_3)\ \partial_{\nu)}\bG(t_4)\to -\tfrac{1}{2}\,
\partial_{(\mu}\big[\bG(t_3)-\bG(t_4)\big]\ 
\partial_{\nu)}\big[\bG(t_3)-\bG(t_4)\big] \ , 
\ee
and we find from \rf{566},\rf{5.6}
\begin{align}
\la{5.9}
 {\mathsf{Q}}^{\rm (tot)}_{xy}= & \int\frac{dtdz}{z^{2}}\Big[
16 \KK_2(t_3) \KK_3(t_1) \KK_3(t_2) t_{13}^2 t_{23}^2
+16 \KK_2(t_4) \KK_3(t_1) \KK_3(t_2) t_{14}^2 t_{24}^2\notag \\
&\qquad \quad -16 \KK_1(t_3) \KK_1(t_4) \KK_3(t_1) \KK_3(t_2) t_{14}^2 t_{23}^2
-8 \KK_1(t_3) \KK_1(t_4) \KK_2(t_1) \KK_2(t_2) t_{34}^2\notag \\
&\qquad \quad  -16  \KK_1(t_3) \KK_1(t_4) \KK_3(t_1) \KK_3(t_2) t_{13}^2 t_{24}^2
+16 \KK_1(t_3) \KK_1(t_4) \KK_3(t_1) \KK_3(t_2) t_{12}^2 t_{34}^2\Big]\notag \\
= &16 t_{13}^2 t_{23}^2 \,T_{2,3,3}(t_{3},t_{1},t_{2})+16 t_{14}^2 \
t_{24}^2 T_{2,3,3}(t_{4},t_{1},t_{2})\notag \\
&\qquad -16 D_{3,3,1,1} t_{14}^2 \, t_{23}^2-8 D_{2,2,1,1} t_{34}^2-16 D_{3,3,1,1} t_{13}^2 \,
t_{24}^2+16 D_{3,3,1,1} t_{12}^2 \, t_{34}^2 \ . 
\end{align}
Here $T_{\de_{1},\de_{2},\de_{3}}(t_{1},t_{2},t_{3})$ is the  standard  AdS    scalar  3-point function   
(see, e.g., \cite{DHoker:2002nbb})
\begin{align}
\la{5.10}
&T_{\de_{1},\de_{2},\de_{3}}(t_{1},t_{2},t_{3})=  \int\frac{dt\,dz}{z^{2}}\, \KK_{\Delta_{1}}(z, t; t_{1})\ 
\KK_{\Delta_{2}}(z, t; t_{2})\ \KK_{\Delta_{3}}(z, t; t_{3}) = 
\frac{A}{t_{12}^{\Delta_{12}}\,t_{23}^{\Delta_{23}}\,t_{31}^{\Delta_{31}}},
 \\
& A = \tfrac{\sqrt\pi}{2}\,\frac{
\Gamma[\tfrac{\Delta_{12}}{2}]\, 
\Gamma[\tfrac{\de_{23} }{2}]\, 
\Gamma[\tfrac{\de_{31}}{2}]
}{\Gamma(\Delta_{1})\Gamma(\Delta_{2})\Gamma(\Delta_{3})}\,
\Gamma[\tfrac{1}{2}(\Delta_{1}+\Delta_{2}+\Delta_{3}-1)]\ , \qquad  \Delta_{12}\equiv \Delta_{1} +\Delta_{2}-\Delta_{3}, \ \ etc.\ , \la{511}
\end{align}
and the $D$-functions are defined in \rf{B.1}. 
Expressing the latter  in terms of $\overline D$ functions
according to (\ref{B.3})  we   may  use that  in the  AdS$_2$   case (cf. \rf{B.5})
\begin{align}
\la{5.110}
\overline D_{2,2,1,1} &= \tfrac{1}{3\,(1-\chi)\,\chi^{2}}-\tfrac{2+\chi}{3\,\chi^{3}}\,\log(1-\chi)
+\tfrac{1}{3\,(1-\chi)^{2}}\,\log\chi, \notag \\
\overline D_{3,3,1,1} &= 
-\tfrac{2 \chi^2+3 \chi -3}{15 (\chi -1)^2 \chi ^4}
-\tfrac{2 (\chi ^2+3 \chi +6) }{15 \chi ^5}\,\log (1-\chi )
-\tfrac{2}{15 (1-\chi)^3}\,\log\chi.
\end{align}
As a result, 
\begin{align}
\la{5.12}
& \,{\mathsf{Q}}^{\rm (tot)}_{xy} =  
\frac{ 6\pi}{t_{12}^{4}} \left[1-\left(\tfrac{1}{2}-\tfrac{1}{\chi}\right)\,\log(1-\chi)\right]  \ . 
\end{align}
We thus   find   for the leading-order  contribution to the $G$-function in \rf{51} 
\begin{align}\la{5.13}
&G (\chi) = 1 +  \tfrac{1}{(\sql)^2}  G^{(2)}(\chi)     + \OO (\tfrac{1}{(\sql)^3}) \ , \qquad \\
 &G^{(2)}(\chi) =  -  5 \,   ({2\pi})^2 \, \mc C_{2}\,  (\mc C_{\rm N})^2 \,  {t_{12}^{4}} \, \mathsf{Q}^{\rm (tot)}_{xy}  
 =   
- {20} \Big[1-\big(\tfrac{1}{2}-\tfrac{1}{\chi}\big)\,\log(1-\chi)\Big]  \ . \la{5113}
\end{align} 
We observe that  the  strong-coupling 
 contribution to the connected part of  $G$  in  \rf{51}     first appears at order $ \tfrac{1}{(\sql)^2}$   and, remarkably,  that 
$G^{(2)} $ is proportional 
to the corresponding  
expression \rf{151},\rf{153}  for the tree-level $ \langle x^{i} x^{j} y^a y^b \rangle$   correlator found in the supersymmetric line case in 
\cite{Giombi:2017cqn}.
Using the  label  D for   the $G$-function   in the  supersymmetric (Dirichlet propagator)  case 
we  thus get   in the  non-supersymmetric   case 
 \be \la{5222}
 G^{(2)} = 5\, G^{(1)}_{\rm D} \ , \qquad \qquad G^{(1)}_{\rm D} = - {4} \Big[1-\big(\tfrac{1}{2}-\tfrac{1}{\chi}\big)\,\log(1-\chi)\Big] \ . \ee
 We will  explain  the reason for this  coincidence  in section \ref{ND} below.


Let us comment on the OPE interpretation of the function $G(\chi)$ in  \rf{51},(\ref{5.13}, \ref{5113}). 
Exchanging   $t_{2}\leftrightarrow t_{3}$  in   (\ref{51})  we get  (cf. \rf{217},\rf{217})
\begin{align}
\la{5.19}
& \llangle { \FF}\indices{_{t}^{i}} (t_1) \,  \P_A(t_2) \,{ \FF}\indices{_{t}^{i}} (t_3)  \, \P_B(t_4) \rrangle 
= \tfrac{1}{6}\delta^{ij} \delta_{AB} \,\frac{C'_x}{ 
(t_{12}\,t_{34})^{2+\D}}\,\left|\frac{t_{24}}{t_{13}}\right|^{2-\D} \, \mG (\chi),\notag \\
& \mG (\chi) \equiv  \chi^{2+\D}\,G(\chi^{-1}) = \chi^{2+\D}\,\Big(
1-\tfrac{20}{(\sql)^{2}}\,\big[1+(\chi -\ha )\,\log\tfrac{1-\chi}{\chi}\big]+  \OO (\tfrac{1}{(\sql)^3}) \Big) \ , 
\end{align}
where $\Delta$ is given by \rf{41}. 
The  corresponding conformal block expansion is\footnote{$c_{h}$ is related  to the  coefficient in the 3-point function 
between $\FF $, $\Phi$, and the exchanged operator $\mc O_{h}$ of  conformal dimension $h$. Let us recall that 
in the supersymmetric case (cf.  \rf{151})   the operator $\mc O_{h}$ takes a schematic 
form $\Phi\partial^{n}_{t}\mathbb F$, and has dimension $h_n=3+n-\frac{1}{2\,\sql}\,(n+1)(n+4)+\cdots$\, \cite{Giombi:2017cqn}.
The normalization of $c_{h}$ in  \rf{521} below takes
into account that in the present case in \rf{5.13} we have $G(\chi) = 1+ \cdots\, $.}
\be
\la{5.20}
\mG(\chi) = \sum_{h}c_{h}\,\chi^{h}\, _{2}F_{1}(h+2-\D, h-2+\D, 2\,h, \chi) \ .
\ee
Comparing (\ref{5.19}) with (\ref{5.20}), and using the expansion (\ref{41}) for $\D$, we find the following 
results for the  corresponding  intermediate operator dimensions and  coefficients $c_h$
consistent content in (\ref{5.20})
\begin{align}
h_{0} &= 2+\tfrac{5}{\sql}-\tfrac{10-d_{2}}{(\sql)^{2}}+\cdots, \qquad 
& c_{h_{0}} &= 1-\tfrac{20}{(\sql)^{2}}+\cdots, \notag \\
h_{1} &= 3+\tfrac{3}{\sql}+\cdots, 
& c_{h_{1}} &= -\tfrac{10}{\sql}+\tfrac{25-2d_{2}}{(\sql)^{2}}+\cdots, \notag \\
h_{2} &= 4+\tfrac{0}{\sql}+\cdots, 
& c_{h_{2}} &= \tfrac{10}{3\,\sql}+(\tfrac{80}{3}+\tfrac{2d_{2}}{3})
\,\tfrac{1}{(\sql)^{2}}+
\cdots, \notag \\
h_{3} &= 5-\tfrac{4}{\sql}+\cdots, 
& c_{h_{3}} &= -\tfrac{25}{21\,\sql}+(-\tfrac{8125}{441}-\tfrac{5d_{2}}{21})\,\tfrac{1}{(\sql)^{2}}+
\cdots, \quad etc. \la{521} 
\end{align}
 For $n\ge 2$ the general expression for  the  leading  order $\frac{1}{\sql}$  correction   is 
\be
h_{n} =2+  n-\tfrac{(n+5)(n-2)}{2}\,\tfrac{1}{\sql}+\cdots, \qquad \qquad 
c_{h_{n}} = \tfrac{20}{3}\,\tfrac{n+2}{n}\,(-\tfrac{1}{4})^{n+2}\,\tfrac{\sqrt{\pi}\,(n+3)!}{\Gamma(n+\frac{3}{2})}\,
\tfrac{1}{\sql}+\cdots\, .\la{522} 
\ee
Notice that   for large $n$    the  dimension $h_n$   of the
intermediate  operator  $\Phi \del^n_t  \FF$   has  the same  universal  behaviour as in
the supersymmetric line case in  \cite{Giombi:2017cqn}:  
$ h_n \to  n - {n^2\ov 2\sql} + ...$  (compared to \rf{2.37},\rf{C.5}  where the  operator 
contains $\del^{2n}_t$   here $n \to \ha n$). 
This  universality    supports  the existence of a semiclassical explanation  of this large $n$ asymptotics
(indeed, possibly  related classical  string solution should not be sensitive  to boundary conditions in $S^5$).

\section{Four-point function $\langle Y^{A}Y^{B}Y^{C}Y^{D}\rangle$}

Given the 2-point function \rf{4.1},   the general  structure of the $SO(6)$ scalar  4-point function 
controlled by   the 1d conformal invariance and crossing should be  as in \rf{2.18},\rf{2.19}, i.e.
\begin{align} 
\la{6.2}
&\langle Y^{A}(t_1) Y^{B}(t_2) Y^{C}(t_3) Y^{D}(t_4) \rangle = \frac{C_Y^{2}}{|t_{12}\ t_{34}|^{2\de}}\,G^{ABCD}(\chi) \  , \\
 G^{ABCD} =
  G_{S}\,\delta^{AB}\delta^{CD}
+ &G_{T}\Big[\delta^{AC}\delta^{BD}+ 
\delta^{BC}\delta^{AD}-\tfrac{1}{3}\,\delta^{AB}\delta^{CD}
\Big]
+G_{A}\Big[\delta^{AC}\delta^{BD}-\delta^{BC}\delta^{AD}\Big].\la{6622}
\end{align}
Here  $G_{S}(\chi)$ is the  basic function   
 with  $G_{T}$ and $G_{A}$  expressed in terms of it  via leg interchange, i.e.  using  the crossing relations 
\rf{2.26},\rf{226}. 
In what follows we shall set $C_Y= {1\ov 6}$  as   in  \rf{402}. 

To compute $G_S$ it is sufficient to consider the singlet correlator as in \rf{2.21}, i.e. 
\be
\la{6.3}
\langle Y^{A}(t_1) Y^{A}(t_2) Y^{B}(t_3) Y^{B}(t_4) \rangle = \frac{1}{|t_{12}t_{34}|^{2\de}} \, G_{S}\  .
\ee
Here $n^A$ dependence drops out (so the  integration over $S^5$ is trivial). Thus \rf{6.3} can be computed in any explicit parametrization of $Y_A$  and we shall again use \rf{3.5}, i.e.
$Y^{A} = n^{A}+\z^{A}-\frac{1}{2}n^{A}\,{\z}^{2}+\mc O(\z^{4})$ with $n_A \z_A=0, \ n_A n_A=1$. 

\subsection{Leading-order  contributions}
\la{sec:gen-struct}

Let us first consider the  simplest  -- leading order --   contributions to \rf{6.3}
 \begin{align}
 \la{6.4}
 &\langle Y^{A}(t_1) Y^{A}(t_2) Y^{B}(t_3) Y^{B}(t_4) \rangle = 1   + \tfrac{1}{\sql}  Q^{(1)}  +  
 \tfrac{1}{(\sql)^2}  Q^{(2)}  +  \tfrac{1}{(\sql)^3}  Q^{(3)} + \cdots\  .\end{align}
  At order $\tfrac{1}{\sql}$ these are  just the tree-level terms 
$\langle \z_A\z_A n_B n_B \rangle + \langle n_A n_A \z_B \z_B\rangle$,   giving  as in \rf{4.1},\rf{4.11}
\be 
      Q^{(1)}  = -  5 \big( \nD_{12}  + \nD_{34} \big) \ , \qquad \qquad \nD_{12} =\log t_{12}^2 
 \ .  \la{64}
 \ee
 $ Q^{(1)}$  thus  corresponds to the leading term in the  expansion of
  the prefactor  $(t_{12}t_{34})^{-2\de}$  in \rf{6.2},\rf{6.3}  with   $\de=\frac{5}{\sql}+\dots$.
 At the next   $\frac{1}{(\sql)^{2}}$ order we will get several contributions
  from tree-level diagrams   with four $\z$  and two contractions
 (see Fig.~\ref{fig:disc2}). Denoting their contribution to $Q^{(2)}$     as $Q^{(2)} _0$   we get 
\begin{align}
 \la{6.5}
 Q^{(2)} _0 &= \tfrac{5}{2} \big( \nD_{12}^2  +   \nD_{34}^2 + \nD_{13}^2 + \nD_{14}^2  + \nD_{23}^2   + \nD_{24}^2   \big)  + 25   \nD_{12}   \nD_{34}  
  \notag \\
&\qquad   -5\big(  \nD_{13}   \nD_{14} +  \nD_{23}   \nD_{24}  +   \nD_{13} \nD_{23}  + \nD_{14} \nD_{24} \big)
    + 5 \big(  \nD_{13} \nD_{24} +   \nD_{14} \nD_{23} \big) \ . 
 \end{align}
 Here the first  group of terms  comes from diagrams like  Fig.~\ref{fig:disc2}(a), the  second from Fig.~\ref{fig:disc2}(b), the third from 
 Fig.~\ref{fig:disc2}(c)  and the forth  from Fig.~\ref{fig:disc2}(d)   and Fig.~\ref{fig:disc2}(e). 
\begin{figure}[ht]
\centering
\begin{tikzpicture}[line width=1 pt, scale=0.35]
\node at (0,-2.8) {(a)};
\coordinate (A1) at (140:2);    \coordinate (A2) at (40:2);
\coordinate (A3) at (-140:2);   \coordinate (A4) at (-40:2);
\draw[densely dashed] (0,0) circle (2);
\draw (A1) to [out=25,in=155] (A2);
\draw (A1) to [out=-20,in=200] (A2);
\node[left]    at (A1) {\footnotesize $-\frac{1}{2}\bm{\z}^{2}\,n^{A}$}; 
\node[right]  at (A2) {\footnotesize $-\frac{1}{2}\bm{\z}^{2}\,n^{A}$}; 
\node[left]    at (A3) {\footnotesize $n^{B}$};
\node[right]  at (A4) {\footnotesize $n^{B}$}; 
\draw[fill=white] (A1) circle (0.15); \draw[fill=white] (A2) circle (0.15);
\draw[fill=white] (A3) circle (0.15); \draw[fill=white] (A4) circle (0.15);
\end{tikzpicture}
\begin{tikzpicture}[line width=1 pt, scale=0.35]
\node at (0,-2.8) {(b)};
\coordinate (A1) at (140:2);    \coordinate (A2) at (40:2);
\coordinate (A3) at (-140:2);   \coordinate (A4) at (-40:2);
\draw[densely dashed] (0,0) circle (2);
\draw (A1)--(A2);
\draw (A3)--(A4);
\node[left]    at (A1) {\footnotesize $\z^{A}$}; 
\node[right]  at (A2) {\footnotesize $\z^{A}$}; 
\node[left]    at (A3) {\footnotesize $\z^{B}$};
\node[right]  at (A4) {\footnotesize $\z^{B}$}; 
\draw[fill=white] (A1) circle (0.15); \draw[fill=white] (A2) circle (0.15);
\draw[fill=white] (A3) circle (0.15); \draw[fill=white] (A4) circle (0.15);
\end{tikzpicture}
\begin{tikzpicture}[line width=1 pt, scale=0.35]
\node at (0,-2.8) {(c)};
\coordinate (A1) at (140:2);    \coordinate (A2) at (40:2);
\coordinate (A3) at (-140:2);   \coordinate (A4) at (-40:2);
\draw[densely dashed] (0,0) circle (2);
\draw (A1)--(A3);
\draw (A1)--(A4);
\node[left]    at (A1) {\footnotesize $-\frac{1}{2}\bm{\z}^{2}n^{A}$}; 
\node[right]  at (A2) {\footnotesize $n^{A}$}; 
\node[left]    at (A3) {\footnotesize $\z^{B}$};
\node[right]  at (A4) {\footnotesize $\z^{B}$}; 
\draw[fill=white] (A1) circle (0.15); \draw[fill=white] (A2) circle (0.15);
\draw[fill=white] (A3) circle (0.15); \draw[fill=white] (A4) circle (0.15);
\end{tikzpicture}
\begin{tikzpicture}[line width=1 pt, scale=0.35]
\node at (0,-2.8) {(d)};
\coordinate (A1) at (140:2);    \coordinate (A2) at (40:2);
\coordinate (A3) at (-140:2);   \coordinate (A4) at (-40:2);
\draw[densely dashed] (0,0) circle (2);
\draw (A1)--(A4);  \draw[fill=white,white] (0,0) circle (0.2);
\draw (A2)--(A3);
\node[left]    at (A1) {\footnotesize $\z^{A}$}; 
\node[right]  at (A2) {\footnotesize $\z^{A}$}; 
\node[left]    at (A3) {\footnotesize $\z^{B}$};
\node[right]  at (A4) {\footnotesize $\z^{B}$}; 
\draw[fill=white] (A1) circle (0.15); \draw[fill=white] (A2) circle (0.15);
\draw[fill=white] (A3) circle (0.15); \draw[fill=white] (A4) circle (0.15);
\end{tikzpicture}
\begin{tikzpicture}[line width=1 pt, scale=0.35]
\node at (0,-2.8) {(e)};
\coordinate (A1) at (140:2);    \coordinate (A2) at (40:2);
\coordinate (A3) at (-140:2);   \coordinate (A4) at (-40:2);
\draw[densely dashed] (0,0) circle (2);
\draw (A1)--(A3);  
\draw (A2)--(A4);
\node[left]    at (A1) {\footnotesize $\z^{A}$}; 
\node[right]  at (A2) {\footnotesize $\z^{A}$}; 
\node[left]    at (A3) {\footnotesize $\z^{B}$};
\node[right]  at (A4) {\footnotesize $\z^{B}$}; 
\draw[fill=white] (A1) circle (0.15); \draw[fill=white] (A2) circle (0.15);
\draw[fill=white] (A3) circle (0.15); \draw[fill=white] (A4) circle (0.15);
\end{tikzpicture}
\caption{Types of diagrams contributing to  (\ref{6.5}). 
Other  diagrams  are obtained 
 by interchanging points. 
}
\label{fig:disc2}
\end{figure}
The terms  $ \tfrac{5}{2} \big( \nD_{12}^2  +  \nD_{34}^2  \big) $ and  $25   \nD_{12}   \nD_{34}    $  
with 12  and 34 propagators  should 
 corresponds again to the $ \log^2$  terms appearing  from  the expansion of  the prefactor
 $(t^2_{12}\, t^2_{34})^{-{5\ov \sql} + \cdots}$ in \rf{6.3}. 

\begin{figure}[ht]
\centering
\begin{tikzpicture}[line width=1 pt, scale=0.35]
\draw[densely dashed] (0,0) circle (2);;
\draw (30:2)--(150:2);
\node[right] at (30:2) {$\z^{A}$}; \node[left] at (150:2) {$\z^{A}$}; 
\node[right] at (-30:2) {$n^{B}$}; \node[left] at (-150:2) {$n^{B}$};
\draw[fill=white] (30:2) circle (0.12);
\draw[fill=white] (-30:2) circle (0.12);
\draw[fill=white] (150:2) circle (0.12);
\draw[fill=white] (-150:2) circle (0.12);
\draw[fill=white] (0,0.96) circle (0.6);
\draw[fill=lightgray] (0,0.96) circle (0.6);
\end{tikzpicture}
\caption{
\la{fig:dash}
A disconnected diagram contributing $\langle Y^{A}Y^{A}Y^{B}Y^{B}\rangle$.
The  $\z$-propagator  includes loop corrections, with  1-loop ones 
corresponding  to   the second and third  diagram in  Fig.~\ref{fig:loop}.}
\end{figure}
In addition,  there  are also   similar  
 terms  coming from  the 1-loop propagator correction 
 diagrams  like in Fig. \ref{fig:dash}. 
As follows from  the structure of the loop-corrected 
 propagator in \rf{4.1} there will  be a $\log $ correction  to $Q^{(2)} $   given by 
\be\la{777}
Q^{(2)} _{\log} = - d_2 \big( \nD_{12}  +  \nD_{34} \big) \ . 
\ee
From the analysis of the $\langle YY \rangle$ correlator in section \ref{sec:2-point}
we know that these  loop diagrams  also contribute the   $\log^{2}(t_{12}) + \log^{2}(t_{34}) $ terms (cf. \rf{4.11}) necessary  to build up 
the  prefactor $ |t_{12}\, t_{34}|^{-2\de}$  as required by conformal invariance. 
The coefficient of  these terms 
is given by  $\g_2^{(1)} = \frac{25}{2}-\frac{5}{2}=10$  in \rf{419}. Thus  we get  for the additional   1-loop   contribution   to $Q^{(2)} $   
\begin{align}
 \la{6.6}
Q^{(2)} _1= 10  \big( \nD_{12}^2  +  \nD_{34}^2 \big)\ .
 \end{align}
 Equivalently,   this term is found from the $ {1\ov \sql} Q^{(1)}$ term in \rf{6.4}   upon the substitution \rf{401}. 
 Thus 
 \begin{align}\no 
Q^{(2)} &= Q^{(2)} _{\log} +  Q^{(2)} _0 +  Q^{(2)} _1 = Q^{(2)} _{\log}    + 
 \tfrac{25}{2}   \big( \nD_{12}  +  \nD_{34} \big)^2 + \bar Q^{(2)} \ ,    \\   \la{6666}  
\bar Q^{(2)} &= \tfrac{5}{2}  \big(   \nD_{13} +\nD_{24} -  \nD_{14}   - \nD_{23}  \big)^2 \ . 
 \end{align}
   Multiplying \rf{6.4}   by $ |t_{12}\, t_{34}|^{2\de}= 1 +  {5\ov \sql}  (\nD_{12}  +  \nD_{34} ) + {25\ov 2(\sql)^2}   (\nD_{12}  +  \nD_{34} )^2    + .... $   (cf.  \rf{6.4},\rf{6.3}) we conclude that  all 
   $\nD_{12}$ and $ \nD_{34}$   dependent terms    cancel  out  (in particular, $\log$ term in \rf{777} does not contribute) 
   so that  the 
 the leading  contribution to $G_S$ is given by 
 \begin{align} 
 \la{6.799}
 G _S (\chi) &= 1   +  \tfrac{1}{(\sql)^{2}} \bar Q^{(2)} 
+ \OO (\tfrac{1}{(\sql)^3}) =1 +\tfrac{1}{(\sql)^2} G _S ^{(2)} (\chi) 
 +\OO (\tfrac{1}{(\sql)^3}) \ ,   \\  & \qquad  G _S ^{(2)} = 10 \log^{2}(1-\chi) \ . \la{6.7}
 \end{align} 
 There is no $\tfrac{1}{\sql}$ term as the  leading-order   correction \rf{64}  correspond  just  to the prefactor in \rf{6.3}.
  As  there is no other ``connected'' contribution  at order $\tfrac{1}{(\sql)^2}$ 
  the expression  in \rf{6.7}   gives the full  conformally invariant expression 
  for $G_S$ to this order. 
 
 
 To  find $G_T$ and $G_A$ in \rf{6622} we may use the general crossing relations\rf{2.26},\rf{226} with $N=6$   and 
  $\Delta$ given by \rf{41}, i.e.  $\de= {5\ov \sql} + {d_2\ov (\sql)^2} + ...$. As a result, 
  \begin{align}
\la{6.9}
G_{T}(\chi) 
&= \tfrac{3}{4} +\tfrac{9}{2\sql}\log\tfrac{\chi^{2}}{1-\chi}
+\tfrac{3}{2(\sql)^{2}}\Big(9\log^{2}\tfrac{\chi^{2}}{1-\chi} 
+8\,\log^{2}(1-\chi)+\tfrac{3}{5}\, d_2 \log\tfrac{\chi^{2}}{1-\chi} \Big)   +\OO (\tfrac{1}{(\sql)^3}) ,  \\
G_{A}(\chi) 
&=  \tfrac{6}{\sql}\,\log(1-\chi)+\tfrac{6}{(\sql)^{2}}\,\log(1-\chi)\,\Big(
4\,\log\tfrac{\chi^{2}}{1-\chi}+\tfrac{1}{5}d_{2}\Big)  +\OO (\tfrac{1}{(\sql)^3}) \ . 
\la{610}
\end{align}
The  $\tfrac{1}{\sql}$  terms  here originated from  the $\D$-dependence  in \rf{2.26},\rf{226}. 
The appearance of the second anomalous dimension coefficient $d_2$ in \rf{6.9},\rf{610} is not surprising: 
 it  means  that  in order to determine the $\tfrac{1}{(\sql)^2}$ terms in $G^{ABCD}(\chi)$ 
  one needs to compute also the 1-loop graphs (bosonic and fermionic ones, cf. Fig. \ref{fig:loop}) 
   that contribute  not only to  $\D$   but effectively  also to   $G_T$ and $G_S$. 
Similarly, the $\tfrac{1}{(\sql)^3}$ terms in \rf{6.9},\rf{610} will depend  not only on the $\tfrac{1}{(\sql)^3}$
correction to \rf{6.7} but also  on the $\tfrac{d_3}{(\sql)^3}$ term in $\Delta$.


It is important to  stress  that  in contrast to the supersymmetric ($SO(5)$ invariant)  case in \ci{Giombi:2017cqn} 
here the presence of the $n_A$  ``condensate'' in $Y_A$  implies  that 
the disconnected graphs  are not  described  just by a generalized free
 field perturbation theory (cf. Appendix \ref{app:freefield}). 
For example, the averages over $S^5$ do not factorize: 
$\langle n^{A}n^{B}n^{C}n^{D}\rangle\neq \langle n^{A}n^{B}\rangle\langle n^{A}n^{B}\rangle$, etc. 
Thus even $\tfrac{1}{\sql} $ corrections  in \rf{6.9},\rf{610}  are not those 
of a free field theory. For example,  setting $\de=\frac{5}{\sql}+\cdots$ in (\ref{A.1}) and expanding does not reproduce  
the single logarithms proportional to (\ref{6.9}).


\subsection{
Order $1\ov (\sql)^3$  contributions:  Dirichlet/Neumann relations \la{ND}}
At the  next $1\ov (\sql)^3$  order we  get    two different  contributions:   (i) ``reducible'' contributions 
 given  by tree level diagrams with possible 1-loop   or 2-loop propagator corrections;
  (ii)  ``irreducible''   connected   tree-level  contributions where all  four points are connected to the bulk vertex. 
  The 3-loop propagator corrections  (like in  Fig. \ref{fig:xxx}(b)) can appear  only in the  disconnected   parts
  $\langle \z^A \z^A n^B n^B\rangle + \langle n^A n^A \z^B \z^B\rangle$   (see Fig. \ref{fig:dash}) 
  and thus  contribute only to the prefactor $|t_{12} t_{34}|^{-2\Delta}$ 
   in \rf{6.3} but not to $G_S$. 
   
  Non-trivial   reducible   contributions  come  from connected  tree diagrams  with 3 propagators like the one  in 
Fig. \ref{fig:111} and also from the leading order  diagrams in Fig. \ref{fig:disc2}   with   one of the propagators   being ``dressed'' by 1-loop 
 correction as   in Fig. \ref{fig:loop} or Fig. \ref{fig:dash}.   
 We will   discuss these  reducible  contributions  in detail in Appendix \ref{app:other}.
\begin{figure}[ht]
\centering
\begin{tikzpicture}[line width=1 pt, scale=0.5]
\coordinate (A1) at (140:2);
\coordinate (A2) at (40:2);
\coordinate (A3) at (-140:2);
\coordinate (A4) at (-40:2);
\draw[densely dashed] (0,0) circle (2);;
\draw (A1)--(A2);
\draw (A1)--(A3);
\draw (A2)--(A3);
\node[left]    at (A1) {$-\tfrac{1}{2}n^{A}{\z}^{2}\ $}; 
\node[right]  at (A2) {$\ -\tfrac{1}{2}n^{A}{\z}^{2}\ $}; 
\node[left]    at (A3) {$-\tfrac{1}{2}n^{B}{\z}^{2}\ $};
\node[right]  at (A4) {$\ n^{B}$}; 
\draw[fill=white] (A1) circle (0.12);
\draw[fill=white] (A2) circle (0.12);
\draw[fill=white] (A3) circle (0.12);
\draw[fill=white] (A4) circle (0.12);
\end{tikzpicture}
\caption{``Reducible'' tree-level diagram contributing   at order $1\ov (\sql)^3$.}\label{fig:111}
\end{figure}
    \begin{figure}[ht]
\centering
\begin{tikzpicture}[line width=1 pt, scale=0.5]
\draw[densely dashed] (0,0) circle (2);;
\draw (40:2)--(-140:2);
\draw (-40:2)--(140:2);
\node[right] at (40:2) {$\z^A$}; \node[left] at (140:2) {$\z^A$};
\node[right] at (-40:2) {$\z^B$}; \node[left] at (-140:2) {$\z^B$};
\draw[fill=white] (40:2) circle (0.12);
\draw[fill=white] (-40:2) circle (0.12);
\draw[fill=white] (140:2) circle (0.12);
\draw[fill=white] (-140:2) circle (0.12);
\draw[fill=black] (0,0) circle (0.12);
\end{tikzpicture}
\caption{Contact diagram contributing  at order $1\ov (\sql)^3$.}\label{fig:zzzz}
\end{figure}

In addition, 
there is also   an    ``irreducible''   connected contribution  to \rf{6.2},\rf{6.3} 
that comes   from   the  contact tree diagram in Fig. \ref{fig:zzzz} 
  where all four fields in $\langle\z^A(t_1)\z^A(t_2)\z^B(t_3)\z^B(t_4) \rangle$ are attached 
 to a quartic   vertex  from  $L_4$ in \rf{3.7} 
 The analog of it  (see Fig. \ref{fig:bps-conn})
 was the only leading  connected 
contribution \rf{2.34} in the   supersymmetric line case with  the Dirichlet  bulk-to-boundary propagators
     \ci{Giombi:2017cqn}. 

Having   one  bulk 4-vertex  in \rf{3.7}    (proportional to $ \sql$) and  four   $\z$-propagators  (each  bringing a  $1\ov \sql$  factor )   this connected   contribution     should  scale as  ${1\ov (\sql)^3} {G_{S, {\rm conn}}^{(3)}}$. 
Note that the normalization   in the  supersymmetric case \rf{2.34} was different,  so    comparing 
 to it  below we shall strip off the $1\ov \sql$ factors. 
  In total,   we should find  (cf.  \rf{6.7}, \rf{G10}) 
\begin{align} 
 \la{677}
& G _S  = 1 + \tfrac{1}{(\sql)^{2}}  G_S^{(2)}    +\tfrac{1}{(\sql)^{3}}  G_S^{(3)} +\OO (\tfrac{1}{(\sql)^4}) \ , \\
 & G_S^{(3)}  =  {G_{S, {\rm red }}^{(3)}} + {G_{S, {\rm conn}}^{(3)}} \ , 
\qquad \qquad    {G_{S, {\rm red }}^{(3)}} = {G_{S, {\log^2}}^{(3)}} + {G_{S, {\log^3}}^{(3)}}  \ , 
 \la{6770} 
 \end{align}
 where  ${G_{S, {\log^2}}^{(3)}}$  and $ {G_{S, {\log^3}}^{(3)}} $ are given in \rf{GG4} and \rf{88}. 
 
 Trying to compute  ${G_{S, {\rm conn}}^{(3)}} $   directly  one observes 
  that   the  logarithmic form of the Neumann  bulk-to-boundary propagator \rf{399}  leads to  complicated  \ads   integrals.  
A  useful observation  is that  applying  boundary-point $\del_{t_i}$   derivatives
to the  contact contribution to the correlator $\langle  YYYY \rangle$  
   it  is possible to relate   the  expressions  for the integrands  with the differentiated Neumann propagators to  the similar ones  
in the Dirichlet   propagator    case. 

Let us define (see \rf{2.30x},\rf{57};  below  $\del_\m= (\del_t, \del_z),  \ \del^\m A \del_\m B = z^2 \del_\m A \del_\m B $,
 $\epsilon_{\m\n} =\pm \epsilon_{tz}=\pm 1$; repeated  low indices  are contracted with $\delta_{\m\n}$)
\begin{align}\la{610y} 
&\bG'(t_{a}) \equiv \del_{t_a} \bG(t_a)= 2{  t_a-t \over (t-t_a)^2 + z^2} =  \tfrac{ 2(t_a-t)}{z} \KK_{1}(t_a) \ , \\
&\bG(t_a) = \log\big[ (t-t_a)^{2} + z^2 \big]  \ , \qquad \qquad 
\la{610xxx}\KK_{1}(t_a) = \frac{z}{(t-t_a)^{2} + z^2 }=\ha \del_z \bG(t_a) ,\\ 
& \del_\m \bG'(t_{a}) = 2\,  \epsilon_{\m\n} \del_\n  \KK_{1}(t_a) \ , \qquad   \qquad \del_\m= (\del_t, \del_z)\ .   \la{H1} 
\end{align} 
Using  \rf{H1}  we may thus relate the  expressions containing  bulk-point derivatives of $\bG'(t_{a})$
to the ones with bulk-point  derivatives of $\KK_{1}(t_a)$. For example, we get 
\begin{align}
\la{6.10}
&\partial_{\mu}\bG'(t_{1})\, 
\partial_{\mu}\bG'(t_{2}) = 
4\, 
\partial_{\mu}\KK_{1}(t_{1})\, 
\partial_{\mu}\KK_{1}(t_{2}) \ .
\end{align}
Equivalently,  (\ref{6.10})  
follows  simply  from  the complex  coordinate   decomposition of $\KK_{1}$ and  $\bG'$
\be
\la{6.12}
\KK_{1}(t_a) = -\tfrac{1}{2\,i}\,\Big(\frac{1}{w}-\frac{1}{\overline w}\Big),\qquad
\bG'(t_a) = -\Big(\frac{1}{w}+\frac{1}{\overline w}\Big), \qquad   w \equiv  t-t_a +i\,z  \ , 
\ee
using that   $\del_\m A \del_\m B=4 \del_w A \del_{\bar w} B $  (cf. \rf{d6}). 

From (\ref{6.10}), we see that the contact diagram associated to the $(\partial\z)^{4}$  term  in \rf{3.7}
contributing to  the 4-point function  in the   Neumann propagator  theory 
 is  simply proportional  to the same diagram in the  theory  with the Dirichlet propagator.
 A similar relation is true   for the  contributions of the  mixed  
$xx YY$  4-derivative vertices in \rf{2.8}.
There is also a close relation  between  the two cases 
for the contribution of the   2-derivative  $\z^{2}(\partial\z)^{2}$  vertex in \rf{3.7}. Explicitly, one  finds   (see  Appendix \ref{app:dual})
\begin{align}
\la{6.13}
& \int {dzdt}\, \bG'(t_{1}) \,\bG'(t_{2})\, 
\partial_{\mu} \bG'(t_{3})\,\partial_{\mu}\bG'(t_{4}) = 
16 \int {dzdt}\,   \KK_1(t_{1})\,  \KK_1(t_{2})\,
\partial_{\mu}\KK_{1}(t_{3})\,\partial_{\mu}\KK_{1}(t_{4}) +  \omega \ , \\
& \qquad  \qquad  \omega (t_1,t_2,t_3, t_4) 
= - \frac{8 \pi}{t^2_{34} } \big({1\ov t_{13} t_{23} }  +   {1\ov t_{14} t_{24} }      \big) \ , 
\ \ \ \ \ \ \  t_{ij} = t_i - t_j  \ . \la{6130}
\end{align}
which  may be proved by using \rf{610y}--(\ref{6.12}) and  performing the integrals. 
The  ``deficit''   $\omega$-term   here   corresponds  to the 
non-zero  boundary  contribution   that survives  upon manipulating one integral into the other using 
 integration  by parts (see \rf{HH5}--\rf{hh2}).

 We then arrive at the following    symbolic relations between the 
 $G$-functions 
 appearing in  the  corresponding   connected contributions   to the 
 correlators in \rf{51} and \rf{6.2}, \rf{6622}  in the Dirichlet  and Neumann  cases\foot{While in the D-case it is natural to strip off  normalization factors of all 2-point functions in the correlator 
 in the N-case this is not natural  as the  2-point function   of $Y^A$   expanded  in   $1\ov \sql$   starts with  constant 
 rather than the tree-level propagator. We may still  formally do this  but  without 
  changing sign, so the factor  associated to the N-propagator in \rf{57}   will be 
  $ { 2 \pi \ov \sql} |\mc C_{N}| = { 1 \ov \sql}$.  Finally, when relating  the 
 expressions in the  D and N cases we  omit  the $1\ov \sql$ factors.  This   formal    identification 
  requires  the -1   factor in \rf{614}.}
\begin{align}
\la{614}
  \langle x^{i}(t_1) x^{j}(t_2) Y^{A}(t_3) Y^{B}(t_4) \rangle:  && \partial_{t_{3}}\partial_{t_{4}}  \widehat G
&=- 2 \, \frac{1}{t_{34}^{2}}\,G_{\rm D}(\chi),  \\
\langle Y^{A}(t_1)Y^{B}(t_2)Y^{C}(t_3)Y^{D}(t_4)\rangle:  &&  \partial_{t_{1}}  \partial_{t_{2}}  \partial_{t_{3}}   \partial_{t_{4}} 
\widehat G
&= 4\, \frac{1}{t_{12}^{2}\,t_{34}^{2}}\, G_{\rm D}(\chi)  +  \Omega  \ .  \la{6.14} 
\end{align}
Here  $\widehat G$   and $G_{\rm D}$ 
stand  for the   contact  diagram  $1\ov (\sql)^3$ contributions  in the Neumann and Dirichlet cases  respectively
with all symmetry group   factors  stripped off 
 before  averaging over $n^A$ in the N-case ($\widehat G$-functions     are related to $G$-functions in \rf{6622} 
   as in \rf{6.22}  below). 
   For simplicity, 
 in this section shall often omit the 
label ``${(3)}$''  on $G^{(3)}$.
$\Omega$   in  \rf{6.14} is the total contribution of the $\omega$-terms in the  relation like \rf{6.13}. 
The  basic idea behind \rf{614},\rf{6.14}  is that   after the differentiation over the boundary points 
the Neumann  propagator contributions  get   related  to the Dirichlet ones as in \rf{6.10},\rf{6.13}. 
To find the conformally-invariant solution  for  the total  $ G $    we will need  to add  also 
the ``reducible'' contribution as  in \rf{6770}    that will cancel  non-invariant terms in $\Omega$. 

More explicitly, to  compare to the   supersymmetric line  case with $SO(5)$ scalars  in \rf{2.34},\rf{151} 
 one is to replace $Y^A$ by $y^a$  and  postpone    the averaging over $n^A$ till the end.
For  the mixed correlator in  \rf{614}   we will have  (cf. \rf{3100})
\be
\la{6.15} G_{\rm D}^{ab} = \delta^{ab}\,G_{\rm D} \ , \qquad \qquad
\widehat G^{AB} =  P^{AB} \,\widehat G\  \ \to\   \ G^{AB} =  \tfrac{1}{6} \delta^{AB} \, G\ , \qquad   \qquad
  G =  {5}\,   \widehat  G \ . 
\ee
In the   massless 4-scalar  correlator  case, starting with the  expression  \rf{2.34}  in the  supersymmetric line case 
we are  first  to  replace  $\delta^{ab}\to P^{AB}= \delta^{AB} - n^A n^B $ 
and $K_1 =\mc C_{1}\KK_{1}\to  \GN=\mc C_{\rm N}\bG$ in  the $SO(5)$  version  of   \rf{2.19}  getting  (cf. \rf{6.14})
\begin{align}
\la{6.17}
\widehat G^{ABCD } = 
\widehat{G}_{ S}(\chi)\,P^{AB}P^{CD}
&+\widehat{G}_{ T}(\chi)\Big[P^{AC}P^{BD}+
P^{BC}P^{AD}-\tfrac{2}{5}\,P^{AB}P^{CD}
\Big]\notag \\
& +\widehat{G}_{ A}(\chi)\Big[P^{AC}P^{BD}
-P^{BC}P^{AD}\Big], \\ 
\la{6.18}
\partial_{t_{1}}  \partial_{t_{2}} \partial_{t_{3}}  \partial_{t_{4}} \widehat{G}_{ c} 
&= 4\, \frac{1}{t_{12}^{2}\,t_{34}^{2}}\,G_{{\rm D}, c}(\chi)   +  \Omega_c\ ,\qquad\ \ \   c = S, T, A \ . 
\end{align}
Here   the functions  $G_{{\rm D}, c}(\chi)$  are given by  the  leading-order connected  expressions  (\ref{2.35}). 
 Averaging \rf{6.17} over $n^A$ according to  \rf{310},\rf{3100} we end up with
  (cf. \rf{6622}  and  \rf{2.20}--\rf{221})
\begin{align}
\la{6.19}  
& \qquad \qquad  \widehat G^{ABCD}  \ \to \  \tfrac{1}{36} G^{ABCD}   \ , \\
 &  G^{ABCD} = 
G_{ S}\,\delta^{AB}\delta^{CD} +G_{ T}\Big[\delta^{AC}\delta^{BD}+
\delta^{BC}\delta^{AD}-\tfrac{2}{6}\,\delta^{AB}\delta^{CD}
\Big]+G_{ A}\Big[\delta^{AC}\delta^{BD}
-\delta^{BC}\delta^{AD}\Big] \ ,\no  \\
&  G_{ S} =   
25 \,\widehat{G}_{ S}, \qquad\qquad
G_{ T} = \tfrac{3}{4}\,\widehat{G}_{ S}+\tfrac{126}{5}\,\widehat{G}_{ T}, \qquad\qquad 
G_{ A} =  
24 \,\widehat{G}_{ A}. \la{6.22}
\end{align}
\iffa
 we obtain, {\em cf.} (\ref{2.20}) with $N=5$ or $6$, 
\begin{align}
\la{6.21}
G^{AABB} &= 25\,\widehat{G}_{ S} = 36\,G_{ S}, \notag \\
G^{ABAB} &= 5\,\widehat{G}_{ S}+28\,\widehat{G}_{ T}+20\,\widehat{G}_{ A}
= 6\,G_{ S}+40\,G_{ T}+30\,G_{ A},\notag \\
G^{ABBA} &= 5\,\widehat{G}_{ S}+28\,\widehat{G}_{ T}-20\,\widehat{G}_{ A}
= 6\,G_{ S}+40\,G_{ T}-30\,G_{ A}.
\end{align}
\fi
 Before turning to the  case of $\langle YYYY \rangle$ 
 let   us  first    demonstrate   how the above   D/N  relation \rf{614} 
explains the  proportionality of 
the expressions   for  the  leading  connected part  
of the mixed  correlator  $ \langle x^{i}(t_1) x^{j}(t_2) Y^{A}(t_3) Y^{B}(t_4) \rangle$ 
in the supersymmetric (D)  \rf{153}  and  non-supersymmetric (N) \rf{5.13},\rf{5222}  cases. 
The  leading order  term in  $G_{\rm D}$ is $ G^{(1)}$ in  \rf{153}. 
To  find the corresponding term in $G_{\rm N}$  we may integrate   the relation in \rf{614}. 

The double derivative   operator in \rf{614}   has a  nice interpretation 
in terms of the quadratic Casimir operator  of the 1d conformal group (i.e. $J^2$  for $SO(1,2)$). 
Indeed,  $t_{34}^{2}\,\partial_{t_{3}}\partial_{t_{4}}$ is  
 invariant under  the scale transformations, translations, and
also the inversion. 
When acting on a function of the cross-ratio 
$\chi= {t_{12}t_{34} \ov t_{13} t_{24}}$  it becomes 
\be
\la{6.24}
t_{34}^{2}\,\partial_{t_{3}}\partial_{t_{4}}  f(\chi) = -\mathscr D  f(\chi) \ , \qquad \qquad 
\mathscr D \equiv  \chi^2\,(1-\chi)\, \partial_{\chi}^{2}- \chi^2\,\partial_{\chi},
\ee
 where $\mathscr D$  is the  conformal Casimir operator
(see, e.g., \cite{Dolan:2011dv,Hogervorst:2017sfd}).
 The  eigenfunctions  of  $\mathscr D$  are the 
$SL(2,R)$ conformal blocks (cf.  (\ref{C.0}))
\be
\la{6.25}
\mathscr D   \FF_h  = h(h-1) \FF_h \ ,  \qquad  \qquad  \FF_h = \chi^{h}\,F_{h}(\chi) \ , \ \ \  F_{h} \equiv   {_{2}F_{1}}(h,h,2h,\chi)  \ . 
\ee
From (\ref{614}),\rf{6.15}  we have (cf. \rf{153}) 
\be
\la{6.26}
{t_{34}^{2}}\,  \partial_{t_{3}}\partial_{t_{4}}   G(\chi) = 
-\mathscr D  G(\chi) =- 10\, G_{\rm D}(\chi) \ . 
\ee 
One can check  that 
\be\la{624}
 \FF_2=  \chi^2\,{}_{2}F_{1}(2,2,4,\chi) =  -12\,\Big[ 1- \big(\tfrac{1}{2}-\tfrac{1}{\chi}\big)\,\log(1-\chi)\Big] =3 
   G_{\rm D}(\chi)  \  , \qquad 
 \mathscr D G_{\rm D}(\chi)  = 2  G_{\rm D}(\chi)  \ . 
\ee
Thus $  G_{\rm D} =   G^{(1)} $ in  \rf{153}   is given just 
 by a  single conformal block   corresponding to  the  dimension $h=2$.
 This means that in the supersymmetric line case the only operator that can appear  in the OPE 
 channel $12\to 34$ (besides the 
identity which contributes to the disconnected part) is the $h=2$ singlet $\sim y^{a}y^{a}$.
Integrating  (\ref{6.26}) for  $G  $   using \rf{624}   we  get       
\be
\la{6.28}
 G(\chi) = 5\, G_{\rm D}(\chi) + c_{1}+c_{2}\,\log(1-\chi) \ ,
\ee
where the last two terms are the  zero  modes of  the Casimir operator $  \mathscr D  $, 
i.e. a   linear combination   of the  $h=0$ and $h=1$ conformal  blocks.

Let us argue that this  ``zero-mode'' part is to be omitted, i.e. 
   one should  set  $c_{1}=c_{2}=0$. 
The  leading order term in the  small $\chi$ expansion of   generic $G(\chi)$ in \rf{2.13}   should be  determined
by the minimal dimension of the fields appearing in the corresponding  OPE. 
In the present   case of  connected part of $G$   this is the $\Delta=2$ operator   suggesting  that 
 $G(0)=0$.
Assuming   the symmetry under
$t_{3}\leftrightarrow t_{4}$, i.e. under $\chi\to -\frac{\chi}{1-\chi}$, we get  also $G'(0)=0$. Then a
  (connected part of)  $G(\chi)$  should have the small $\chi $ expansion\foot{This  will  also apply to the 
singlet  part of  the 4-scalar correlator  below but will  not   be true in general in the  T- and A- channels.}
\be 
G(0)=G'(0)=0 \ . \la{629}\ee
This   property is  readily checked   for  $G_{\rm D}=G^{(1)}$  in \rf{153} 
  and should  hold  also for $G$ in \rf{6.28},   implying  that
$c_{1}=c_{2}=0$. As a result,    we find that      $ G$   in \rf{6.28}
 coincides  with  the expression in \rf{5113},\rf{5222}   that we found above  
 by  the direct computation in the Neumann propagator case.

\iffa 
\paragraph{Remark:} 
It is unclear why in this case $G_{\rm D}(\chi)$ is an eigenfunctions of the 
conformal Casimir, {\em i.e.} a single conformal block with $h=2$. This could be 
related to underlying integrability 
(``particle production'' should be highly constrained in integrable theories, and in 
CFT$_{1}$ language that might translate on constraints on the OPE content) or 
be a specific feature of the strong coupling limit we are taking -- similar to decoupling of operators
in other cases.
\fi 

\subsection{Contact diagram contribution and 
 $G_{S,T,A}$ functions at order $1\ov (\sql)^3$
} 
  \la{conne}

The    four-point function $\langle Y^{A}Y^{B}Y^{C}Y^{D}\rangle$  in the $SO(6)$ Neumann theory 
 \rf{6.2},\rf{6622}    is expressed in terms 
 of the three functions
$G_{ c}$  ($c= S,T,A$).
The main task is to determine $G_S$   as  then $G_T$ and $G_A$   can be found   using  the crossing relations
 \rf{2.26},\rf{226} (with $N=6$)
 \begin{align}
{G}_{T}(\chi) &= -\tfrac{3}{20}\Big[ \,{G}_{S}(\chi)
-  3 \,\chi^{2\de} {G}_{S}\big(\tfrac{1}{1-\chi}\big)
-  3 \,\big(\tfrac{\chi}{\chi-1}\big)^{2\de}{G}_{S}(1-\chi)\Big] \ ,  \la{6300} \\
{G}_{A}(\chi) &= 
\tfrac{3}{5} \Big[ \,\chi^{2\de}\,\widehat{G}_{S}\big(\tfrac{1}{1-\chi}\big)
-\,\big(\tfrac{\chi}{\chi-1}\big)^{2\de}\widehat{G}_{S}(1-\chi)\Big] \ , \qquad \qquad
\de=\tfrac{5}{\sql}+ \tfrac{d_2}{(\sql)^2} + \tfrac{d_3}{(\sql)^3}   + ...    \la{6.30}
\end{align}
\iffa    
 or equivalently, using  (\ref{6.22}), 
in terms of   $\widehat G_{ c}$.
As  \rf{6.17} applies before $n^A$ averaging, i.e.  formally to an $SO(5)$ theory (cf. \rf{2.19}),  
$\widehat G_{ c}$   are related   by  (\ref{2.26})  with  $N=5$,  i.e.\foot{These
general  relations  are  exact  in the coupling   so expanding them 
in  $1\ov \sql$   leads to a combination of contributions from different kinds of diagrams as $\Delta$  depends on  $1\ov \sql$
(cf. \rf{6.9},\rf{610}).}
\begin{align}
\widehat{G}_{T}(\chi) &= -\tfrac{5}{28}\,\widehat{G}_{S}(\chi)
+\tfrac{25}{56}\,\chi^{2\de}\widehat{G}_{S}\big(\tfrac{1}{1-\chi}\big)
+\tfrac{25}{56}\,\big(\tfrac{\chi}{\chi-1}\big)^{2\de}\widehat{G}_{S}(1-\chi),  \la{6300} \\
\widehat{G}_{A}(\chi) &= 
\tfrac{5}{8}\,\chi^{2\de}\,\widehat{G}_{S}\big(\tfrac{1}{1-\chi}\big)
-\tfrac{5}{8}\,\big(\tfrac{\chi}{\chi-1}\big)^{2\de}\widehat{G}_{S}(1-\chi), \qquad \qquad
\de=\tfrac{5}{\sql}+\mc O(\tfrac{1}{(\sql)^{2}}). \la{6.30}
\end{align}
This, in principle, allows  one to determine ${G}_{T}$  and ${G}_{A}$ 
from   ${G}_{S}$ and the expansion of $\Delta$ in \rf{41}.\foot{These
general  relations  are  exact  in the coupling   so expanding them 
in  $1\ov \sql$   leads to a combination of contributions from different kinds of diagrams as $\Delta$  depends on  $1\ov \sql$
(cf. \rf{6.9},\rf{610}).}
Our  main task will thus be to compute $\widehat{G}_{S}$  and thus  the $1\ov (\sql)^3$ term in $  G_{S} $. 
As  $\widehat{G}_{S}(\chi)$    will  have  a non-trivial   structure 
the  use  of \rf{6300},\rf{6.30} will require a   particular choice of  its   analytic   continuation away from the small $\chi$ region. 
\fi
One   may try   to determine ${G}_{S} $ 
 by integrating   the   relation  \rf{6.18}  of its connected part 
  to the   corresponding function  in the Dirichlet theory \rf{2.35}
  \be
\la{6.31}
t_{12}^{2}\,t_{34}^{2}\,\partial_{t_{1}}  \partial_{t_{2}}\partial_{t_{3}} \partial_{t_{4}} \, ({G}_{ S})_{\rm conn} 
 = 100\,G_{{\rm D}, S}(\chi) +  
U_{S}  \ , \qquad   \qquad U_{S}= t^2_{12} t^2_{34} \Omega_{S} \ . 
\ee
The normalization of the 
 $U_{S}$ contribution is chosen such that it directly contributes
 to $G_{S}$.  Here we   restored   the  label  ``conn''  on  $ {G}_{ S}$    to indicate that   this  contribution comes 
from the contact  connected diagram. 
By the explicit computation from the 4-vertex in first term in \rf{3.7} 
 one  finds that   in the S-channel   the total  combination $\Omega_{S}$ of $\omega$-terms  coming from relations like \rf{6.13} 
is  such that  
\begin{align}
U_{S}(t_{1},t_{2},t_{3},t_{4}) =   40\, t^2_{12} t^2_{34}\,\Big(  &\te   \frac{1}{t_{12}^2 t_{23} t_{24}}-\frac{5}{t_{12} t_{23}^2 t_{24}}-
\frac{6}{t_{12} t_{23} t_{24}^2}+\frac{1}{t_{12}^2 t_{14} t_{13}}-
\frac{1}{t_{23}^2 t_{34} t_{13}}+\frac{6}{t_{12} t_{14}^2 t_{13}}\notag \\
& +\te 
\frac{2}{t_{23} t_{34}^2 t_{13}}+\frac{5}{t_{12} t_{14} t_{13}^2}-
\frac{1}{t_{23} t_{34} t_{13}^2} \Big)\ . \la{639}
\end{align}
Since   $U_{S}$ is   not conformally invariant,   the contact diagram  contribution   to ${G}_{ S} $ 
is also not  just a function  of $\chi$  
so we cannot  simply  replace $t_{12}^{2}\,t_{34}^{2}\,\partial_{t_{1}}  \partial_{t_{2}}\partial_{t_{3}} \partial_{t_{4}} $ 
in \rf{6.31}  by the square of the Casimir  operator $\mathscr D$ \rf{6.24}. 
However, the conformal invariance is restored  in the total expression for   $G_S$, i.e.  
 once we add the ``reduced'' diagram   contributions as in \rf{6770}. 
  Indeed,  the expression for $t_{12}^{2}\,t_{34}^{2}\,\partial_{t_{1}}  \partial_{t_{2}}\partial_{t_{3}} \partial_{t_{4}}$ 
  applied to the reduced part $({G}_{ S})_{\rm red}  $  is given by the sum of \rf{G9}  and \rf{G121}. 
  As a result, we find that non-invariant terms in \rf{G121}   cancel against those in \rf{639}   and we are left with 
  \begin{align}
\la{6315}
&t_{12}^{2}\,t_{34}^{2}\,\partial_{t_{1}}  \partial_{t_{2}}\partial_{t_{3}} \partial_{t_{4}} \,  ({G}_{ S})_{\rm conn}
= \mathscr D^2 ({G}_{ S})_{\rm conn}
 = 100\,G_{{\rm D}, S}(\chi) +  \V_{S}  (\chi)   \ , \\
 &    {G}_{ S}  = ({G}_{ S})_{\rm conn}  + ({G}_{ S})_{\rm red}\ ,    \qquad  \qquad   ({G}_{ S})_{\rm red} 
 = ({G}_{ S})_{\rm \log^2} + ({G}_{ S})_{\rm \log^3} \ , \la{6615}  \\
 &  \V_S=   \te   { 8 d_2   } \Big[  \chi^2 +  {\chi^2 \ov (1-\chi)^2} \Big]  + 
 320  \frac{\chi ^2}{(1-\chi)^2} \Big(  [ 1 + (1-\chi)^2]  (1  + \ha \log \chi )     -  {1\ov 2}  \log (1-\chi ) \Big) \ . \la{6400}
 \end{align}
Here  $\V_S$   is the  combination of   $U_S$   with  the contributions \rf{G9}, \rf{G121}
of the ``reduced''  terms      in which all non-invariant terms happen to  cancel  out. 
The $d_2$  term in \rf{6400} is the contribution of the  $\log^2$    reduced  term  in \rf{G9};  
as its   contribution to  the invariant part of ${G}_{ S} $   is known already (see \rf{G89}) in what follows we will 
simply omit it,   concentrating  on other  invariant terms in  ${G}_{ S} $    solving \rf{6315}. 

We may formally split $({G}_{ S})_{\rm conn} $  into the  sum $  \widebar{G}_{ S}  + \widetilde  {G}_{ S} $ of the   solution  of 
$ \mathscr D^2 \widebar{G}_{ S} = 100\,G_{{\rm D}, S}(\chi) $  where $G_{{\rm D}, S}= G^{(1)}_S $ in \rf{2.35} 
and the solution of 
 $ \mathscr D^2 \widetilde  {G}_{ S} =  \V_{S}  (\chi) $  where $\V_S$ is  given by \rf{6400}, 
\be  \la{671}
({G}_{ S})_{\rm conn} =  \widebar{G}_{ S}  + \widetilde  {G}_{ S} \ , \qquad 
\mathscr D^2 \widebar   {G}_{ S} = 100\,G_{{\rm D}, S}(\chi) \ , \qquad 
\mathscr D^2 \widetilde   {G}_{ S} = \V_{S}  (\chi)  \ . \ee
Explicitly, one finds  that  the most general  solution  for $\widetilde   {G}_{ S} $  may be written as\foot{The appearance
of  $\text{Li}_n$  functions here (absent in the ``reduced'' $\log^3$   contribution in \rf{88})  should be attributed to the contribution of the 
$\Omega$-part of the contact diagram  contribution   to $ \V_{S}  $: for example, the  4 times  integrated   expression 
of the $\omega$ in \rf{6130}  can be seen to be given by a combination of the polylogarithmic  functions.}
\begin{align}
  \widetilde   {G}_{ S} = & -320\, \text{Li}_3(1-\chi )+320\,
\text{Li}_2(1-\chi ) \log (1-\chi )+160\, \text{Li}_2(\chi ) \log (1-\chi)\notag \\
& \te
-\tfrac{80}{3}\, \log ^3(1-\chi )+240\, \log \chi  \log ^2(1-\chi )  + \sum^4_{n=1} c_n \psi_n(\chi)  \ , \la{6719} \\
& \psi_{1}=1, \ \ \ \  \psi_{2}=\log(1-\chi), \quad \ \ 
\psi_{3}=\log \chi\ , \quad \ \ 
\psi_{4}= 
 \,\text{Li}_{2}(\chi)+ \tfrac{1}{2} \log\chi\,\log(1-\chi)   \la{631}
\ ,  \end{align} 
where $c_n$  are constants   multiplying the zero modes $\psi_n(\chi)$ of the $\mathscr D^2$  operator  (cf. \rf{6.28}).
Expanding \rf{6719} for small $\chi$ we get 
\be \la{6720}
 \widetilde   {G}_{ S} = \big[ c_3 \log \chi  + c_1 - 320 \zeta_R(3)\big] + (c_4-c_2  - \ha  c_4 \log\chi ) \chi + 
  \big[\tfrac{1}{ 4}  c_4 - \ha  c_2- 80   + (80- \tfrac{1}{ 4} c_4) \log \chi \big] \chi^2 + \OO(\chi^3) \ . 
\ee
Imposing the condition \rf{629}  fixes 
\be  \la{6721}
c_1 = 320 \zeta_R(3)\  , \qquad \qquad c_2=c_3=c_4=0 \ . \ee
Similarly, we may  attempt to    solve the  equation for $\widebar {G}_{ S}(\chi) $  in \rf{671}   which has a more complicated source term
(cf.  \rf{2.35})  and  try to constrain   the zero-mode freedom   by imposing 
the $3\leftrightarrow 4$  crossing symmetry  condition on the  total function    (cf. \rf{2119},\rf{2.23}) 
\be
\la{6.32}
{G}_{ S}(\chi) =  {G}_{ S}(\tfrac{\chi}{\chi-1}) \ , 
\ee
and also  the condition  \rf{629}. 
A somewhat  complicated  structure of $\psi_{4}$ in \rf{631} 
 suggests that  
 finding a  correct  analytic continuation  of   ${G}_{ S}(\chi) $ 
  out of the perturbative region $\chi\to 0$  may be non-trivial.\footnote{A possible solution of the analytic 
  continuation problem may be based on the following relations
\be
\notag 
\mathscr{D}^{2}\,[f(1-\chi)] = \tfrac{\chi^{2}}{(1-\chi)^{2}}\,\left[\mathscr{D}^{2}f(\chi)\right]_{\chi\to 1-\chi},
\qquad
\mathscr{D}^{2}\,[f(\tfrac{1}{1-\chi})] = \chi^{2}\,\left[\mathscr{D}^{2}f(\chi)\right]_{\chi\to \frac{1}{1-\chi}}.
\ee
Indeed, to determine, for instance, $f(\tfrac{1}{1-\chi})$ from the solution to $\mathscr D^{2}f = g$, one simply
writes  $
\mathscr D^{2}[f(\tfrac{1}{1-\chi})] = \chi^{2}\,g(\tfrac{1}{1-\chi}). $
If the r.h.s. admits a simple analytic continuation (e.g.  using  the $\log(\cdots)\to \log|\cdots|$ rule)  under which 
it keeps essentially the same complexity, this will  then readily  give an expression for $f(\tfrac{1}{1-\chi})$
after  the integration.
}

To   avoid  these issues 
let   us  start   from the very beginning 
and   consider  not   the fourth derivative (as in \rf{6.18}), but just the second     derivative  of the  singlet  correlator 
\be
\la{6.33}
\partial_{t_{1}}\partial_{t_{2}}\langle Y^{A}(t_{1}) Y^{A}(t_{2}) Y^{B}(t_{3}) Y^{B}(t_{4}) \rangle \ . 
\ee
 Computing it  using  the  relations  between the N and D  propagators  like \rf{H1}
    we   may then  integrate the resulting   analog of \rf{614}, i.e. 
  follow the same approach as  described above   in the   case of the mixed   correlator $\langle xxYY\rangle$.
  
  Our strategy will be  to find the invariant contribution   to  $\widebar  {G}_{ S} $  (freely doing 
     integrations by parts and assuming that 
  all non-invariant terms  from  boundary terms  cancel  against the  ``reduced''  contributions as  discussed above).
  A  consistency test will   be that   the resulting function will indeed satisfy the  correct 4-derivative equation 
 $ \mathscr D^2 \widebar  {G}_{ S} = 100\,G_{{\rm D}, S}(\chi)$  in \rf{671}. 
  
Given the connected  correlator   with   4-vertices  from  (\ref{3.7})  (see Fig. \ref{fig:zzzz}),  
 applying $\partial_{t_{1}}\partial_{t_{2}}$ to it  we will get  various   contractions  with two of the four 
 bulk-to-boundary  Neumann propagators  \rf{57},\rf{610}  differentiated over the boundary point.
 For example, the 4-derivative vertices  in (\ref{3.7})  will lead to 
  (cf. \rf{6.13}) 
\be\la{632}
\partial_{\m}\bG'(t_{1})\partial_{\m}\bG'(t_{2})
\partial_{\n}\bG(t_{3})\partial_{\n}\bG(t_{4}), \ \  {\rm etc.} 
\ee
 Using \rf{H1} or  $ \partial_{\mu}\bG'= 2\, \eps_{\mu\nu} \partial_{\nu} \KK_1 $  
we  can replace $\bG'$  with  $\KK_{1}$   and also  apply the relations 
 similar to (\ref{5.6}), i.e.
\begin{align}
\partial &\KK_{1}(t_1)\cdot \partial \KK_{1}(t_2) = 
\KK_{1}(t_1)\KK_{1}(t_2)-2\,t_{12}^{2}\,
\KK_{2}(t_1)\KK_{2}(t_2)\ ,
\la{633} \\ 
\partial &\KK_{1}(t_1)\cdot \partial \bG(t_2) = 
-2\,z\,\KK_{2}(t_1)+2\,t_{12}^{2}\,\KK_{2}(t_1)\,\KK_{1}
(t_2). \la{634}
\end{align}
This   allows us  to effectively  
replace all  logarithmic  $\bG$  factors  by  the Dirichlet functions $\KK_n(t,t_a;z)= \big[ {z \ov (t-t_a )^2 + z^2}\big]^n$, 
(cf. \rf{2.32}) so that the resulting  integrals  over   the AdS$_2$   bulk point   become  the  standard ones (see Appendix \ref{app:Dfunctions}).  

There is also  another  type of  contractions   coming   from  the  2-derivative vertex   in (\ref{3.7}):
 after applying 
$\partial_{t_{1}}\partial_{t_{2}}$ to them  we get  integrals $\int dt dz (\cdots)$  like \rf{6.13}    with  the  integrands 
of the  three types
\begin{align}
\la{6.36}
V_{1} &= \bG\,\bG\,\partial_{\m}\bG'\,\partial_{\m}\bG', \qquad 
V_{2} = \bG'\,\bG'\,\partial_{\m}\bG\,\partial_{\m}\bG, \qquad 
V_{3} = \bG\,\bG'\,\partial_{\m}\bG\,\partial_{\m}\bG'.
\end{align}
We can   simplify these   using  
$\Box \bG = \Box \bG' = \Box \KK_{1}=0$ (here $\Box=\del_\m \del_\m = \del_t^2 + \del_z^2$)
and  formal integration by parts. 
Then we get\footnote{Here we use  that for  the 
harmonic  functions  ($\Box H_{i}=0$)
 one has $H_{1}H_{2}\partial_{\m}H_{3}\partial_{\m}H_{4}
= \frac{1}{2}H_{1}H_{2}\Box(H_{3}H_{4}) \to  \frac{1}{2}\Box(H_{1}H_{2})H_{3}H_{4}  = 
\partial_{\m}H_{1}\partial_{\m}H_{2}H_{3}H_{4}$   where we  dropped a total derivative term.}
$V_{1} = 4 \KN \KN \del_\m K_{1} \del_\m K_{1} \to  
4 \del_\m\KN \del_\m\KN  K_{1} K_{1},$
and    we can use  (\ref{5.6}) to eliminate $\bG$ in terms of $\KK_1$.
 $V_2$  in \rf{6.36} can be  also reduced to  the $V_1$-type term:  
  $ V_{2} = \bG' \bG' \partial_{\m} \bG \partial_{\m} \bG \to   \partial_{\m}\bG' \partial_{\m}
\bG'  \bG \bG $. 
The same is also true for  $
V_{3} = \bG \bG'  \partial_{\m}\bG \partial_{\m} \bG'$
(using the $1\leftrightarrow 2$ and  $3\leftrightarrow 4$ symmetry).  

As a result, 
we find  that the second derivative of   ${G}_{S}$   appearing in \rf{6.33}  is given by 
(see (\ref{5.9}) and  \rf{B.5}  for   the  expressions for the $T$ and $\overline D$ functions)\footnote{
In obtaining the  expression (\ref{6336})  we   included the contributions  of 
  diagrams with the  $-\ha n^{A}\, \z^2$  term in $Y^A$ 
at the points $t_{3}$ or $t_{4}$ (like in   Fig.  \ref{fig:XXZZbis} where points $x^i$ are  now replaced by $Y^A$). 
This amounts to a subtraction  of contributions at the coinciding points 
 completely analogous to that in (\ref{5.7}).
}
\begin{align}
&\partial_{t_{1}}\partial_{t_{2}} \widebar {G}_{S} = -\tfrac{1}{2\pi} 
\Big[
-400 t_{13}^2 t_{23}^2 \, T_{2,2,2}(t_{1},t_{2},t_{3})-400 t_{14}^2 t_{24}^2 
\, T_{2,2,2}(t_{1},t_{2},t_{4})\notag \\
& +\tfrac{150 \pi  t_{34}^2 }{t_{13}^2 t_{24}^2} \overline{D}_{1,1,1,1} -\tfrac{60 \pi  \big[(t_{13}^2-7 
t_{14} t_{13}+t_{14}^2) t_{12}^2+5 t_{13} t_{14} (t_{13}+t_{14}) 
t_{12}-5 t_{13}^2 t_{14}^2\big] t_{34}^2}{t_{13}^4 
t_{24}^4}  \overline{D}_{2,2,1,1}
\Big] . \la{6336}
\end{align}
Here   we  again put   bar on ${G}_{S}$ to indicate   that this  connected contribution of the contact diagram is computed 
by formally discarding   boundary terms while integrating by parts. 
Using (\ref{5.10}), (\ref{B.5}) gives (cf. \rf{6.24}) 
\begin{align}
-\mathscr D   \widebar{G}_{ S}(\chi) &= - 10 
 \Big[ \tfrac{\chi ^2-10 \chi +10}{\chi -1}-\tfrac{ (\chi^2-10 \chi +10) \chi ^2}{ (\chi -1)^2}\log\chi  +\tfrac{\chi^3-8 \chi ^2+5 \chi -10 }{ \chi } \log (1-\chi )\Big]  \ \no\\
& = 25  (4 \log \chi -\tfrac{47}{15})\chi ^2 + 25 (4 \log \
\chi -\tfrac{17}{15})\chi ^3 +\OO(\chi^4) \ . \la{6.41}
\end{align}
Integrating this  as in \rf{6.26},\rf{6.28} and  applying the  crossing constraint (\ref{6.32})
and the condition  \rf{629} of regularity at  $\chi\to 0$     we get\footnote{\la{f44}
Let us note two useful relations: 
$\text{Li}_{2}(1-\chi) = \tfrac{\pi^{2}}{6}-\log(1-\chi)\log\chi-\text{Li}_{2}(\chi),$

$
\text{Li}_{3}(1-\chi) = \tfrac{\pi^{2}}{6}\log(1-\chi)+\tfrac{1}{6}\log^{3}(1-\chi)-\tfrac{1}{2}
\log^{2}(1-\chi)\log\chi+\zeta_R(3)-\text{Li}_{3}(\chi)-\text{Li}_{3}\big(
\tfrac{\chi}{\chi-1}\big).\notag
$
}
\begin{align}
\la{6.43}
\widebar {G}_{ S}(\chi) = &  - 240 \,\Big[ \text{Li}_{3}(\chi) + \,\text{Li}_{3}\big(\tfrac{\chi}{\chi-1}\big)\Big]
+50\,\Big[\tfrac{1}{2}-\tfrac{1}{\chi}-\tfrac{1}{5}\chi+\tfrac{8}{5}\,
\text{Li}_{2}(\chi)\Big]\,\log(1-\chi)\notag \\
& + 40 \Big[ \,\log^{3}(1-\chi)-\tfrac{1}{4}\tfrac{\chi^{2}}{1-\chi}
\log \chi -\log^{2}(1-\chi)\,\log \chi \Big]  - 50  \\
  =& 
- 50  \big(\log \chi - \tfrac{137}{60} \big)\,\chi^{2}+  \OO(\chi^3)\ . \no 
\end{align}
We have fixed the  integration constant   to zero  using   \rf{629}.\foot{This condition is natural  as  the connected part of   $\langle Y^A (t_1) Y^A (t_2) Y^B(t_3)  Y^B(t_4) $  
should vanish   for $t_{12}\to 0$ and $t_{34}\to 0$   (implying $\chi\to 0$)
as  we have $Y^A  Y^A=1$  at the coincident points. Thus  in \rf{6.2} we should have $G_S (\chi\to 0) \to 1$, 
i.e.    the  connected part of $G_S$ should vanish  at $\chi=0$.}
A non-trivial check  of \rf{6.43}   is that applying $\mathscr{D}^{2}$  it does satisfy  the  second  equation in \rf{671}   with 
$ G_{{\rm D}, S}(\chi)$ given  by \rf{2.35}. 

 It is   interesting to note that  despite  the relative simplicity of the  $\z^2 (\del \z)^2$  vertex contribution 
 to the   \rf{6.41}  (given by the term $-  80  \Big[ \tfrac{ \chi ^2}{ \chi -1}\log\chi   -  { \chi } \log (1-\chi )\Big]$ 
 on the r.h.s.)
it is this vertex that produces the most  complicated $  \text{Li}_{n}$ dependent part  in $\widebar {G}_{ S}(\chi) $  in \rf{6.43}
while the contribution   $(\widebar {G}_{ S})_{(\del \z)^4}$   of the $(\del \z)^4$   vertex is 
  similar in structure to the  expression in \rf{2.35} in the Dirichlet theory case: 
\be \la{6.411}
(\widebar {G}_{ S})_{(\del \z)^4}=-50 
+50\,\big(\tfrac{1}{2}-\tfrac{1}{\chi}-\tfrac{1}{5}\chi \big)\,\log(1-\chi)  -10\tfrac{\chi^{2}}{1-\chi}
\log \chi 
\  .
\ee
  \iffa 
  to  separate  in \rf{6.43} 
  the contributions of the $\z^2 (\del \z)^2$  and $(\del \z)^2$  vertices  in $L_4$ in \rf{3.7}. 
 Introducing a  marker coefficient $\br$ in front of $\z^2 (\del \z)^2$   vertex (or the contributions of \rf{6.36})
 we get the following generalization  of \rf{6.41} 
\begin{align}
-\mathscr D   \widebar{G}_{ S}(\chi) = & - 10 \Big[ \tfrac{\chi^2-10 \chi +10}{\chi -1}-
\tfrac{ (\chi^2-  2 \chi+ 2  ) \chi ^2}{ (\chi -1)^2}\log\chi  +\tfrac{\chi^3+5 \chi -10 }{ \chi } \log (1-\chi )\Big] \no \\
& -  80 \br  \Big[ \tfrac{ \chi ^2}{ \chi -1}\log\chi   -  { \chi } \log (1-\chi )\Big]
 \ . \la{6.411}
\end{align}
The integrated expression (\ref{6.43}) is then generalized as follows (we set $c_S=0$) 
\begin{align}
&\widebar {G}_{ S}(\chi) =  -50 
+50\,\big(\tfrac{1}{2}-\tfrac{1}{\chi}-\tfrac{1}{5}\chi \big)\,\log(1-\chi)  -10\tfrac{\chi^{2}}{1-\chi}
\log \chi  \notag \\
&   - 240\,\bm{\rho} \Big[ 
 \,\text{Li}_{3}(\chi) +  \,\text{Li}_{3}\big(\tfrac{\chi}{\chi-1}\big)
   - \tfrac{1}{3}   \text{Li}_{2}(\chi) \,\log(1-\chi) - \tfrac{1}{6}  \log^{3}(1-\chi)
 +   \tfrac{1}{6}   \log^{2}(1-\chi)\,\log \chi \Big]
 \  \la{6.431}\\
 &\qquad \quad\ \ 
  =  - 10\ \big(  \log \chi  - \tfrac{17}{12}      \big)\,\chi^{2} -   40\, \br   \big(     \log \chi  -  \tfrac{5}{2}   \big)\,\chi^{2}    +    \OO(\chi^3) \ . \no
\end{align}
Despite the relative simplicity of the  $\z^2 (\del \z)^2$  vertex contribution  in \rf{6.411} (the $\br$-term) 
it is this vertex that produces the most  complicated $  \text{Li}_{n}$ dependent part  in $\widebar {G}_{ S}(\chi) $, 
while the contribution of the $(\del \z)^2$   vertex is 
  similar in structure to the  expression in \rf{2.35} in the Dirichlet theory case. 
 \fi

 The total  expression for  the $1\ov (\sql)^3$ term in $ {G}_{ S}(\chi) $  in \rf{677} 
   is given  by the sum of $ \widetilde {G}_{ S}(\chi) $  in   \rf{6719},\rf{6721}   and  $ \widebar{G}_{ S}(\chi)$  in \rf{6.43} 
   and also  the reducible $d_2$-contribution  in    \rf{G89} (cf. also \rf{6400}), i.e.   explicitly  
   \begin{align} 
   G_S=&1 + \tfrac{10}{(\sql)^2} \log^2(1-\chi) +  \tfrac{1}{(\sql)^3}  G^{(3)}_S +   \OO (  \tfrac{1}{(\sql)^4} ) \ ,  \la{1000}\\
 G^{(3)}_{ S} =&  80\,\Big[\text{Li}_{3}(\chi)+\text{Li}_{3}\big(\tfrac{\chi}{\chi-1}\big)  -    \text{Li}_{2}(\chi) \, \log(1-\chi)   \Big]
+40\,\log\tfrac{\chi}{1-\chi}\,\log^{2}(1-\chi)\notag \\
&\ \ 
-10\,\tfrac{\chi^{2}}{1-\chi}\,\log\chi+5\,\big(5-\tfrac{10}{\chi}-2\,\chi \big)\, \log(1-\chi)
-50  + 4\, d_2 \log^2(1-\chi)  \ 
 \la{6599}\\
=& \ \  (30 \log \chi + \tfrac{205}{6}  + 4d_2 ) \chi^2  +  \OO(\chi^3)\ .  \no 
   \end{align}
   Let us now compute the   $\tfrac{1}{(\sql)^3}$   terms in the    $G_{ T}$ and $G_{ A}$ functions  
   (complementing the  order $\tfrac{1}{(\sql)^2}$  expressions in \rf{6.9},\rf{610}) using 
   the crossing relations \rf{6300},\rf{6.30}.   
As a first step,  let us   replace (\ref{6599})  by the following improved form that is equivalent to (\ref{6599})
for $0<\chi<1$ and represents its real continuation for $\chi>1$ (cf. footnote \ref{f44})
\begin{align}
\la{cross1}
G_{ S}^{(3)} =& -80 \, \text{Li}_3(1-\chi ) + \Big[80 \text{Li}_2(1-\chi )-\tfrac{5 (2 \chi ^2-5 \chi +10)}{\chi }\Big] \log 
|1-\chi| +10\,\tfrac{\chi ^2 }{\chi 
-1}\log\chi\notag \\
& -\tfrac{80}{3} \log ^3|1-\chi|+80 \log \chi  \log^2|1-\chi |+10\, [8 \zeta_R (3)-5] +4\,d_{2}\,\log^{2}|1-\chi|.
\end{align}
Note that using this expression we can consider the analytic continuation to the thermal out of time order correlators, following the procedure described in section \ref{chaos}. It is easy to see that the dominant contribution in the limit relevant for chaos comes again from the term $\sim  \chi^{-1}\log(1-\chi)$ in \rf{cross1}, leading to a maximal Lyapunov exponent. 
This  term  originates, in fact,  just from the  ``Nambu string''   $(\del \z)^4$ vertex   contribution  \rf{6.411}
(and  not from the $S^5$ sigma model vertex  $\z^2 (\del \z)^2$ in \rf{3.7}),   in full analogy 
with   what  happened  also in the supersymmetric line case
(cf.  last term in $G_{ S}^{(1)} $ in \rf{2.35}).

Applying the crossing relations \rf{6300},\rf{6.30}  
we can use (\ref{cross1}) to  get  the  following  (real) expressions for $G_{ T}$ and $G_{ A}$
that are valid in the range $0<\chi<1$  and depend also 
 on the subleading  coefficients in  $\Delta$ in  (\ref{41})
\begin{align}
\la{cross2}
G_{ T}^{(3)} =& 48 
\text{Li}_3(1-\chi ) + 
\log (1-\chi ) \Big[-12 \text{Li}_2(1-\chi )+36 \text{Li}_2(\chi 
)-\tfrac{3 (17 \chi ^2-11 \chi +1)}{2 \chi }-324 \log ^2\chi \Big]\notag \\
& +\tfrac{3 (17 \chi ^2-21 \chi +21) }{2 (
\chi -1)}\,\log \chi -83 \log ^3(1-\chi )+216 \log ^3\chi +294 \log \chi  
\log ^2(1-\chi ) \\
& +\tfrac{3}{2} [16 \zeta_R (3)-25]
-6\pi^{2}\log(1-\chi)
+\tfrac{3}{5}d_{2} \Big[9 \log ^2(\tfrac{\chi ^2}{1-\chi })+8 \log^2(1-\chi )\Big]
+ \tfrac{9}{10} d_{3}  \log (\tfrac{\chi ^2}{1-\chi })\ , \no \\
\la{cross3}
G_{ A}^{(3)} =&
48\Big[  \text{Li}_3(1-\chi )+ 2 \text{Li}_3(\chi )\Big] +\log (1-\chi ) \Big[48 
\text{Li}_2(\chi )-24 \chi +192 \log ^2\chi +3\Big]\notag\\
& +\Big[\tfrac{24 (\chi -2) 
\chi }{\chi -1}-96 \text{Li}_2(\chi )\Big] \log \chi +84 \log ^3(1-\chi 
)-192 \log \chi  \log ^2(1-\chi )-48 \zeta_R (3)
\notag \\
& +\tfrac{48}{5} d_{2} \log (1-\chi ) \log (\tfrac{\chi ^2}{1-\chi })
+\tfrac{6}{5}d_{3} \log (1-\chi ).
\end{align}
The small $\chi$ expansions of these expressions read (cf. \rf{6599})
\begin{align}
\la{cross4}
G_{ T}^{(3)}  = &
216\,\log^{3}\chi+\tfrac{108}{5} d_2\,\log^{2}\chi+ (-\tfrac{63}{2}+\tfrac{9}{5}\,d_{3})\,\log\chi+72 \zeta_R(3)-36\notag \\
&+ \Big[324\,\log^{2}\chi+\tfrac{108}{5} d_2\,\log\chi-\tfrac{63}{4}+\tfrac{9}{10}d_{3}\Big]\,\chi\notag \\
&+\Big[162\,\log^{2}\chi+(\tfrac{54}{5} d_2+\tfrac{513}{2})\,\log\chi+\tfrac{51}{5} d_2 +\tfrac{23}{4}
+\tfrac{9}{20}d_{3}
\Big]\,\chi^{2}+   \OO(\chi^3)\ ,  \\
G_{ A}^{(3)}  = & \Big[
-192\log^{2}\chi+(-\tfrac{96}{5} d_2-48)\log\chi+93-\tfrac{6}{5}\,d_{3}
\Big]\,\chi\notag \\
&+\Big[
-96\log^{2}\chi+(-\tfrac{48}{5}d_2-216)\log\chi-\tfrac{48}{5} d_2+\tfrac{45}{2}-\tfrac{3}{5}d_{3}
\Big]\,\chi^{2}+  \OO(\chi^3)\ . \la{6645}
\end{align}
A  direct computation of  $G_{ T}^{(3)} $ and $G_{ A}^{(3)} $ 
which is not based on the crossing relations  but follows the same  approach as used above  to find 
   $G^{(3)} _{ S}$ is presented in Appendix  \ref{app:TS}.
 Up to  the Casimir operator  zero mode
terms  (cf. \rf{631})   that are not, in general, determined in  the approach based 
on integrating the relations like \rf{6.14}
 the resulting  expressions are  found  to be   equivalent to (\ref{cross2}) and (\ref{cross3}).
 
The reason   why   this   ambiguity  was  not present in the  case of  $G^{(3)} _{ S}$ 
(or, equivalently,   was fixed by the  condition \rf{629})  can be understood from the  OPE constraints: 
  in the     singlet channel 
 the  only non-derivative    operator (with  dimension $\mc O(\frac{1}{\sql})$)   that can appear in the exchange  
 is the identity
(due to $Y^{A}Y^{A}=1$), implying $G_{S}(\chi) = \mc O(\chi^{2})$.
At the same time,   non-singlet $Y^A Y^B$  operators can appear in the OPE  of  $G_{ T} $ and $G_{ A} $. 

Finally, let us note that  the resulting  expressions  for  $G_{ S,T,A}^{(3)} $  in \rf{cross1},\rf{cross2},\rf{cross3}
depend  on  two   subleading coefficients $d_2$ and $d_3$ in the  scalar anomalous dimension \rf{41}   that
receive   contributions from the fermion loops    and  are yet to be 
determined.


\subsection{OPE and   anomalous dimensions}
\la{sec:OPE}
\def \dd {{\gamma}}

Let us now  discuss   the   consistency   of the expressions for  the $G_{S,T,A}$ functions   with the OPE 
  and extract  the anomalous dimensions 
 of  composite operators  appearing  in the intermediate   channels 
  as was done  in  the supersymmetric case in \ci{Giombi:2017cqn}
  (see \rf{2.346}--\rf{2.37} and Appendix \ref{app:BPSanom}). 
Let us assume the following  conformal-block expansion (cf. \rf{2.15},\rf{C.0}) 
\be
\la{6.53}
G_{ c} = \begin{cases}
c_{0}\,\chi^{h_{0}}\,F_{h_{0}}+c_{2}\,\chi^{h_{2}}\,F_{h_{2}}+\dots, & \qquad c=S, T, \\
c_{1}\,\chi^{h_{1}}\,F_{h_{1}}+c_{3}\,\chi^{h_{3}}\,F_{h_{3}}+\dots, & \qquad c=A, 
\end{cases}
\ee
where
\be
c_{n} = c^{(0)} _{n}+c^{(1)}_{n}\tfrac{1}{\sql}+c^{(2)}_{n}\tfrac{1}{(\sql)^{2}}+\dots,\qquad\quad 
h_{n} = n+  \dd^{(1)}_{n}\tfrac{1}{\sql}+  \dd^{(2)}_{n}\tfrac{1}{(\sql)^{2}}+\dots.\la{6653}
\ee
Comparing (\ref{6.799}),\rf{6.7},\rf{6599}     with (\ref{6.53}) we find  in the S-channel:
\begin{align}\la{6541}
h_{0, S} &= 0\ ,  & c_{0,S} &= 1+ \mc O(\tfrac{1}{(\sql)^{4}}) ,\no  \\
h_{2,S} & = 2+\tfrac{3}{\sql}+\cdots \ ,  & 
c_{2,S}  &= \tfrac{10}{(\sql)^{2}}+(\tfrac{205}{6}+4\,d_{2})\tfrac{1}{(\sql)^{3}}+ \cdots  , \notag \\
h_{4,S} &= 4-\tfrac{2}{\sql}+\cdots \ , & 
c_{4,S} &= \tfrac{1}{6\,(\sql)^{2}}+(\tfrac{24}{5}+\tfrac{1}{15}\,d_{2})\,\tfrac{1}{(\sql)^{3}}+\cdots. 
\end{align}
Here $d_{2}$  is  the  subleading coefficient in $\Delta$ in \rf{41}.
$h_0=0$ should correspond to the identity operator ($Y_A Y_A=1$), 
while $h_2$  to   the $Y_A \del^2_t Y_A$ operator.  
Similarly,  we get from \rf{cross4} 
\begin{align}\la{6542}
h_{0,T} &= \tfrac{12}{\sql}+\tfrac{12\,d_2}{5}\,\tfrac{1}{(\sql)^{2}}+ (-42+\tfrac{12}{5}\,d_{3})\,\tfrac{1}
{(\sql)^{3}}+\cdots\ , \qquad  
c_{0,T} = \tfrac{3}{4}-36[1-2\zeta_R(3)]\,\tfrac{1}{(\sql)^{3}}+ \cdots , \no  \\ 
h_{2,T} & = 2+\tfrac{171}{17}\,\tfrac{1}{\sql}+\cdots \ , \qquad \qquad \qquad 
c_{2,T} = \tfrac{51}{2} \tfrac{1}{(\sql)^{2}}+(-\tfrac{2483}{8}+\tfrac{51}{5}\,d_{2})\,\tfrac{1}{(\sql)^{3}}+  \dots , \\
h_{4,T} &= 4+\tfrac{86}{17}\,\tfrac{1}{\sql}+\cdots \ , \qquad  \qquad \ \qquad 
c_{4,T} = \tfrac{17}{40}\tfrac{1}{(\sql)^{2}}+(\tfrac{1893}{200}+\tfrac{17}{100}\,d_{2})\tfrac{1}{(\sql)^{3}}
+\cdots , \no 
\end{align}
and from \rf{6645}
\begin{align}\la{6560}
h_{1,A} & = 1+\tfrac{8}{\sql}+(8+\tfrac{8}{5}\,d_{2})\,\tfrac{1}{(\sql)^{2}}+\cdots\ ,\quad  &
c_{1,A} &= -\tfrac{6}{\,\sql}-\tfrac{6d_{2}}{5}\,\tfrac{1}{(\sql)^{2}}+ (93-\tfrac{6}{5}\,d_{3})
\tfrac{1}{(\sql)^{3}}+\cdots \ ,\no  \\
h_{3,A} & = 3+\tfrac{6}{\sql}+\cdots   \ , & 
c_{3,A} &= -\tfrac{8}{3}\tfrac{1}{(\sql)^{2}}-(\tfrac{368}{9}+\tfrac{16}{15}\,d_{2})\,\tfrac{1}{(\sql)^{3}}+\cdots \ ,  
\\
h_{5,A} &= 5-\tfrac{1}{\sql}+\cdots \ , & 
c_{5,A} &= -\tfrac{12}{175}\,\tfrac{1}{(\sql)^{2}}-(\tfrac{16997}{6125}+\tfrac{24}{875}\,d_{2})\,\tfrac{1}
{(\sql)^{3}}+\cdots\ .\no 
\end{align}
We have found  that the general form of the $1\ov \sql$  term in the  anomalous dimensions in 
\rf{6541},\rf{6542},\rf{6560} is as  in \rf{6653}, i.e. 
$
h_{n, \rm c} = n+ \dd_{n, \rm c}^{(1)} \tfrac{1}{\sql}+\cdots,
$  
with 
\begin{align}\la{6.71}
\dd_{n, S}^{(1)} &= \begin{cases}
0, & n=0, \\
4-\frac{1}{2}\,n\,(n-1), & n=2, 4, 6, \dots,
\end{cases}
\qquad
\dd_{n, T}^{(1)} = \begin{cases}
12, & n=0, \\
\tfrac{188}{17}-\frac{1}{2}\,n\,(n-1), & n=2, 4, 6, \dots,
\end{cases}\notag \\
\dd_{n, A}^{(1)} &= \begin{cases}
8, & n=1, \\
9-\frac{1}{2}\,n\,(n-1), & n=3, 5, 7,  \dots,
\end{cases}
\end{align}
The dependence on $n$ is the same in  all three channels (apart from  the 
``special''  bottom states $n=0$ for $c=S,T$ and $n=1$ for $c=A$). 
This was also true in the supersymmetric case   \ci{Giombi:2017cqn}. 
In fact, the large $n$   behaviour  of $\dd_{n}^{(1)} $  is the same 
in the    supersymmetric and the  non-supersymmetric   case 
\be
h_{n \gg 1}   = n   -    \tfrac{n^2}{2\sql}+\dots\ .  \la{6.721} 
\ee
This     universality (independence of  boundary conditions    on $S^5$ scalars)   should be 
consistent with possible semiclassical explanation of  this   large $n$ scaling 
(see also \rf{522} and comments below it).

 In the T-channel the OPE coefficients  with  $n>0$ are subleading in $1\ov \sql$, 
because they come from 3-point functions
  like $\langle Y(t_1)  Y(t_2)  (Y^{\{A} \del^n_t Y^{B\}})(t_3)\rangle$  which  for $n>0$  do not have an order-zero part. 
At order $1\ov \sql$  
 all the higher powers of  $\chi$  in $G_{T}$  in   \rf{6.9} 
 agree with the OPE    $c_{n} \chi^{\Delta_{n}}  F_n $
 containing  only the $n=0$ term (see also Appendix \ref{app:cube}). 
This means that the  OPE coefficients  with  $n>0$ start at $1\ov (\sql)^2$.
Indeed, we would  get   at least two $\z$-propagators (each  with  $1\ov \sql$) in 
  $\langle Y(t_1)  Y(t_2)  (Y\del^n_t Y)(t_3)   \rangle$ for $n\ge 2$.
   The anomalous dimension   of $ 
    Y^{\{A} \del^n_t Y^{B\}}$ should be 
$\frac{12}{\sql}+\cdots$  as expected from the analysis of the two point function (cf.  \rf{4.5}).  
Note  that the $d_{n}$ corrections to the anomalous dimensions in the previous results are always encoded
by a factor $1+\frac{d_{2}}{5}\frac{1}{\sql}+\frac{d_{3}}{5}
\frac{1}{(\sql)^{2}}+\dots$ correcting the leading order.  This is equal to the relative subleading
corrections to $\Delta$ in \rf{41}.
This follows from  the  universal ``dressing''  of the $\z$-propagator (cf. also  the expression for  $h_2$  in \rf{521})
at leading order in the coefficients $d_{n}$ and is a feature that is not expected to hold at higher orders. 
%
%
%
%

Similar comments  apply  to the S and A channels. 
The lowest-dimension   operator appearing in the A channel is 
$ Y^{[A} \del_t Y^{B]}$  which,  according to \rf{6560},  
  has  $h_{1,A}=1+ {8\ov \sql}+\dots$   and $c_{1,A}=- {6\ov \sql}+\dots$.\foot{
  The coefficient $+8$ in the anomalous dimension of $ Y^{[A} \del_t Y^{B]}$
may be understood  using  the same logic as  leading to $J(J+4)=12$ for
$Y^{\{A} Y^{B\}}$  with $J=2$  in the T channel,
cf. also \rf{4.5}.
In the latter  case,  one  may consider perturbing  the  string action  by the
boundary  interaction  term $\int dt\, T(Y)$,  $T(Y) =C_{A_1\dots
A_J}\,Y^{A_1}\cdots Y^{A_J}$,
and  then  its  anomalous dimension  operator  (entering  the condition
of this   being a marginal perturbation)  is the  scalar Laplacian on
$S^5$   (see, e.g.,  \cite{Tseytlin:2003ac,Beccaria:2017rbe}).
If instead one   considers the perturbation by the   operator $ Y^{[A}
\del_t Y^{B]}$, i.e.
$\int dt    F_{AB}  Y^A  \del_t Y^B \equiv \int dt   V_A(Y)  \del_t Y^A$,
where $F_{AB}$ is antisymmetric then one gets  a special case  of the
boundary vector   perturbation
for which the anomalous dimension operator  is the Maxwell  one or the 
vector Laplacian  on $S^5$ (the above $V_A$ is transverse).
The  eigenvalues of the  latter on $S^d$ are   $\lambda_{J} =
J\,(J+d-1)+d-2$  giving  $+8$
for $d=5$ and $J=1$.}

Another  consistency check   is possible using  the expressions \rf{6739},\rf{6767}   for the ``reducible''  
$\frac{1}{(\sql)^{3}}$  contributions in the T and A channels. 
The lowest order the  operator contributing to  the OPE
expansion (\ref{6.53}) in the  T-channel 
has $h_{0,T}=\frac{12}{\sql}+\dots$. This means that we should find a peculiar 
contribution $\frac{216}{(\sql)^{3}}\log^{3}\chi$  coming from the expansion of $c_{0,T}\,\chi^{h_{0,T}} = \frac{3}{4}\,
\chi^{{12\ov \sql}+\dots}$. 
  There are no such $\log^{3}\chi$ terms in  the  $\widebar{G}_{ c}$  functions  corresponding to 
  the  connected diagram contribution. 
In fact,  this   contribution   is provided by the $\widetilde{G}_{ T}$  function \rf{6739} 
 that complements  $\widebar{G}_{ T}$  
 to the full  ${G}_{ T}$    like   in the S-channel in  \rf{671}: it contains the required $\log^3\chi $ term 
 in its $\chi\to 0$ expansion in \rf{cross4}.
Similarly, the A-channel  expression \rf{6645},\rf{6767}   contains 
the  term $-\frac{192}{(\sql)^{3}}\,\chi\,\log^{2}\chi$  which   is  precisely the one 
 appearing  in  the expansion of 
  $c_{1,A}  \chi^{h_{1,A}} = -\frac{6}{\sql}\chi^{1+{8\ov \sql}+\dots}$,   see  (\ref{6560}).


\section*{Acknowledgments}
We are grateful to 
 S. Dubovsky, V. Gorbenko, 
S. Komatsu, J. Maldacena,  R. Roiban and E. Vescovi for  useful discussions on related topics.
We also  thank   R. Roiban and E. Vescovi for  comments on the draft. 
 The work of SG is supported in part by the US NSF under Grant No. PHY-1620542. 
AAT was supported by the STFC grant ST/P000762/1.

\appendix

\section{Four-point  correlators  of  generalized   free fields}
\la{app:freefield}

Assuming   that $\OO_{\Delta}(t)$ is represented by a  free field  and   normalized so that 
\be \la{A0}
\llangle \OO_{\Delta} (t_1) \ \OO_{\Delta'}(t_2) \rrangle =  {\delta_{\D,\D'} \over (t_{12})^{2 \Delta}} \ , \ee
and  doing three  separate contractions  one finds for the correlator in \rf{2.13} 
\be
\la{A.1}
\llangle \mc O_{\Delta}(t_{1})\,\mc O_{\Delta}(t_{2})\,
\mc O_{\Delta}(t_{3})\,\mc O_{\Delta}(t_{4})\rrangle = \frac{1}{(t_{12}t_{34})^{2\Delta}}
\,  G (\chi) \ , \qquad G = 1+\chi^{2\Delta}+\frac{\chi^{2\Delta}}{(1-\chi)^{2\Delta}}
\ . 
\ee
This can be checked against (\ref{2.15}) by taking into account that the exchanged fields are 
the identity operator  and the composites
\be\la{A.2}
[\mc O_{\Delta}\mc O_{\Delta}]_{2n}\sim \mc O_{\Delta}\partial_{t}^{2n}\mc O_{\Delta},\qquad 
\qquad h = 2\Delta+2n, \quad n=0, 1, \dots,
\ee
with the OPE coefficients given by (see, e.g.,  \cite{Heemskerk:2009pn,Fitzpatrick:2011dm})
\be
\la{A.3}
c_{\Delta,\Delta; 2\Delta+2n} = \frac{2\,[\Gamma(2n+2\Delta)]^{2}\Gamma(2n+4\Delta-1)}
{[\Gamma(2\Delta)]^{2}\Gamma(2n+1)\Gamma(4n+4\Delta-1)}.
\ee
One can show  that the +1 in (\ref{A.1}) comes from the identity, while the rest comes from the 
tower of  operators in \rf{A.2}. Also,  
$1+\frac{1}{(1-\chi)^{2\Delta}}= \sum_{n=0}^{\infty}
c_{\Delta, \Delta; 2\Delta+2n}\,\chi^{2n}\,
_{2}F_{1}(2\Delta+2n,2\Delta+2n,4\Delta+4n,\chi).$
Similarly, in the case of  two different  dimensions \rf{2.17} one gets 
\be
\llangle \mc O_{\Delta_{1}}(t_{1})\,\mc O_{\Delta_{2}}(t_{2})\,
\mc O_{\Delta_{1}}(t_{3})\,\mc O_{\Delta_{2}}(t_{4})\rrangle = 
\frac{1}{t_{13}^{2\Delta_{1}}\,t_{24}^{2\Delta_{2}}} = 
\frac{1}{(t_{12}t_{34})^{\Delta_{1}+\Delta_{2}}}\,\left|
\frac{t_{24}}{t_{13}}\right|^{\Delta_{12}}   \, G(\chi) \ , \qquad  G= \chi^{\Delta_{1}+\Delta_{2}}.\la{A.4}
\ee
Here we assumed that $\Delta_1\not=\Delta_2$   so that \rf{A.1} is a not a limit of \rf{A.4}. 
The form  of $\mc G = \chi^{\Delta_{1}+\Delta_{2}}$   here can again be explained in terms of the fusion
$\mc O_{\Delta_{1}}+\mc O_{\Delta_{2}}\stackrel{h}{\to}
\mc O_{\Delta_{1}}+\mc O_{\Delta_{2}}$ 
leading to the composite operators 
\be
[\mc O_{\Delta_{1}}\mc O_{\Delta_{2}}]_{n}\sim 
\mc O_{\Delta_{1}}\partial_{t}^{n}\mc O_{\Delta_{2}},
\qquad h = \Delta_{1}+\Delta_{2}+n, \quad n=0, 1, \dots,\la{A.5}
\ee
with the OPE coefficients
\be
\la{A.6}
c_{\Delta_{1},\Delta_{2}; \Delta_{1}+\Delta_{2}+2n} = 
\frac{(-1)^{n}\Gamma(n+2\Delta_{1})\Gamma(n+2\Delta_{2})
\Gamma(n+2\Delta_{1}+2\Delta_{2}-1)}
{\Gamma(2\Delta_{1})\Gamma(2\Delta_{2})\Gamma(n+1)
\Gamma(2n+2\Delta_{1}+2\Delta_{2}-1)} \ .
\ee
  
\section{Anomalous dimensions from OPE in   supersymmetric  case }
\la{app:BPSanom}

Here  we recall  how  the anomalous dimensions may be extracted from the  OPE   expansion of the 
$G(\chi)$ function in \rf{2.15} on the example  of  the symmetric traceless tensor  part 
in the  supersymmetric line case  following  \cite{Giombi:2017cqn}. 
The strong-coupling expansion of the 5-scalar  four-point function  \rf{2.34} leads to  (cf. \rf{2.15}) 
\begin{align}
& G_{T}^{(0)}(\chi)+\tfrac{1}{\sqrt\lambda}\,G_{T}^{(1)}(\chi)+\dots = \sum_{h}c_{h}\chi^{h}F_{h}(\chi) ,
\qquad \qquad F_{h}(\chi) = {_{2}F_{1}}(h,h,2h,\chi),\la{C.0} \\
& h_n = 2+2n+\tfrac{1}{\sqrt\lambda}\,\gamma^{(1)}_{[\Phi\Phi]^{T}_{2n}}+\dots,\qquad
c_{h}= c^{(0)}_{\Phi\Phi[\Phi\Phi]_{2n}^{T}}+\tfrac{1}{\sqrt\lambda}c^{(1)}_{\Phi\Phi[\Phi\Phi]_{2n}^{T}}
+\dots. \la{C.1}
\end{align}
Comparing the  leading order  term   
 with the free-field result (\ref{A.1}),(\ref{A.3}), we obtain 
\be
\la{C.2}
c^{(0)}_{\Phi\Phi[\Phi\Phi]_{2n}^{T}} = \frac{[\Gamma(2n+2)]^{2}\Gamma(2n+3)}
{\Gamma(2n+1)\Gamma(4n+3)}.
\ee 
To get the subleading order correction   we use 
\be\la{C4}
\chi^{h}=\chi^{2+2n+\frac{1}{\sqrt\lambda}\gamma^{(1)}}=\chi^{2+2n}\big(1+\tfrac{1}{\sqrt\lambda}
\gamma^{(1)}\log\chi+\dots\big),
\ee
and the general inversion formula 
\be	
\la{C.4}
\sum_{n=0}^{\infty}c_{n}\,\chi^{2+2n}\,F_{2+2n}(\chi) = f(\chi)\quad\to\quad
c_{n} = \oint\frac{d\chi}{2\,\pi\,i}\chi^{-3-2n}F_{-1-2n}(\chi)\,f(\chi) \ . 
\ee
As a result, 
\be
\la{C.5}
\gamma^{(1)}_{[\Phi\Phi]^{T}_{2n}} = \big[ c^{(0)}_{   \Phi\Phi[\Phi\Phi]_{2n}^{T}} \big]^{-1} 
\oint\frac{d\chi}{2\,\pi\,i}\chi^{-3-2n}F_{-1-2n}(\chi)\,[G_{T}^{(1)}(\chi)]_{\log\chi} = -3\,n-2\,n^{2}.
\ee
One can  compute in a similar way  the  correction to 
the OPE coefficients \cite{Giombi:2017cqn}.

\section{AdS contact integrals}
\la{app:Dfunctions}

The building block for AdS$_{d+1}$ diagrams with a 4-point contact term like in \rf{2.34} 
 is the $D$-function (see, e.g.,   \cite{DHoker:1999kzh,Dolan:2000ut,Arutyunov:2002fh})
\be\la{B.1} 
D_{\Delta_{1},\Delta_{2},\Delta_{3},\Delta_{4}}(x_{1}, x_{2}, x_{3}, x_{4}) = 
\int\frac{dz\,d^{d}x}{z^{d+1}}\,\prod_{n=1}^{4}\KK_{\Delta_{n}}(z, x; x_{n}) \ , 
\ee
where $\KK_\D$ was  defined in \rf{2.32}. 
A useful identity is 
\begin{align}
\la{B.2}
& g^{\m\n} \partial_\mu \KK_{\Delta_{1}}(z,x;x_{1})\,
\partial_\n  \KK_{\Delta_{2}}(z,x;x_{2})  \notag \\
&=\Delta_{1}\Delta_{2}\,\Big[\KK_{\Delta_{1}}(z,x;x_{1})
\KK_{\Delta_{2}}(z,x;x_{2})-2\,x_{12}^{2}\,
\KK_{\Delta_{1}+1}(z,x;x_{1})\KK_{\Delta_{2}+1}(z,x;x_{2})
\Big],
\end{align}
where $\partial_{\mu}=(\partial_{z},\partial_r)$ ($r=(0,i)$)  and $g^{\mu\nu}=z^{2}\delta^{\mu\nu}$.
It is useful to replace $D$ functions by $\overline{D}$ functions that depend on the conformally
invariant ratios $u=\frac{x_{12}x_{34}}{x_{13}x_{24}}$, $v=\frac{x_{14}x_{23}}{x_{13}x_{24}}$ 
 ($\Sigma\equiv \frac{1}{2}\sum_{n}\Delta_{n})$
\begin{align}
\la{B.3}
 D_{\Delta_{1}, \Delta_{2}, \Delta_{3}, \Delta_{4}} &= 
\frac{\pi^{d/2}\Gamma(\Sigma-\frac{d}{2})}{2\Gamma(\Delta_{1})\Gamma(\Delta_{2})
\Gamma(\Delta_{3})\Gamma(\Delta_{4})}\frac{
x_{14}^{2(\Sigma-\Delta_{1}-\Delta_{4})}\,
x_{34}^{2(\Sigma-\Delta_{3}-\Delta_{4})}}
{x_{13}^{2(\Sigma-\Delta_{4})}\,x_{24}^{2\Delta_{2}}}\,\overline D_{\Delta_{1}, \Delta_{2}, \Delta_{3}, \Delta_{4}}(u,v),\notag \\
\overline D_{\Delta_{1}, \Delta_{2}, \Delta_{3}, \Delta_{4}}(u,v) &=
\int d^{3}\alpha\,\delta(\sum^3_{i=1}\alpha_i-1)\,\alpha_{1}^{\Delta_{1}-1}
\alpha_{2}^{\Delta_{2}-1}\alpha_{3}^{\Delta_{3}-1}\,\frac{\Gamma(\Sigma-\Delta_{4})\Gamma(\Delta_{4})}
{(\alpha_{1}\alpha_{3}+\alpha_{1}\alpha_{2}\,u+\alpha_{2}\alpha_{3}v)^{\Sigma-\Delta_{4}}}.
\end{align}
Specializing to AdS$_2$ or   $d=1$   where  $u=\chi^{2}$ ,  $v=(1-\chi)^{2}$,  
one can prove that 
\begin{align}
\la{B.4}
& \overline{D}_{\Delta_{1}, \Delta_{2}, \Delta_{3}, \Delta_{4}} = \tfrac{\Gamma(\Delta_{1})\,\Gamma(\Delta_{4})\,
\Gamma(\frac{\Delta_{1}+\Delta_{2}+\Delta_{3}-\Delta_{4}}{2})
\Gamma(\frac{-\Delta_{1}+\Delta_{2}+\Delta_{3}+\Delta_{4}}{2})}
{\Gamma(\frac{\Delta_{1}+\Delta_{2}+\Delta_{3}+\Delta_{4}}{2})}\
\chi^{-\frac{\Delta_{1}+\Delta_{2}-\Delta_{3}-\Delta_{4}}{2}}
(1-\chi)^{-\Delta_{1}-\Delta_{2}-\Delta_{3}+\Delta_{4}}
\notag \\
&\times \int_{-\infty}^{\infty}d\tau\,e^{-\tau\,\frac{\Delta_{1}-\Delta_{2}+\Delta_{3}-\Delta_{4}}{2}}\,
(e^{\tau}+\chi)^{\Delta_{1}-\Delta_{4}}\, {}_{2}F_{1}\Big(\Delta_{1},\tfrac{\Delta_{1}+\Delta_{2}+\Delta_{3}-\Delta_{4}}{2},
\tfrac{\Delta_{1}+\Delta_{2}+\Delta_{3}+\Delta_{4}}{2}, - \tfrac{4\chi}{(1-\chi)^{2}}\,\cosh^2 \tfrac{\tau}{2} \Big) \ .  
\end{align}
In particular, we get   (assuming  $0<\chi<1$)
\begin{align}
\la{B.5}
\overline{D}_{1,1,1,1} &\tet  = -\frac{2\,\log\chi}{1-\chi}-\frac{2\,\log(1-\chi)}{\chi},\qquad 
\overline{D}_{1,1,2,2} = \frac{\chi ^2 \log (\chi )}{3 (\chi -1)^2}-\frac{1}{3 (\chi \
-1)}-\frac{(\chi +2) \log (1-\chi )}{3 \chi },\notag \\
\overline{D}_{1,2,2,1} &\tet = \frac{\log (1-\chi )}{3 \chi ^2}+\frac{1}{3 (\chi -1)^2 \chi \
}-\frac{(\chi -3) \log (\chi )}{3 (\chi -1)^3},\qquad 
\overline{D}_{2,2,1,1} = -\frac{(\chi +2) \log (1-\chi )}{3 \chi ^3}-\frac{1}{3 (\chi -1) \chi \
^2}+\frac{\log (\chi )}{3 (\chi -1)^2},\notag \\
\overline{D}_{1,2,1,2} &\tet = -\frac{(2 \chi +1) \log (1-\chi )}{3 \chi ^2}+\frac{1}{3 (\chi -1) \
\chi }+\frac{(2 \chi -3) \log (\chi )}{3 (\chi -1)^2},\no \\
\overline{D}_{2,1,2,1}& \tet = -\frac{(2 \chi +1) \log (1-\chi )}{3 \chi ^2}+\frac{1}{3 (\chi -1) \
\chi }+\frac{(2 \chi -3) \log (\chi )}{3 (\chi -1)^2},\notag \\
\overline{D}_{2,1,1,2} &\tet = \frac{(\chi -1)^2 \log (1-\chi )}{3 \chi ^2}+\frac{1}{3 \chi \
}-\frac{(\chi -3) \log (\chi )}{3 (\chi -1)}\ , \no \\ 
\overline{D}_{2,2,2,2} &\tet = -\frac{2 (\chi ^2-\chi +1)}{15 (\chi -1)^2 \chi ^2}+\frac{(2 \chi \
^2-5 \chi +5) \log (\chi )}{15 (\chi -1)^3}-\frac{(2 \chi ^2+\chi +2) \
\log (1-\chi )}{15 \chi ^3}.
\end{align}

\section{Green's    functions  for   2d massless scalar }  
\la{app:propagator}

In this Appendix we 
discuss    the form  of  2d massless  scalar propagator 
with Neumann boundary conditions on a  space  with  half-plane   or  disc topology
(with AdS$_{2}$ being a special case).

It is useful first  to  recall  the  case 
of compact 2d surface with no boundary (i.e. sphere  topology). 
\iffa 
Starting with the massless   scalar action 
\be
S = \frac{1}{4\pi}\int d^{2}\sigma\sqrt{g}\,g^{ab}\partial_{a}X\partial_{b}X 
\ee
and expand in  (slowly varying) background plus fluctuation
$X^{A} = x^{A}+\sqrt{2\pi}\,\xi^{A}$. At leading order in the background expansion
the fluctuations propagator is 
\be
\la{D.2a}
G^{AB}(\sigma, \sigma') = \int \mc D\xi\,\xi^{A}(\sigma)\xi^{B}(\sigma')\,e^{-S_{2}}
=\delta^{AB}\,G(\sigma, \sigma'),\qquad
S_{2}=\frac{1}{2}\int d^{2}\sigma\,\sqrt{g}\,g^{ab}\partial_{a}\xi^{A}\partial_{b}\xi^{A}.
\ee
\fi
 The Laplace-Beltrami
operator $-D^{2} = -\frac{1}{\sqrt{g}}\,\partial_{\m}(\sqrt{g}\,g^{\m\n}\partial_{\n})$
has eigenvectors
$-D^{2}u_{n}=\lambda^{2}_{n}\,u_{n}$ with 
$\int d^{2}\sigma\,\sqrt{g}\,u_{n}u_{m} = \delta_{nm}$ and 
$\sum_{n}u_{n}(\sigma)u_{n}(\sigma') = \frac{1}{\sqrt{g}}\,\delta^{(2)}(\sigma-\sigma')$.
Separating the constant zero mode 
$u_{0}(\sigma) = \frac{1}{\sqrt V}$   we get for the Green's function (see, e.g., \cite{RandjbarDaemi:1987aj})
\begin{align}
\la{D.1}
&  \GG(\sigma,\sigma') = \sum_{n>0}\frac{1}{\lambda_{n}^{2}}u_{n}(\sigma)u_{n}(\sigma'),
\qquad  
 -D^{2}\GG(\sigma, \sigma') = \sum_{n>0}u_{n}(\sigma)u_{n}(\sigma') = 
\delta^{(2)}(\sigma,\sigma')-\frac{1}{V} \ , 
\end{align}
where $\delta^{(2)}(\sigma,\sigma') = \frac{1}{\sqrt{g}}\delta^{(2)}(\sigma-\sigma')$.

In conformally flat coordinates $ds^{2} = e^{2\rho}\,dw\,d\overline{w}$    the Green's function formally 
should not depend on the conformal factor; assuming plane  topology  it may still enter  via a (covariant) UV cutoff $\eps\equiv \eps_{\rm UV}$ 
introduced as 
\be
\la{D.2}
\GG(w,w')=-\frac{1}{4\pi}\log \big(|w-w'|^2 + \eps^2 \, e^{- \rho(w)- \rho(w')}\big) \ .
\ee
For a sphere topology a natural counterpart of this expression  is 
\be \la{D.2x}
\GG(w,w')= -\frac{1}{4\pi}\log [s^2(w,w') + \eps^2] \ , \qquad \qquad
 s^2(w,w') = e^{ \rho(w)+\rho(w')}\, |w-w'|^2 \ . \ee
   In  critical string theory  in Polyakov approach \cite{Polyakov:1981rd} the
  dependence on conformal  factor  should   completely  cancel out in  the expressions 
  for on-shell scattering amplitudes (see, e.g., \ci{Nepomechie:1981ne}).\foot{ It survives in  general in   sigma model partition function
     \cite{Fradkin:1985ys}.}
  
 Similarly,  for  the  critical string on a world sheet 
  with a  boundary (with, e.g.,  half-plane topology) 
   the standard  massless propagator  can   be  found  using the method of
images 
\be
\GG_{\rm D,N}(w,w') = -\frac{1}{4\pi}\Big[\log |w-w'|^2 \mp\log |w-\overline{w}'|^2 \Big],
\ee
where  the $\mp$ signs  correspond to the 
 Dirichlet (D)  and Neumann (N) boundary conditions.
Introducing   a  covariant UV cutoff  like in  (\ref{D.2})  gives 
(see also \cite{Durhuus:1981gu})
\be
\la{D.5}
\GG_{\rm D,N}(w,w') = -\frac{1}{4\pi}\Big[
\log\big( |w-w'|^2 +\eps^{2}\, e^{- \rho(w)- \rho(w')} \big)  \mp  \log |w-\overline{w}'|^2 \Big].
\ee
   This is true  on a half-plane with any conformal factor.
   In the  special case of  AdS$_{2}$  
 \be\la{d6}
  ds^2 =  \frac{1}{z^{2}} ( dt^2 + dz^2)= {d w d \overline{w} \ov { ({\rm Im} \, w)^2}}\ , \ \ \ \   \ \ \ \ \ \ \ \ \ w=t+i\,z  \ , \ee
   we get   from  (\ref{D.5})  (cf. (\ref{3.9}))
\be
\la{D.7}
\text{AdS}_{2}:\quad 
\GG_{\rm D,N}(t,z; t', z') = -\frac{1}{4\pi}\Big[\log\big[  (t-t')^{2}+(z-z')^{2}+\eps^{2}z\,z'\big] 
\mp \log\big[  (t-t')^{2}+(z+z')^{2}\big] \Big] \ . 
\ee
The bulk-to-boundary propagators 
are obtained  by 
taking $z=\epsilon \to 0$ ($\epsilon=\eps_{\rm IR}$ is an IR cutoff): 
\begin{align}
\GG_{\rm  D}(t, z; t', \eps) &=\epsilon\,   K (t,z;t')  +\mc O(\epsilon^2 ), \qquad \qquad
   K (t,z;t' ) = \,\frac{1}{\pi}\, \frac{z}{(t-t')^{2}+z^{2}} \ , \la{D.8}  \\ 
\GG_{\rm N}(t, z; t',\eps) &= -\frac{1}{2\pi}\,\log\big[ (t-t')^{2}+z^{2}\big]+\mc O(\epsilon). \la{D.9}
\end{align}
In the Dirichlet case  we  obtain the standard bulk-to-boundary propagator (\ref{2.32})  in AdS$_{2}$;  
 the  extra $\epsilon$  factor may be  absorbed into a  rescaling of boundary  fields. 
 In the Neumann case  the rescaling  is not  needed    
 (consistently with the free  boundary fields 
being dimensionless) and we recover (\ref{57}). 

If  the Weyl  invariance  of the  theory is not  manifest (like   in the expansion of the Nambu action) we
may use instead a  covariant approach specific to  a particular 2-space. For
a homogeneous space like a  half-sphere or AdS$_2$  it is natural  to represent the
propagator in terms of  the geodesic distance $s(\sigma, \sigma')$. Then   in conformally  flat  coordinates 
  for a half-plane topology  ($ds^2 = e^{2 \rho} d w d \overline{w}'$)  we get  
\be\la{d99}
\GG_{\rm D, N}(w,w') =  -\frac{1}{4\pi}\big[ \log s^2(w,w')   \mp  \log   s^2(w,\overline{w}')\big]  \ , 
\ee
where the  covariant bulk UV cutoff    may be  introduced  as in \rf{D.2x} 
 by $  s^2(w,w')\to  s^2(w,w') + \eps^2$. 
In the AdS$_2$  case       we then get  explicitly  for the Neumann case\footnote{Notice that here 
 for the  separated points (when the delta-function  is zero) we get 
$-D^{2}\GG_{\rm N} = -z^{2}\partial_{a}\partial_{a}\GG_{\rm N} = \frac{1}{2\pi}$. This may be interpreted as in (\ref{D.1})
as a  consequence of projecting out the  constant zero-mode contribution present  for  the Neumann  boundary conditions:
indeed, this   expression  is  
in agreement with (\ref{D.1}) after   taking into account  that the regularized  volume  of AdS$_{2}$  with  the $S^1$ boundary 
is  $V=-2\pi$. 
} 
\be
\la{D.11}
\GG_{\rm N}(t, z; t', z') = -\frac{1}{4\pi}\,\left[
\log\frac{(t-t')^{2}+(z-z')^{2}}{2zz'}+\log\frac{(t-t')^{2}+(z+z')^{2}}{2zz'}
\right].
\ee
Note that   the normal derivative of $\GG_{\rm N}$ is constant at  $z=0$, instead of 
being zero  as  for  the naive   Neumann boundary conditions. In fact, the 
natural Neumann boundary condition  on a  massless  scalar  here 
 is $\partial_{n}\varphi\Big|_{z=0} = h=\text{constant}$: 
near the boundary   $\varphi(z\to 0) = h\log z + \dots$ which is  consistent with 
$\varphi\sim a z^{\de}+\cdots$ when $\de\to 0$. 
A  closely related  discussion of the  Neumann   function for AdS$_2$  may be found in \cite{Solodukhin:1998ec}.\footnote{In    \cite{Solodukhin:1998ec}
 one finds  equivalent expressions:    in global  AdS$_2$   coordinates $ds^2 = dr^2 + \sinh^2 r d \phi^2$ 
 the  geodesic distance $s(r, \phi; r', \phi')$   satisfies $\cosh s   = \cosh r \, \cosh r' - \sinh r \, \sinh r' \cos (\phi - \phi')$  and then 
    $\GG_{\rm D}= - \frac{1}{4\pi}  \log { u\over u+1}$ and 
$\GG_{\rm N}= - \frac{1}{4\pi} \log [ u (u+1)]$
where  $u= \sinh^2 {s\ov 2} $,  $u+1=\cosh^{2}{s\ov 2}$,  etc.
Here again   $\partial_{n}\GG_{\rm N}$
tends to a constant at  the boundary, see   Eq.(5.21) of \cite{Solodukhin:1998ec}.}


In  bosonic model  where   power  divergences do not automatically  cancel out results for correlators 
involving derivatives of the Green's function at coinciding points in general depend  on regularization scheme. 
In the case of string sigma model  that    scheme should be fixed so  that to preserve underlying (target-space) 
symmetries of the theory.  
For example, 
the  second derivative  at coinciding points   $g^{\m\n}D_{\m}D_{\n}'
\GG(\sigma, \sigma')\big|_{\sigma=\sigma'}$
 depends on the choice
of UV  regularization  as discussed, e.g., in   \cite{RandjbarDaemi:1987aj,Tseytlin:1988jp}. 
Using   spectral or heat kernel regularization 
$
\GG(\sigma, \sigma'; \eps) =  \sum_{n>0}\frac{1}{\lambda_{n}^{2}}u_{n}(\sigma)u_{n}(\sigma')\,
e^{-\eps\,\lambda_{n}^{2}},
$
one finds  in  conformally-flat  coordinates (in the absence of the boundary) 
\begin{align}
&\partial_{\m}\GG(\sigma, \sigma'; \eps)\big|_{\sigma=\sigma'} = \frac{1}{4\pi}\partial_{\m}\rho(\sigma),\quad\la{D123} \\
&\partial_{\m}\partial_{\m}'\GG(\sigma, \sigma'; \eps)\big|_{\sigma=\sigma'} =  \frac{e^{2\rho(\sigma)}}{4\pi\,\eps}
+  \frac{a}{4\pi}\partial^{2}\rho(\sigma) -e^{2\rho(\sigma)}\,u_{0}^{2}\ , \la{D.12}
\end{align} 
where $u_{0}^{2}= {1\ov V}$   and $a= {2\ov 3}$. 
The coefficient  $a$  of  the $\partial^{2}\rho$  in $\partial_{a}\partial_{a}'G$ 
  is regularization dependent:  it  becomes $a=1$ in dimensional
regularization   and     is   $a=0$    if one uses  the 
 covariant Green's function on $S^2$  (see  \cite{RandjbarDaemi:1987aj}). 
It is a  particular (dimensional regularization or equivalent)
  scheme that leads to results consistent  with
string  theory symmetries  in the 2-sphere  case  (see, e.g., \cite{deAlwis:1985zy,Tseytlin:1988ne}). 

Similar  expressions    are found in the presence of the boundary. 
Using  that  for  AdS$_2$   $\rho =- \log z$, $z^2 \del_\m \del_\m \rho = 1$ (and ignoring the first  and  the 
 last term in \rf{D.12})
 the   coefficient  $a=1$ of $\partial^{2}\rho$  term 
in \rf{D.12}  corresponds to $k=1$   choice in \rf{4.15}.\foot{This follows also 
directly from  (\ref{D.7}) as well as (\ref{D.11}) supplemented with the $\eps$-regularization term.}  
At the same time, 
as   discussed   in \ci{Tseytlin:1988ne}, in  the boundary  case a   more  natural 
  option is  to  keep   only the last  term in the analog of \rf{D.12}. Then 
(with $V_{{\rm AdS}_2} = - 2\pi$)  we get 
\be \la{ddd} 
e^{-2\rho} \partial_{\m}\partial_{\m}'  \GG_{\rm N} (\sigma, \sigma'; \eps)\big|_{\sigma=\sigma'} =   { 1 \ov 2 \pi} \ , \ee 
which corresponds to   $k=2$ choice in \rf{4.15}  that we used in \rf{419}.

\section{Equivalence of different parametrizations of $S^5$  } 
\la{app:alternative}

The quartic  Lagrangian \rf{2.99}  used in  the supersymmetric   line case 
\ci{Giombi:2017cqn}  corresponds to the parametrization of $S^5$ 
defined in \rf{8}. At the same time, in the discussion of the  non-supersymmetric case in section \ref{sec:main} we used a different parametrization \rf{3.5}    with the corresponding  Lagrangian in \rf{3.7}.  
Choosing  there $n^a=0, n^6=1 (a=1, \dots, 5)$
and renaming  $\z^a\to y^a$   the two  Lagrangians   become special cases  of the  following family
\begin{align}
\la{E.1}
L_{4}  = r_1  \,y^{b}y^{b} \,(\partial y^{a}\cdot\partial y^{a})
+   r_2 \, y^{a}\,y^{b}\,(\partial y^{a}\cdot\partial y^{b}) + \OO\big( (\del y)^4)) 
\end{align}
where  \rf{2.99}  corresponds to   $r_1= -{1\ov 4}, r_2=0$ and  \rf{3.7} -- to 
$r_1= 0, r_2={1\ov 2} .$ 
That  the two cases are  related by a field redefinition  is reflected 
in the fact that if we integrate by parts and  ignore the term proportional to the  $y^a$   equations  of motion ($\Box y^a=0$) 
then the quartic  Lagrangian becomes the same -- depending on the  combination $r_1- \ha r_2 $ which is  equal to $-{1\ov 4}$ in both cases:
\begin{align}\la{E2}
L_{4y} = (r_1-\tfrac{1}{2}\, r_2)\, y^{b}y^{b} \,(\partial y^{a}\cdot\partial y^{a}) + \OO\big( (\del y)^4)) \ . 
\end{align}
Explicitly, 
$
y^{a}\,y^{b}\,(\partial y^{a}\cdot\partial y^{b}) = \tfrac{1}{4} \partial(y^{2})\cdot\partial(y^{2})
\to -\tfrac{1}{4} y^{2} \Box(y^{2}) = -\tfrac{1}{2}\,(y^{2})(\partial y\cdot \partial y) + \OO(\Box y)$.
One can check  that  field redefinitions leave  boundary (``on-shel'')  AdS correlators invariant: 
 the  correlator \rf{2.34}   computed starting  directly with \rf{E.1} 
depends only on $ r_1-\tfrac{1}{2}\, r_2$, i.e. is the same as  the one corresponding to \rf{E2}. 
\iffa
In   particular, 
 the anomalous dimension  for $\D=1$  operator in  (\ref{2.37}) extracted from the 
4-scalar   correlator  computed  using \rf{E.1} as a starting point   becomes 
\be\la{E3} 
\de=1: \qquad \gamma_{n}^{S} = -1+16\, (r_1-\tfrac{1}{2}\, r_2  )-3n-2n^{2} \ .
\ee
\fi 

\iffa 
\section{An alternative point splitting regularization of tadpole contributions {\it Remove?}}
\la{app:altern-ps}

As a check of the \underline{naive bulk} vertex contribution in (\ref{419}), i.e. to recover the coefficient
\be
\la{F.0}
\frac{1}{6}\,\left(-\frac{5}{4}k+15-\frac{5}{2}\right)_{k=1} = \frac{15}{8},
\ee
we  
discuss an alternative idea is based on maintaining the $SO(6)$ covariant combination $Y^{A}$
as the basic composite operator in terms of $\z$. We start by rewriting the first term of (\ref{3.7})
as 
\begin{align}
L_{4}^{(I)} = \tfrac{1}{2}\,\zeta^{m}\zeta^{n}\,(\partial \zeta^{m}\cdot\partial\zeta^{n}) = 
\tfrac{1}{8}\,\partial^{\mu}(\z^{m}\z^{m})\,\partial_{\mu}(\z^{n}\z^{n}) \to 
-\tfrac{1}{8}\,\z^{m}\z^{m}\,\partial^{\mu}\partial_{\mu}(\z^{n}\z^{n}). 
\end{align}
Let us denote $\z^{A}_{1}=\z^{A}(\sigma)$ and $\z^{A}_{2}=\z^{A}(\sigma')$ where $\sigma, \sigma'$
denote the split points. We need first to compute
\begin{align}
\langle\z^{A}(t_{1}) & \z^{B}(t_{2})\,\z^{m}_{1}\z^{m}_{1}\,\Box_{2}(\z^{n}_{2}\z^{n}_{2})
\rangle\notag \\
& = 4\,\langle \z^{A} \z^{m}_{1}\rangle \, \langle \z^{B} \z^{m}_{1}\rangle
\,\Box_{2}\,\langle\z^{n}_{2}\z^{n}_{2}\rangle+
4\Big\{
\langle \z^{A} \z^{m}_{1}\rangle \,\Box_{2}\,\big[
\langle \z^{B} \z^{n}_{2}\rangle\langle\z^{m}_{1}\z^{n}_{2}\rangle\big]
+\langle \z^{B} \z^{m}_{1}\rangle \,\Box_{2}\,\big[
\langle \z^{A} \z^{n}_{2}\rangle\langle\z^{m}_{1}\z^{n}_{2}\rangle\big]
\Big\}
\notag \\
&=4\,\underbrace{\langle P^{AB}P^{nn}}_{\frac{25}{6} \delta^{AB}}\rangle\,
\langle \z(t_{1}) \z_{1}\rangle  \langle \z(t_{2}) \z_{1}\rangle\ \Box_{2}
\langle\z_{2}\z_{2}\rangle\notag \\
&+4\,\underbrace{\langle P^{AB}\rangle}_{\frac{5}{6} \delta^{AB}}\,\Big\{
\langle \z(t_{1}) \z_{1}\rangle \,\Box_{2}[\langle \z(t_{2}) \z_{2}\rangle\langle\z_{1}\z_{2}\rangle]
+
\langle \z(t_{2}) \z_{1}\rangle \,\Box_{2}[\langle \z(t_{1}) \z_{2}\rangle\langle\z_{1}\z_{2}\rangle]
\Big\}.
\end{align}
Taking at the end the limit $\sigma'\to \sigma$ and denoting $\sigma^{\mu}=(t,z)$ we have 
(we write the pole and finite parts)
\begin{align}
\lambda^{3/2}\ & \ \langle\z^{A}(t_{1})  \z^{B}(t_{2})\,\bm{\z}^{2}\,\Box\bm{\z}^{2} 
\rangle = \frac{C(t,z,t_{1},t_{2})}{\eps^{2}}\notag \\
& -\frac{40\,z^{2}}{3}\Big[
\frac{\log((t-t_{1})^{2}+z^{2})}{(t-t_{2})^{2}+z^{2}}+
\frac{\log((t-t_{2})^{2}+z^{2})}{(t-t_{1})^{2}+z^{2}}\Big]+
\frac{110}{3}\,\log((t-t_{1})^{2}+z^{2})\ \log((t-t_{2})^{2}+z^{2}).
\end{align}
The last term is nothing but the old $\int\frac{dzdt}{z^{2}}GG$ term. The first two terms can be integrated
by parts. If $L_{12} = \log((t-t_{12})^{2}+z^{2})$, then (including $z^{2}$ from measure)
\be
\frac{1}{z^{2}}z^{2}\Big[
\frac{\log((t-t_{1})^{2}+z^{2})}{(t-t_{2})^{2}+z^{2}}+
\frac{\log((t-t_{2})^{2}+z^{2})}{(t-t_{1})^{2}+z^{2}}\Big] = 
(L_{1}\tfrac{1}{2z}\partial_{z}L_{2}+L_{2}\tfrac{1}{2z}\partial_{z}L_{1}) = 
\frac{1}{2z}\partial_{z}(L_{1}L_{2})\to \frac{1}{2z^{2}}L_{1}L_{2}.
\ee
Thus, with this integration by parts understood,
\begin{align}
\lambda^{3/2}\ & \ \langle\z^{A}(t_{1})  \z^{B}(t_{2})\,\bm{\z}^{2}\,\Box\bm{\z}^{2} 
\rangle_{\rm finite}=
\frac{90}{3}\,\log((t-t_{1})^{2}+z^{2})\ \log((t-t_{2})^{2}+z^{2}).
\end{align}
So, using (\ref{4.16}) and adding coupling 
we get total
\be
(-\tfrac{\sql}{2\pi})\times (-\tfrac{1}{8})\times\tfrac{1}{\lambda^{3/2}}
\tfrac{90}{3}\,\pi\,\log^{2}(t_{12}^{2})
= \frac{15}{8\,\sql}\,\log^{2}(t_{12}^{2}).
\ee
which is equal (\ref{F.0}).

\fi

\section{Neumann/Dirichlet  relations for  bulk integrals}
\la{app:dual}

Let us  provide some details  of the proof of the  relations leading to   \rf{614},(\ref{6.14}).
 
 To show  \rf{614}  let us note that   the  contribution  of the contact vertex in \rf{2.8} to the mixed correlator   involves 
 the following integral
 (here contractions are with flat  metric in $(t,z)$ space and  $\int \equiv  \int^\infty_0{dz}\int^\infty_{-\infty}{dt}$)
\begin{align}
\la{H.4}
I_{\rm N} = \tfrac{1}{4}\,I^{(1)}_{\rm N}-\tfrac{1}{2}\,I^{(2)}_{\rm N},\\ 
I^{(1)}_{\rm N} = \int \partial_{\mu}\KK_{2}(t_{1})\,\partial_{\mu}
\KK_{2}(t_{2})\,\partial_{\nu}\bG'(t_{3})\,
\partial_{\nu}\bG'(t_{4}), &\qquad  I^{(2)}_{\rm N} &=  \int\partial_{\mu}\KK_{2}(t_{1})\,\partial_{\nu}
\KK_{2}(t_{2})\,\partial_{\mu}\bG'(t_{3})\no 
\partial_{\nu}\bG'(t_{4}).
\end{align}
Denoting by $I^{(k)}_{\rm D}$  similar  integrals with  $\bG'\to \KK_{1}$, we  using  the identity in  (\ref{H1})
\begin{align}
I^{(1)}_{\rm N} = 4\,I^{(1)}_{\rm D}\ , 
\qquad 
I^{(2)}_{\rm N} =  4\,\int\partial_{\mu}\KK_{2}(t_{1})\,\partial_{\nu}
\KK_{2}(t_{2})\,\eps_{\mu\rho}\eps_{\nu\lambda}\partial_{\rho}\KK_{1}(t_{3})\,
\partial_{\lambda}\KK_{1}(t_{4})  
 = 4\,(I^{(1)}_{\rm D}-I^{(2)}_{\rm D}).\la{G6}
\end{align}
As a result, \rf{H.4} becomes 
\be
I_{\rm N} = 
-I^{(1)}_{\rm D}+2\,I^{(1)}_{\rm D} = -4\,I_{\rm D}. \la{H7}
\ee
This gives  the relation in  (\ref{614}) after dividing  by  the 
ratio of the  factors in the propagators $\mc C_{1}/\mc C_{\rm N} = -2$ (cf. \rf{2.30x},\rf{57}). 

 The contribution  of the 
 $(\partial\z)^{4}$ vertex in \rf{3.7}  to $\langle YYYY\rangle$  in the  Neumann   case  involves the integral 
 \be \la{HH5}
 J_{\rm N} = \int\bG'(t_{1}) \,\bG'(t_{2})\,  \partial_{\mu} \bG'(t_{3})\,\partial_{\mu}\bG'(t_{4})
 = 4\,
\int\bG'(t_{1}) \,\bG'(t_{2})\, 
\partial_{\mu} \KK_{1}(t_{3})\,\partial_{\mu}\KK_{1}(t_{4}) \ , \ee
where  the second equality follows  again from   \rf{6.10}.  Now using  that 
$\Box \bG'=0$ and $\Box \KK_{1}=0$  ($\Box=\del_\m\del_\m$)  and  formally integrating by parts  one finds that 
\begin{align}
 J_{\rm N} &= 2\,\int\bG'(t_{1}) \,\bG'(t_{2})\,\Box[ 
\KK_{1}(t_{3})\,\KK_{1}(t_{4})] \to  2\,\int\Box[\bG'(t_{1}) \,\bG'
(t_{2})]\, 
\KK_{1}(t_{3})\,\KK_{1}(t_{4}) \notag \\
&= 4\,\int\partial_{\mu}\bG'(t_{1}) \,\partial_{\mu}\bG'
(t_{2})\, 
\KK_{1}(t_{3})\,\KK_{1}(t_{4})  = 
16\,\int\partial_{\mu}\KK_{1}(t_{1}) \,\partial_{\mu}\KK_{1}
(t_{2})\, 
\KK_{1}(t_{3})\,\KK_{1}(t_{4}) \notag\\
& \to  16\,\int\KK_{1}(t_{1}) \,\KK_{1}
(t_{2})\, \partial_{\mu}
\KK_{1}(t_{3})\,\partial_{\mu}\KK_{1}(t_{4}) \ .  \la{G3}
\end{align}
It turns out that, in fact,  the $z=0$ boundary term  is non-zero  and  is given by $\Omega$ in \rf{6130}.
Namely,   using AdS$_2$   covariant form of the integrands  we have for the difference of \rf{HH5} and \rf{G3} 
\begin{align}
\Omega 
 = \int   ^\infty_0  {dz\ov z^2} \int^\infty_{-\infty} dt  \Big[\bG'(t_{1}) \,\bG'(t_{2})\, 
\partial^{\mu} \bG'(t_{3})\,\partial_{\mu}\bG'(t_{4}) - 
16 \,   \KK_1(t_{1})\,  \KK_1(t_{2})\,
\partial^{\mu}\KK_{1}(t_{3})\,\partial_{\mu}\KK_{1}(t_{4})\Big] \ . \la{F6}
\end{align}
 The integrand  here is a rational function of $z, t$.
The integral over $t$ can be done by computing the residues at $t=t_{a}+i\,z$ ($a=1,2,3,4$). The result
is a  rational function of $z$ (and $t_{a}$) that can be integrated over $z$  explicitly. 
Finally, we get\footnote{For  example,  choosing  $t_{1}=0$, $t_{2}=1$, $t_{3}=-1$, $t_{4}=2$, one finds 
\begin{align} \notag
\Omega &= \int_{0}^{\infty}dz\int_{-\infty}^{\infty}dt\, 
\frac{16 (t^2-t-z^2) \big[t^4-2 t^3+t^2 (2 z^2-3)-2 t (z^2-2)+z^4-13 \
z^2+4\big]}{(t^2+z^2) (t^2-4 t+z^2+4)^2 (t^2-2 t+z^2+1) (t^2+2 t+z^2+1)^2} \\
& = -16 \pi \int_{0}^{\infty}dz\, \frac{  z (256 z^8-2624 z^6-4192 z^4+812 z^2+999)}{(z^2+1)^2 \
(4 z^2+1)^2 (4 z^2+9)^3} = -\frac{8\pi}{9}. \no
\end{align}
}
\be
\la{hh2}
\Omega(t_{1}, t_{2}, t_{3}, t_{4}) = -8\pi\,
\frac{t_{13}\,t_{23}+t_{14}\,t_{24}}{t_{13}\, t_{23}\,  t_{14}\ t_{24}\, t_{34}^{2}}\ . 
\ee

\section{``Reducible''   contributions to $G_{S}$  at order  ${1\ov (\sql)^{3}}$  \la{app:other}}

Here we shall consider  the  ${ 1\ov (\sql)^{3}} $  correction  to  $G_{S}\equiv G_{{\rm N},S}$ in \rf{6.3},\rf{677}
coming from the   ``reducible''   diagrams  (tree level plus  loop  corrections to the $\z$-propagators, cf.  Fig.
 \ref{fig:dash} and Fig.  \ref{fig:111}). This  is   part of    the total $G^{(3)}_S$ in \rf{677} 
which     is the direct  analog of the  ${ 1\ov (\sql)^{2}}$   term in \rf{6.7}.

According to the definition in   (\ref{6.3}),\rf{6.4}
\begin{align}
\la{G11}
&G_{S}  = \,|t_{12}t_{34}|^{2\de}\,\langle Y^{A}(t_1) Y^{A}(t_2) Y^{B}(t_3) Y^{B}(t_4) \rangle =
1 +  \sum_{n=1}^{\infty}
\tfrac{1}{(\sql)^{n}}\,G_{S}^{(n)} \ , \\
&\langle Y^{A}(t_1) Y^{A}(t_2) Y^{B}(t_3) Y^{B}(t_4) \rangle = 1  + \sum_{n=1}^{\infty}
\tfrac{1}{(\sql)^{n}}\,Q^{(n)} \ . \la{G22}
\end{align}
At order $\frac{1}{(\sql)^{2}}$, the contributions to $\langle Y^{A}Y^{A}Y^{B}Y^{B}\rangle$
are  given   by   the sum of the expressions in (\ref{6.4}),(\ref{6.5}),(\ref{6.6}), \rf{6666} 
and after the extracting the contribution of   the prefactor   $|t_{12}t_{34}|^{-2\de}$ 
  we have   found $G_{S}  ^{(2)}$ in \rf{6.7}. 
  
 In general,  the total expression $G_{S}^{(3)} $  will be given by the sum on the ``reducible''   and  ``connected'' (bulk contact, see Fig. \ref{fig:zzzz})  diagram contributions
\be 
{G_{S}^{(3)}} = {G_{S, {\rm red }}^{(3)}} + {G_{S, {\rm conn}}^{(3)}} \ , 
\qquad \qquad    {G_{S, {\rm red }}^{(3)}} = {G_{S, {\log^2}}^{(3)}} + {G_{S, {\log^3}}^{(3)}}  \ , 
 \la{G10} 
\ee
with  the  ``reducible''    contribution being the sum of  the terms  $ {G_{S, {\log^2}}^{(3)}} $ and $ 
 {G_{S, {\log^3}}^{(3)}}  $ containing  products  of two  and three  $\log t_{ij}$  factors  respectively.
   It is the  total expression  ${G_{S}^{(3)}} $ that should be conformally invariant. 
  Our aim below will be to compute ${G_{S, {\rm red }}^{(3)}} $. 

The   ${ 1\ov (\sql)^{3}} $  contributions  to \rf{G22}  will  come from:  (i)  
 tree diagrams (given by products of  three $\z$-propagators as in Fig.  \ref{fig:111}),   and (ii) 
  diagrams  with loops  corresponding to the $\z$-propagator ``self-energy'' 
  corrections (cf. Fig. \ref{fig:loop}).\foot{One can see  that a diagram   with a bulk  fermionic loop 
  and  three   $\z$-propagators  attached to it    gives zero contribution 
    as one of  the legs will be contracted with $n_A$.}   
  The  tree diagrams  will give ${ 1\ov (\sql)^{3}}  \log^3$ terms   while the   ones with loop  corrections 
    will give also ${ 1\ov (\sql)^{3}}  \log^2 $ terms. 
    
  To find   $G_{S}  ^{(3)}$  will  then need  to multiply  the resulting expression for \rf{G22} 
  by  (see    \rf{41}) 
\begin{align}\la{G2} 
|t_{12}t_{34}|^{2\de} = 1 &+ \big[\tfrac{5}{\sql}  + \tfrac{d_2}{(\sql)^2} +\tfrac{d_3}{(\sql)^3} + ...\big] \,   (\nD_{12}+\nD_{34})
+   \big[ \tfrac{25}{2\,(\sql)^{2}}      +   \tfrac{5 d_2}{(\sql)^{3}}  + ... \big]\,    (\nD_{12}+\nD_{34})^{2}
\no \\
& +\big[\tfrac{125}{6\,(\sql)^{3}} + ... \big]\, (\nD_{12}+\nD_{34})^{3}+\dots, \qquad \qquad \nD_{ij}\equiv \log(t_{ij}^{2}) \ , 
\end{align}
and extract  the order $\tfrac{1}{(\sql)^{3}} $ term.  Multiplying 
$1  + \tfrac{1}{ \sql}\, Q^{(1)} +    \tfrac{1}{( \sql)^2 }\,Q^{(2)}
+  \tfrac{1}{( \sql)^3}\,Q^{(3)} +... $    by   \rf{G2}  and 
 using the expressions in \rf{64},\rf{6666}  gives 
\begin{align}
G_{S}^{(3)} = &{d_2} (\nD_{12}+\nD_{34}) Q^{(1)}  + 5 d_2 (\nD_{12}+\nD_{34})^{2} \no \\
&+ \tfrac{125}{6}(\nD_{12}+\nD_{34})^3 +  \tfrac{25}{2}(\nD_{12}+\nD_{34})^2  Q^{(1)}
  +  5 (\nD_{12}+\nD_{34}) Q^{(2)}  
  + Q^{(3)} \ \la{G4} \\
  =&  -5\,  d_2 (\nD_{12}+\nD_{34})^{2} +
   \tfrac{125}{6}(\nD_{12}+\nD_{34})^3  
  +  5 (\nD_{12}+\nD_{34}) \bar Q^{(2)}  
  + Q^{(3)} \ ,  \la{GG3}
\end{align}
where  we used  that $Q^{(1)}  = - 5(\nD_{12}+\nD_{34})$  in \rf{64} 
  and $Q^{(2)}  =Q^{(2)}_{\log}  +   \tfrac{25}{2}(\nD_{12}+\nD_{34})^2 + \bar Q^{(2)}  $  where 
  $Q^{(2)}_{\log}   = - d_2 (\nD_{12}+\nD_{34})$, see  \rf{777}, \rf{6666}.  Note that  the 
    $d_2$-dependent terms in  the first line  of \rf{G4} cancelled  out. 
Also,     some terms in 
$ Q^{(3)}$    which come from  disconnected diagrams  involving  ``dressed'' 12 and 34  propagators will cancel in \rf{GG3}
(like  that happened   in  $G_{S}^{(2)}$  in \rf{6666},\rf{6.7}).

As was  mentioned  at the beginning of section \ref{ND}, the $\log$   contribution  to $ Q^{(3)}$   from  the 2-loop  propagator 
correction  should  cancel against  the $\tfrac{d_3}{(\sql)^3}$ term in \rf{G2} so let us 
consider the $\log^2$  contributions to $ Q^{(3)}$. 
These may come from the  tree diagrams  with two propagators in Fig. \ref{fig:disc2}  where  one of the propagators   replaced by the 1-loop corrected one corresponding to the $d_2 \log$ term in \rf{4.1}.  The  $\log^2$ terms may  thus be obtained  by the replacement 
\be 
N_{ij} \ \to \  \big( 1 + \tfrac{ d_2}{5\sql}\big) N_{ij} \la{GG2}\ee
in  the expression for $ Q^{(2)}_0$ in \rf{6.5}.  At the same time, no  additional contributions should come from  diagrams  in Fig. \ref{fig:dash} 
as they are already  accounted for in the $\Delta$-dependent terms. 
As a result, we get  the following  $\log^2$  contribution to $ Q^{(3)}$  
\be
 Q^{(3)}_{\log^2} =  
d_2 \Big[     5  (\nD_{12}^{2}+\nD_{34}^{2})   + 10 \nD_{12} \nD_{34}   +     \big(   \nD_{13} +\nD_{24} -  \nD_{14}   - \nD_{23}  \big)^2\Big]  \ . 
\la{GG35}
\ee 
Substituting this into \rf{GG3}  we end up with   the  $\log^2$ term in $G_{S}^{(3)} $
\be \la{GG4}
 {G_{S, {\log^2}}^{(3)}} = d_2 \Big[ -4\, (\nD_{12}^{2}+\nD_{34}^{2}) +     \big(   \nD_{13} +\nD_{24} -  \nD_{14}   - \nD_{23}  \big)^2\Big]
=    d_2 \Big[ -4\, (\nD_{12}^{2}+\nD_{34}^{2}) +   4 \log^2(1-\chi) \Big]   \  . 
 \ee
 While  the  first term  here  depending separately on 12 and 34  pairs of  points  is   not conformally invariant, the second  is  -- it is, in fact,  
  the same  as in \rf{6666},\rf{6.7}.   While the non-invariant part  of \rf{GG4} should cancel in   the total  combination in \rf{G10}, 
  this invariant part will simply combine with ${G_{S}^{(2)}}$  in \rf{6.7} as 
  \be \la{G89}
  \tfrac{1}{(\sql)^2} {G_{S}^{(2)}}  +  \tfrac{1}{(\sql)^3}  {G_{S, {\log^2}}^{(3)}}    \ \to \      \big[ \tfrac{10}{(\sql)^2}    +   \tfrac{4d_2}{(\sql)^3} \big]     \log^2(1-\chi) \ . \ee
Note also that  under the  four   derivatives  over $t_i$  only the second term in  \rf{GG4} contributes, i.e.  
\begin{align}
 t^2_{12} t^2_{34} \,  \partial_{t_{1}}\del_{t_2} \del_{t_3} \partial_{t_{4}}{G_{S,{\log^2}}^{(3)}} &= 8\,d_2\,\Big(
\frac{ t^2_{12} t^2_{34} }{t_{13}^{2}\,t_{24}^{2}} + \frac{ t^2_{12} t^2_{34} }{t_{23}^{2}\,t_{14}^{2}}
\Big) =  { 8 d_2   } \Big[  \chi^2 +  {\chi^2 \ov (1-\chi)^2} \Big]      \la{G9}
\end{align}
is  conformally invariant.

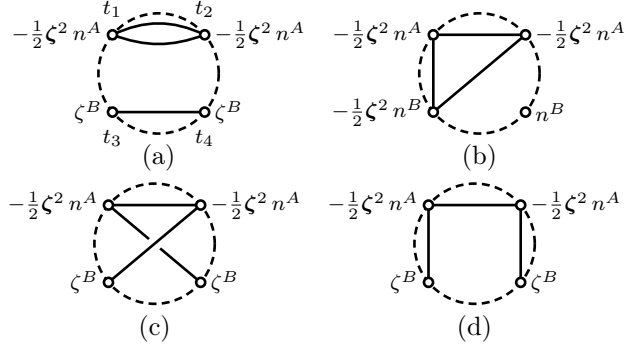
\begin{figure}[ht]
\centering
\begin{tikzpicture}[line width=1 pt, scale=0.4]
\node at (0,-2.8) {(a)};
\coordinate (A1) at (140:2);    \coordinate (A2) at (40:2);
\coordinate (A3) at (-140:2);   \coordinate (A4) at (-40:2);
\draw[densely dashed] (0,0) circle (2);
\draw (A1) to [out=25,in=155] (A2);
\draw (A1) to [out=-20,in=200] (A2);
\draw (A3) to (A4);
\node[left]    at (A1) {\footnotesize $-\frac{1}{2}\bm{\z}^{2}\,n^{A}$}; 
\node[right]  at (A2) {\footnotesize $-\frac{1}{2}\bm{\z}^{2}\,n^{A}$}; 
\node[left]    at (A3) {\footnotesize $\z^{B}$};
\node[right]  at (A4) {\footnotesize $\z^{B}$}; 
\draw[fill=white] (A1) circle (0.15); \draw[fill=white] (A2) circle (0.15);
\draw[fill=white] (A3) circle (0.15); \draw[fill=white] (A4) circle (0.15);
\node[above=3] at (A1) {\footnotesize $t_{1}$};
\node[above=3] at (A2) {\footnotesize $t_{2}$};
\node[below=3] at (A3) {\footnotesize $t_{3}$};
\node[below=3] at (A4) {\footnotesize $t_{4}$};
\end{tikzpicture}
\begin{tikzpicture}[line width=1 pt, scale=0.4]
\node at (0,-2.8) {(b)};
\coordinate (A1) at (140:2);    \coordinate (A2) at (40:2);
\coordinate (A3) at (-140:2);   \coordinate (A4) at (-40:2);
\draw[densely dashed] (0,0) circle (2);
\draw (A1) -- (A3);
\draw (A2) -- (A3);
\draw (A1) -- (A2);
\node[left]    at (A1) {\footnotesize $-\frac{1}{2}\bm{\z}^{2}\,n^{A}$}; 
\node[right]  at (A2) {\footnotesize $-\frac{1}{2}\bm{\z}^{2}\,n^{A}$}; 
\node[left]    at (A3) {\footnotesize $-\frac{1}{2}\bm{\z}^{2}\,n^{B}$};
\node[right]  at (A4) {\footnotesize $n^{B}$}; 
\draw[fill=white] (A1) circle (0.15); \draw[fill=white] (A2) circle (0.15);
\draw[fill=white] (A3) circle (0.15); \draw[fill=white] (A4) circle (0.15);
\end{tikzpicture}

\begin{tikzpicture}[line width=1 pt, scale=0.4]
\node at (0,-2.8) {(c)};
\coordinate (A1) at (140:2);    \coordinate (A2) at (40:2);
\coordinate (A3) at (-140:2);   \coordinate (A4) at (-40:2);
\draw[densely dashed] (0,0) circle (2);
\draw (A1) -- (A4); \draw[fill=white,white] (0,0) circle (0.2);
\draw (A2) -- (A3);
\draw (A1) -- (A2);
\node[left]    at (A1) {\footnotesize $-\frac{1}{2}\bm{\z}^{2}\,n^{A}$}; 
\node[right]  at (A2) {\footnotesize $-\frac{1}{2}\bm{\z}^{2}\,n^{A}$}; 
\node[left]    at (A3) {\footnotesize $\z^{B}$};
\node[right]  at (A4) {\footnotesize $\z^{B}$}; 
\draw[fill=white] (A1) circle (0.15); \draw[fill=white] (A2) circle (0.15);
\draw[fill=white] (A3) circle (0.15); \draw[fill=white] (A4) circle (0.15);
\end{tikzpicture}
\begin{tikzpicture}[line width=1 pt, scale=0.4]
\node at (0,-2.8) {(d)};
\coordinate (A1) at (140:2);    \coordinate (A2) at (40:2);
\coordinate (A3) at (-140:2);   \coordinate (A4) at (-40:2);
\draw[densely dashed] (0,0) circle (2);
\draw (A1) -- (A2);
\draw (A1) -- (A3);
\draw (A2) -- (A4);
\node[left]    at (A1) {\footnotesize $-\frac{1}{2}\bm{\z}^{2}\,n^{A}$}; 
\node[right]  at (A2) {\footnotesize $-\frac{1}{2}\bm{\z}^{2}\,n^{A}$}; 
\node[left]    at (A3) {\footnotesize $\z^{B}$};
\node[right]  at (A4) {\footnotesize $\z^{B}$}; 
\draw[fill=white] (A1) circle (0.15); \draw[fill=white] (A2) circle (0.15);
\draw[fill=white] (A3) circle (0.15); \draw[fill=white] (A4) circle (0.15);
\end{tikzpicture}
\caption{Tree-level  diagrams (and similar ones obtained by permutations) 
contributing  to  (\ref{G33}). 
}
\label{fig:disc3}
\end{figure}

Let us now turn to the more non-trivial
 $\log^3$ 
 contribution to      \rf{G22}   at    order $\frac{1}{(\sql)^{3}}$.
  One finds from 
    the tree diagrams in  Fig. \ref{fig:disc3}  (cf. \rf{6.4}--\rf{6.5}) 
\begin{align}
Q^{(3)}_{1} =&  -\tfrac{25}{2} \big( \nD_{12}^2\nD_{34} +  \nD_{12}\nD_{34}^2\big) +5 \big[
\nD_{12} ( \nD_{13} \nD_{23}  +  \nD_{14} \nD_{24})
+ \nD_{34} ( \nD_{13} \nD_{14} 
+ \nD_{23} \nD_{24})\big] \no\\
&-5\big( \nD_{12}  \nD_{14} \nD_{23}  + \nD_{34} \nD_{13} \nD_{24} ) - 5(  \nD_{12}  \nD_{13} \nD_{24} +  \nD_{34} \nD_{14} 
\nD_{23}    \big). \la{G33}
\end{align}
The first bracket   comes  from  (a) in  Fig. \ref{fig:disc3} and its  analog;  the second bracket 
  comes from  4 diagrams of type (b) (which  is same as 
Fig.~\ref{fig:111}); 
the third   bracket  comes from (c) and its analog; the fourth comes from (d)  and its analog. 

To \rf{G33}  we  should  add  also the contributions of loop diagrams,  i.e. the 
 terms  coming from   the   same diagrams as  the  lower  order terms in \rf{6.4}   where the $\z$-propagators 
are replaced  by the ones containing  ``self-energy'' corrections (see Fig. \ref{fig:loop}, Fig. \ref{fig:xxx},  Fig. \ref{fig:dash}). 
The  2-loop   corrections  to the propagator in  Fig.  \ref{fig:xxx}  
should produce  the  analog of the $\g_3^{(2)}$ term in \rf{4.11},\rf{4022}
\be\la{G61}
Q^{(3)}_{2} = -\tfrac{65}{6}(\nD_{12}^{3}+\nD_{34}^{3}) \ .
\ee
Other    $1\ov (\sql)^3$ terms  coming from 1-loop corrections to propagators in the order $1\ov (\sql)^2$ tree diagrams in Fig. \ref{fig:disc2} 
can be   generated  from $Q^{(2)}$ in \rf{6.5}  by  the substitution \rf{401} or   $\nD_{ij} \to \nD_{ij} - {2\ov \sql} \nD^2_{ij}$:
\begin{align}
 \la{G62}
 Q^{(3)} _3 =& -10  \big( \nD_{12}^3  +   \nD_{34}^3\big) - 50 (  \nD_{12}^2   \nD_{34}  +  \nD_{12}   \nD_{34}^2) 
 - 10( \nD_{13}^3 + \nD_{14}^3  + \nD_{23}^3   + \nD_{24}^3   \big) 
  \notag \\
&\ + 10 \big(  \nD_{13}^2   \nD_{14} +  \nD_{13}   \nD_{14}^2  +      \nD_{23}^2   \nD_{24}  + \nD_{23}   \nD_{24}^2
  + \nD_{13}^2 \nD_{23}   +   \nD_{13}\nD_{23}^2      + \nD_{14}^2 \nD_{24}  +   \nD_{14} \nD_{24}^2 \big)\no \\
& \     -10  \big(  \nD_{13}^2 \nD_{24} +  \nD_{13} \nD_{24}^2 +         \nD_{14}^2  \nD_{23}  +  \nD_{14}  \nD_{23}^2   \big) \ . 
 \end{align}
The total  $1\ov (\sql)^3$ term in  \rf{G22} is then given by the sum of \rf{G33},\rf{G61} and \rf{G62}
\begin{align}
Q^{(3)}_{\log^3} = & Q^{(3)}_{1}+ Q^{(3)}_{2}+Q^{(3)}_{3}= - \tfrac{125}{6} (\nD_{12} + \nD_{34})^3 + \bar Q^{(3)} \ , \la{77799} \\
\bar Q^{(3)}  =
&
5 \big[ \nD_{12} ( \nD_{13}- \nD_{14})(  \nD_{23}- \nD_{24}) 
+ \nD_{34} ( \nD_{13} - \nD_{23})    (\nD_{14} -   \nD_{24} )             \big] \no\\
 &- 10\big[  ( \nD_{13} - \nD_{14}) ( \nD_{13} + \nD_{14})   + ( \nD_{24}   -  \nD_{23})  (   \nD_{24} +\nD_{23}) \big]
 ( \nD_{13} - \nD_{14}  + \nD_{24}   - \nD_{23}) 
  \ .\la{G7}
\end{align}
To compute  the corresponding $\log^3$  term in $G_{S}^{(3)}$  we  need to  substitute this into \rf{GG3}. As a result  
 (using  \rf{64},\rf{6666})
\begin{align} 
G_{S, \log^3}^{(3)} =  
     5 (\nD_{12}+\nD_{34}) \bar Q^{(2)}  + \bar Q^{(3)} 
  = \tfrac{25}{2}  (\nD_{12}+\nD_{34})  (   \nD_{13} +\nD_{24} -  \nD_{14}   - \nD_{23} )^2    + \bar Q^{(3)}              \ . \la{88}
\end{align}
Thus   most of the terms with $\nD_{12}$  and $\nD_{34}$  cancelled   out (as expected as  they correspond
to ``factorized''   contributions of dressed propagators connecting points 12 and 34) but 
in contrast to their complete cancellation   at  order $ \tfrac{1}{(\sql)^{2}}$   in    \rf{6.7}    here the terms  linear  in 
  $\nD_{12}$  and $\nD_{34}$  survive. This  is not surprising as $\bar Q^{(3)}  $   contains
   them in the  ``irreducible''    contributions 
  of  diagrams  like (b),(c),(d) in Fig. \ref{fig:disc3} and  as they  may appear also in the product of the 
  linear term $(\nD_{12}+\nD_{34})$  in the expansion of the prefactor  and the ``irreducible''  $\bar Q^{(2)} $ 
  part of $ \tfrac{1}{(\sql)^{2}}$  term corresponding to the diagrams (c),(d),(e) in Fig. \ref{fig:disc2}. 

Like  $G_{S, \log^2}^{(3)}$  in \rf{GG4} 
 the expression for $G_{S, \log^3 }^{(3)}$  in \rf{88}  is not  conformally invariant by itself.\foot{For example,  
it  is  easy to see the absence of scale invariance: under $\nD_{ij} \to \nD_{ij}  + \ell $  the second line in \rf{G7} is invariant while the first changes by $5 \ell [  ( \nD_{13}- \nD_{14})(  \nD_{23}- \nD_{24})  + ( \nD_{13} - \nD_{23})    (\nD_{14} -   \nD_{24} )  ]$;  
the second by 
$- 20 \ell   ( \nD_{13}  + \nD_{24}  - \nD_{14}    - \nD_{23})^2 $;
 the variation of   the term with $\bar Q^{(2)} $
in \rf{88}  is  $25 \ell   (   \nD_{13} +\nD_{24} -  \nD_{14}   - \nD_{23} )^2 $ and  in total 
$\delta  G_{S, \log^3}^{(3)} =   5 \ell \Big[  ( \nD_{13}- \nD_{14})(  \nD_{23}- \nD_{24})  + ( \nD_{13} - \nD_{23})    (\nD_{14} -   \nD_{24} )  
+ (   \nD_{13} +\nD_{24} -  \nD_{14}   - \nD_{23} )^2 \Big]$.
}
The  conformal invariance should be restored  in  the total  expression \rf{G10}, i.e. after 
 adding the contribution $G_{S, \rm cont}^{(3)}$  of the contact bulk contribution 
    discussed in sections \ref{ND} and \ref{conne}.
An indication that this is   indeed what happens   is that  the  operator 
$   t_{12}^2 t^2_{34} \del_{t_1} \del_{t_2} \del_{t_3} \del_{t_4} $    applied to  $G_{S, \rm cont}^{(3)}  + G_{S, \log^3 }^{(3)} $  
gives  indeed a conformally   invariant expression depending only on $\chi$. To demonstrate this (see section \ref{ND}) 
we will  need the following expression that  follows directly   from \rf{G7},\rf{88}  (cf. \rf{G9}) 
\begin{align}
 t_{12}^2 t^2_{34}\,  \del_{t_1} \del_{t_2} \del_{t_3} \del_{t_4} \,  G_{S, \log^3}^{(3)}  = 
&\te  40  t_{12}^2 t^2_{34} \,\Big(\frac{4}{t_{12}^2 t_{14} t_{23}}+\frac{4}{t_{12} t_{14}^2 
t_{23}}-\frac{5}{t_{12}^2 t_{23} t_{24}}+\frac{5}{t_{12} t_{23}^2 
t_{24}}+\frac{4}{t_{14}^2 t_{23} t_{34}}\notag \\
&\te  -\frac{5}{t_{14}^2 t_{24} 
t_{34}}-\frac{4}{t_{14} t_{23}^2 t_{34}}-\frac{5}{t_{14} t_{24}^2 
t_{34}}-\frac{4}{t_{12} t_{14} t_{23}^2}+\frac{8}{t_{14}^2 t_{23}^2}+
\frac{5}{t_{12} t_{23} t_{24}^2}+\frac{4}{t_{14} t_{23} 
t_{34}^2}\notag \\
&\te  -\frac{5}{t_{14} t_{24} t_{34}^2}-\frac{5}{t_{12}^2 t_{14} 
t_{13}}+\frac{4}{t_{12}^2 t_{24} t_{13}}+\frac{5}{t_{23}^2 t_{34} 
t_{13}}+\frac{4}{t_{24}^2 t_{34} t_{13}}-\frac{5}{t_{12} t_{14}^2 
t_{13}}-\frac{4}{t_{12} t_{24}^2 t_{13}}\notag \\
&\te  -\frac{5}{t_{23} t_{34}^2 
t_{13}}+\frac{4}{t_{24} t_{34}^2 t_{13}}-\frac{5}{t_{12} t_{14} 
t_{13}^2}+\frac{4}{t_{12} t_{24} t_{13}^2}+\frac{5}{t_{23} t_{34} 
t_{13}^2}-\frac{4}{t_{24} t_{34} t_{13}^2}\Big)\notag \\
 &\te +  160 \Big[  \chi^2    (1   +    \log \chi  \big)  + \tfrac{\chi^2}{(1-\chi)^2} \log  \tfrac{\chi}{1-\chi }
\Big]\ . \la{G121} 
\end{align}
The   terms in the last line  are conformally invariant  
while  other non-invariant  parts of other terms will cancel against   non-invariant terms  coming from 
$G_{S, \rm cont}^{(3)}  $.

\section{Direct  computation  of   $G_T$   and $G_A$ functions at  order   $1\ov (\sql)^3$ }
\la{app:TS}

In  section \ref{conne}  we computed the function $G_{ S}^{(3)}$ by a direct diagram computation
combined with   integration of the relation \rf{6.31} or \rf{6.41} 
and obtained the final result (\ref{6599}). 
The  expressions for  $G_{ T}^{(3)}$  and  $G_{ A }^{(3)}$ in \rf{6622}  we found 
 using the  crossing  symmetry relations \rf{6300},\rf{6.30} 
 and led to (\ref{cross2}), (\ref{cross3}).
In this Appendix, we  will discuss  a direct    computation of $G_{ T}^{(3)}$  and  $G_{ A }^{(3)}$   based on the same approach as used for 
$G_{ S}^{(3)}$.  This  will provide a useful check of (\ref{cross2}), (\ref{cross3})   and is also of   technical interest.

To find the $\widehat{G}_{ T}$   function  (related to  ${G}_{ T}$ 
as in \rf{6.22}) 
 we start from  the  corresponding  combination of contractions of $\widehat G_{\rm N}^{ABCD}$  (see \rf{2.21}) 
\be\la{6394} 
\widehat{G}_{ T} = \tfrac{1}{56}\,\Big(\widehat G_{\rm N}^{ABAB}+\widehat G_{\rm N}^{ABBA}-
\tfrac{2}{5}\,\widehat G_{\rm N}^{AABB}\Big).
\ee
We have  checked that as in the  case of $G_{ S}$   in  section \ref{conne}    the  expression for the  square of the conformal Casimir operator 
 $\mathscr{D}^{2}$ 
 applied to the total (contact diagram  plus ``reducible'')  contribution to  $\widehat{G}_{ T}$ 
or  $\widehat{G}_{ A}$   is conformally invariant, i.e. 
 non-invariant parts of boundary terms  from  integrating by parts  in   bulk integrals 
 cancel against  the non-invariant parts of  ``reducible''  diagram  contributions. 

A straightforward
computation of $\partial_{t_{1}}\partial_{t_{2}}\widehat{G}_{ T}$ gives (adding   bar  as in \rf{6336},\rf{6.41} 
to indicate that we have used formal integration by parts)\footnote{Note that the expression for $\mathscr D\widehat{\widebar G}_{ T}$ is correctly invariant under
$3\leftrightarrow 4$ exchange ($\chi\to \frac{\chi}{\chi-1}$) 
 upon the 
 assumed  replacement $\log(1-\chi)\to \log|1-\chi|, \ \log \chi \to \log |\chi|$.
}
\begin{align}
\la{6.46}
t_{12}^{2}  \,\partial_{t_{1}}\partial_{t_{2}}\,\widehat{\widebar G}_{ T} = -\mathscr D\,
\widehat{\widebar G}_{ T} & =  
\tfrac{\chi^{2}}{1-\chi}-\chi\,(\chi+2)\,\log(1-\chi)+\tfrac{\chi^{4}}{(1-\chi)^{2}}\,\log\chi  
 \ . 
\end{align}
Integrating, we obtain 
\begin{align}
\la{6.47}
\widehat{\widebar{G}}_{ T} =& c_{1T}  + c_{2T} \log (1-\chi) 
 + 6\,\text{Li}_{3}(\chi)+6\,\text{Li}_{3}\big(\tfrac{\chi}{\chi-1}\big) 
  -2\,\text{Li}_{2}(\chi)\, \log(1-\chi) 
\notag \\
&-\log^{3}(1-\chi)  +\log\chi\, \log^{2}(1-\chi)   -\tfrac{\chi^{2}}{1-\chi}\log\chi - \chi \log(1-\chi) \ . 
\end{align}
Here  the first two terms are  the possible  0-mode  contribution  as, e.g.,  in \rf{6.28}. 
\iffa 
In general, the result of integration  is fixed up to the zero-mode terms 
$c_1 + c_2 \log (1-\chi)$  as, e.g.,  in \rf{6.28}. 
Here  $c_T=c_1 $  is an  integration  constant
that  reflects  the contribution of  the operator 
$Y^{\{A} Y^{B\}}$  (with $ \Delta\sim {1\ov \sql}$) in the OPE. 
At  the same time,   $ \log(1-\chi)=\chi+ ... $ zero mode contribution  is 
constrained to be zero due to the crossing relation 
analogous to (\ref{6.32}), cf. also (\ref{2.23}).  {\bf ???? this is to be omitted as this OPE constraint should apply to total result ??? 
we should   just keep 0-mode freedom 
}
\fi
A non-trivial consistency check is that  the analog of \rf{671} (cf. \rf{6.18}, \rf{6.31})  is satisfied, i.e. 
$ \mathscr{D}^{2}\,\widehat{\widebar G}_{ T}(\chi) = 4\,G_{{\rm D}, T}(\chi)\ $
where 
$G_{{\rm D}, T}(\chi)$ is given  by   \rf{2.35}. 

In the case of $\mathscr D\,\widehat{G}_{ A}$    it turns out   that we cannot  
express   the structure $V_{3}$ in (\ref{6.36})
in terms of  the  Dirichlet $\KK_n$ functions only so 
 we  go back  to  solving  the analog of \rf{6.18},\rf{6.31}, 
i.e.   $\mathscr D^2 \widehat {\widebar G}_{\rm N,A} = 4\, G_{\rm D,A}$   with  $G_{\rm D,A}$ from  \rf{2.35}.
 From the 
explicit expression for  the operator $\mathscr D$ in \rf{6.24}, one 
 can check that the solution $f$  of  the equation $\mathscr D f = g$
obeys  
\be
\la{6.49}
f'(\chi) = \frac{c_2}{1-\chi}+\frac{1}{1-\chi}\int_{0}^{\chi}\frac{d\chi'}{\chi'^{2}}\,g(\chi') \ . 
\ee
 Integrating  \rf{6.49}   for    $f= \mathscr D \widehat {\widebar G}_{\rm N}$   with $g= 4\, G_{{\rm D},A}$ or $ 4 G^{(1)}_{A} (\chi)$  from \rf{2.35}
  and including the zero-mode terms  we  get\footnote{The application of (\ref{6.49}) in the case
of $g=4\, { G}_{{ D}, S}$ or $g=4\,  { G}_{{\rm D}, T}$ requires  only elementary integrations and the result 
is precisely (\ref{6.41}) and (\ref{6.46}).} 
\begin{align}
\la{6.50}
\mathscr D\,\widehat{\widebar G}_{ A} = &
8\,  \text{Li}_2(\chi )-\tfrac{(\chi -2) (\chi ^2-2 \chi +2)\chi }{(\chi -1)^2}   \log \chi 
+ \big[3 + (1-\chi)^2   +4 \log \chi\big] \log (1-\chi ) +\tfrac{(\chi -2) \chi }{\chi -1} \notag \\
& 
+c_{1A}+c_{2A}\,\log(1-\chi)\ .
\end{align}
We can impose the last relation  in (\ref{2.23}) to show that $c_{1A}=0$
(the operator $\mathscr D$
commutes with the crossing transformation $\chi\to \frac{\chi}{\chi-1}$). 
Instead of directly integrating   \rf{6.50}    we may find  $\widehat{G}_{ A} $  order  by order in small $\chi$ expansion 
 (and again applying
also (\ref{2.23})).  This gives  the  following expression 
 depending on   the  two free  zero-mode  parameters $c_{2A}$, $c_{3A}$    
\begin{align}
&\widehat{\widebar G}_{ A} =  \chi  \big[(6-c_{2A}) \log \chi +c_{3A}\big]+\chi ^2 \Big[(3-\tfrac{c_{2A}}{2}) \log \chi -\tfrac{c_{2A}}{2}
+\tfrac{c_{3A}}{2}
+3\Big]\no  \\
& +\chi^3 
\Big[\big(\tfrac{22}{9}-\tfrac{c_{2A}}{3}\big) \log \chi -\tfrac{4 c_{2A}}{9}
+\tfrac{c_{3A}}{3}+\tfrac{37}{18}\Big]+\chi ^4 \
\Big[\big(\tfrac{13}{6}-\tfrac{c_{2A}}{4}\big) \log \chi-\tfrac{3 c_{2A}}{8}
+\tfrac{c_{3A}}{4}
+\tfrac{14}{9}\Big]+\dots\la{6146}
\end{align}
\iffa 
\begin{align}
\widehat{\widebar {G}}_{ A} =  &  \big(\tfrac{4}{9} \log \chi 
-\tfrac{11}{18}\big)\, \chi^3  +
\big(\tfrac{2}{3} \log \chi-\tfrac{25}{36}\big)\, \chi ^4 +\dots\ . \la{647}
\end{align}
\fi
The total expressions for the $\frac{1}{(\sql)^{3}}$ terms in the  functions ${G}_{ T}$  and ${ G}_{ A}$ 
are given by the sums   of the ``connected''    $\widebar G$-expressions   computed using \rf{6.22}  added to 
the analogs  of $\widetilde G_{ S}$ in \rf{671},\rf{6719}.
The explicit expressions  for the   $\frac{1}{(\sql)^{3}}$ terms   in the latter  are 
found to be
\begin{align}
\widetilde{G}_{ T} =& 96\big[ 2\, \text{Li}_3(1-\chi )+ \, \text{Li}_2(\chi )\, \log (1-\chi )\big] 
\,      -83 \log ^3(1-\chi ) +216\, \log ^3\chi \no  \\
& +354\, \log \chi \, \log ^2(1-\chi )   -324\, \log ^2\chi\, \log (1-\chi)  \la{6739}    \\
=&216\,\log^{3}\chi+192\,\zeta_R(3)+(324\,\log^{2}\chi-32\,\pi^{2})\,\chi
+(162\,\log^{2}\chi+258\,\log\chi+48-16\pi^{2})\,\chi^{2}+\cdots\ ,\notag \\
\widetilde{G}_{ A} = &
96 \,\big[  2 \text{Li}_3(1-\chi )  + 4 \, \text{Li}_3(\chi ) - 2 \, \text{Li}_2(\chi ) \log \chi +  \, \text{Li}_2(\chi )  \log (1-\chi )\big]
\notag \\
& +84\, \log ^3(1-\chi )-144\, \log\chi  \log ^2(1-\chi )    +192\, \log^2\chi\, \log (1-\chi )\notag \\
=& 192\,\zeta_R(3)+(-192\,\log^{2}\chi-192\,\log\chi+384-32\,\pi^{2})\,\chi + \dots,   \la{6767}
\end{align}
where we omitted for simplicity the $d_2$ and $d_3$ dependent   contributions coming 
from the   loop corrections to the propagators in the ``reducible'' contributions.\foot{Note that in contrast to 
$G^{(3)}_S$   the  functions $G^{(3)}_T$   and $G^{(3)}_A$  can receive (in agreement  with (\ref{cross2}),(\ref{cross3}))
 the   contributions 
proportional  to the   coefficient $d_3$  of the 2-loop   correction (cf.   Fig.\ref{fig:xxx}(b))   in the 2-point function  or $\Delta$ in \rf{41}: 
these  come from  diagrams  like in Fig. \ref{fig:dash}   where 
the two points    carry indices  other than  $A$ and $B$, i.e. from   contractions like $\langle \z^A  n^B  \z^C  n^D\rangle$, etc., 
that  do  not contribute to the prefactor  in \rf{6.2}.}

The final  expressions  for    $G^{(3)}_T$   and $G^{(3)}_A$   can be shown to be equivalent to (\ref{cross2}) and (\ref{cross3})
up to the  zero mode contributions. In particular,  the  latter account   for the fact that the small $\chi$
expansions in (\ref{cross4}),\rf{6645}  do not start at order $\mc O(\chi^{2})$, consistently with the OPE 
analysis in  section  \ref{sec:OPE}.

\section{3-point function $\langle Y\,   Y\,  [YY]  \rangle$
\la{app:cube}}

In considering the  OPE decomposition of 4-point  $Y$-scalar  correlator  in \rf{6.53}   in the T-channel \rf{6542} 
one   finds the  contribution  of the  traceless  symmetric  operator $Y^{\{A} \del^n_t Y^{B\}}$
(cf. also Appendix \ref{app:BPSanom}). 
For $n=0$ its dimension is 
$\Delta_0 = {12\ov \sql} + ...$  and  the OPE   coefficients should   be proportional to the   square of the 
coefficients in the  3-point function $ \langle Y^A(t_{1})\, Y^B(t_{2})\,  [ Y^{\{ C } Y^{D \} } ](t_3)   \rangle$
Introducing a  complex  null 6-vector $u^A$ ($u^2=0$)  we  have  $ (u \cdot Y)^2 = Y^{\{ A } Y^{B \} } u_A  u_B $ 
so that  we may consider  the  equivalent  correlator 
\begin{align}
&  \langle Y^A(t_{1})\, Y^B(t_{2})\, [{u}\cdot {Y}(t_{3})]^2\rangle  = T(t_1,t_2,t_3) \, u^{A}\,u^{B} \  , \qquad 
T(t_1,t_2,t_3) = \tfrac{c_{112}\,
}{|t_{12}|^{2\Delta-\Delta_{0}}\,|t_{23}|^{\Delta_{0}}\,|t_{13}|^{\Delta_{0}}}   \ , 
\la{J.1} \\
&\Delta = \tfrac{5}{\sql}+\dots,\qquad 
\Delta_{0} = \tfrac{12}{\sql}+\dots\ , \qquad  c_{112} = c^{(0)} _{112} +  \tfrac{1}{\sql}  c^{(1)} _{112} + ... \ . \la{J2} 
\end{align}
Its  form  is fixed   by  the  $SO(6)$   and conformal invariance. 
To order $\tfrac{1}{(\sql)^{2}}$  we find  from tree-level  contributions  with one or  two  boundary-to-boundary 
propagators  $\nD_{12}= \log (t_{12})^2  $  (cf. \rf{57}) 
\begin{align}
\la{J.4}
  T_{\rm tree} = &  
\tfrac{1}{24}\,
\Big[
1+\tfrac{1}{\sql}\,  (\nD_{12}-6 \nD_{13}-6 \nD_{23})\notag \\
&\qquad +\tfrac{1}{(\sql)^{2}}\,\Big(\tfrac{5}{2} \nD_{12}^2 -6 \nD_{13} \nD_{12}-6 \nD_{23} \nD_{12}+6 \nD_{13}^2+6 \
\nD_{23}^2+36 \nD_{13} \nD_{23}\Big)+\dots\Big].
\end{align}
1-loop  ``self-energy'' corrections (as  in Fig. \ref{fig:loop})  to the $\z$-propagator 
are taken into account (to  the leading log order) by the replacement   \rf{401}. This gives  
\begin{align}
\la{J.6}
T_{\rm tree+ loop} = &
\tfrac{1}{24}\,
\Big[
1+\tfrac{1}{\sql}\,  (\nD_{12}-6 \nD_{13}-6 \nD_{23})\notag \\
&\qquad +\tfrac{1}{(\sql)^{2}}\,\Big(
\tfrac{1}{2}\nD_{12}^2-6 \nD_{13} \nD_{12}-6 \nD_{23} \nD_{12}+18 \
\nD_{13}^2+18 \nD_{23}^2+36 \nD_{13} \nD_{23}
\Big)+\dots\Big].
\end{align}
Using that 
\begin{align}
\la{J.7}
& |t_{12}|^{2\Delta-\Delta_{0}}\,|t_{23}|^{\Delta_{0}}\,|t_{13}|^{\Delta_{0}} =  
1+\tfrac{1}{\sql}  (-\nD_{12}+6 \nD_{13}+6 \nD_{23})\notag \\
& +\tfrac{1}{(\sql)^{2}}\,\Big(
\tfrac{\nD_{12}^2}{2}-6 \nD_{13} \nD_{12}-6 \nD_{23} \nD_{12}+18 \
\nD_{13}^2+18 \nD_{23}^2+36 \nD_{13} \nD_{23}\Big)+\dots\Big],
\end{align}
we find that  $c_{112} $   does not receive $\tfrac{1}{\sql} $ and $\tfrac{1}{(\sql)^2}$    corrections
\be
c_{112} = \tfrac{1}{24}+ \OO (\tfrac{1}{(\sql)^3})\ .\la{J8}   
\ee
Notice that (\ref{J.6}) differs from (\ref{J.7}) just in the sign of the 
$1\ov \sql$ correction, i.e.   it is the inverse of the exponential expansion in (\ref{J.7}).

\newpage

\bibliography{BT-Biblio}
\bibliographystyle{JHEP}

\end{document}